\documentclass[12pt]{iopart}
\usepackage{iopams,amsthm}
\usepackage[dvips]{graphicx}
\usepackage{epsfig}

\begin{document}

\title[Quantum gravitationally decohered quantum mechanics]{Expectation values,
experimental predictions, events and entropy in quantum gravitationally
decohered quantum mechanics}

\author{Bernard S. Kay and Varqa Abyaneh}

\address{Department of Mathematics, University of York, York YO10 5DD, UK}
 
\ead{({\rm address for correspondence}) bsk2@york.ac.uk}

\begin{abstract}
We restate Kay's 1998 hypothesis which simultaneously offers an
objective definition for the entropy of a closed system, a microscopic
foundation for the Second Law, a resolution of the Information Loss (and
other) Black-Hole Puzzle(s) as well as an objective mechanism for
decoherence.  The hypothesis presupposes a conventional unitary theory
of low-energy quantum gravity and offers all this by taking the physical
density operator of a closed system to be the partial trace of its total
density operator (assumed pure) over gravity and by defining its
physical entropy to be its `matter-gravity entanglement entropy'.  We
further recall Kay's 1998 modification of non-relativistic (many-body)
quantum mechanics based on Kay's hypothesis with a Newtonian
approximation to quantum gravity.  In this modification, we find formal
expectation values for certain `observables' such as momentum-squared
and Parity are altered but those for functions of positions are
unaltered.  However, by arguing that every real measurement can 
ultimately be taken to be a position measurement,  we prove that, in
practice, it is impossible to detect any alteration at all and, in
particular, we predict no alteration for an experiment recently proposed
by Roger Penrose.  Nevertheless, Kay's modification contains no
Schr\"odinger Cat-like states, and also allows a possible `events'
interpretation which we tentatively propose and begin to explore.  We
also obtain a first result of Second-Law type for a non-relativistic
toy-model closed system and argue that similar results will apply for a
wide class of model Newtonian and post-Newtonian closed systems although
we argue that ordinary actual laboratory-sized systems can never be
treated as closed for the purpose of calculating their entropy. 
Compared with `collapse models' such as GRW, Kay's Newtonian theory does
a similar job while being free from ad hoc assumptions. 
\end{abstract}


\section{Introduction}
\label{Sect:intro}

\subsection{Background}
\label{Sect:background}

In 1998, \cite{kay1}, one of us (BSK) argued that several of the
problems and puzzles concerning decoherence and thermodynamic behaviour,
which are inherent in our current understanding of quantum physics in
general, as well as several problems and puzzles specifically related to
quantum black holes $<$\ref{Note:puzzles}$>$\footnote{The small  Roman
numerals in angle-brackets refer to the section entitled  `Notes'
(Section \ref{Sect:notes}) at the end of this article.}, would appear to
find natural resolutions if one makes the following hypothesis (which is
in three parts):  First, that for the resolution of these problems and
puzzles, a quantum gravitational setting is required; more precisely, a
low-energy quantum gravitational setting, where, `low' here refers to 
energies well below the Planck energy. Second, that low energy  quantum
gravity is a quantum theory of a conventional type with a given closed
quantum gravitational system described by a  total Hilbert space ${\cal
H}_{\hbox{\small{total}}}$ which arises as a tensor product
\begin{equation}
\label{tensorprod}
{\cal H}_{\hbox{\small{total}}}={\cal H}_{\hbox{\small{matter}}}\otimes 
{\cal H}_{\hbox{\small{gravity}}}
\end{equation}
of a matter and a gravity Hilbert space $<$\ref{Note:low}$>$, and with a total
time-evolution which is unitary, but that, while, as would usually be
assumed in a standard quantum theory, one still assumes that there is an
ever-pure time-evolving `underlying' state, modelled by a density
operator of form
\begin{equation}
\label{total}
\rho_{\hbox{\small{total}}}=|\Psi\rangle\langle\Psi|,
\end{equation}
$\Psi\in {\cal H}_{\hbox{\small{total}}}$, at each `instant of time',  one should
add the new assumption that the {\it physically relevant} density
operator is not this underlying density operator, but rather its partial
trace, $\rho_{\hbox{\small{matter}}}$, over
${\cal H}_{\hbox{\small{gravity}}}$.  Third, that the physical entropy of a
closed quantum gravitational system is to be identified with the von
Neumann entropy of $\rho_{\hbox{\small{matter}}}$,
\[
S_{\hbox{\small{physical}}}=-k\tr(\rho_{\hbox{\small{matter}}}\ln
\rho_{\hbox{\small{matter}}}).
\]
In other words, $<$\ref{Note:entangle}$>$:

\medskip

{\sl The physical entropy of a closed system is its 
matter-gravity entanglement entropy.}

\medskip

With this hypothesis, an initial underlying state of a closed quantum
gravitational system with a low degree of matter-gravity entanglement
would be expected to become more and  more entangled as time increases
and thus the physical entropy, as we have defined it above, would be
expected to increase monotonically,  thus (when the theory is applied to
a model for the universe as a whole) offering the possibility of an
objective microscopic explanation for the Second Law of Thermodynamics
and (when the theory is applied to a model closed system consisting of a
black hole sitting in an otherwise empty universe) offering a resolution
to the Information Loss Puzzle.  It offers both of these things in that,
by defining the physical entropy of $\rho_{\hbox{\small{total}}}$ to be
the von-Neumann entropy of $\rho_{\hbox{\small{matter}}}$, one
reconciles an underlying unitary time-evolution on
${\cal H}_{\hbox{\small{total}}}$ with an entropy which varies/increases in
time  $<$\ref{Note:resolutions}$>$.

A second paper, \cite{kay2} (also by BSK) also in 1998, investigated the
implications of the hypothesis of \cite{kay1} for the decoherence
of ordinary matter in the `Newtonian' -- i.e. non-relativistic,
weak-gravitational-field -- regime.  With some further assumptions, the
conclusions of this work were that, if, in ordinary non-relativistic
quantum mechanics, the centre-of-mass degree of freedom of a
uniform-mass-density ball of mass $M$ and radius $R$ would be described
by a Schr\"odinger wave function $\psi\in L^2({\mathbb R}^3)$ then the
physically relevant density operator $\rho_{\hbox{\small{matter}}}$
(from now on, where it can cause no confusion,  we will often just call
this $\rho$) is given by 
\begin{equation} 
\label{1.1} 
\rho({\bf x}, {\bf x'}) = \rho_0({\bf x}, {\bf x'})e^{-D({\bf x}, {\bf
x'})}, 
\end{equation} 
where $\rho_0({\bf x}, {\bf x}')=\psi({\bf x})\psi^*({\bf
x'})$ is the  position-space density matrix one would expect to have
were gravitationally-induced-decoherence effects to be ignored, and $D$,
which is called the \textit{decoherence exponent}, is a certain function
$<$\ref{Note:D}$>$ of
${\bf x}$ and ${\bf x'}$ which depends on ${\bf x}$ and ${\bf x}'$ only
through the square of their difference $({\bf x}-{\bf x}')^2$ and which
vanishes when ${\bf x}={\bf x'}$.   Explicitly, $D$ is given (below and
from now on, $a$ stands for $|{\bf x}-{\bf x}'|$ and we regard $D$ as a
function of $a$) by \[ D(a) = 216 M^2 \int_0^\infty
\frac{(\sin(\kappa)-\kappa \cos(\kappa))^2}{\kappa^7} \left(\frac{\kappa
a/R-\sin(\kappa a/R)}{\kappa a/R} \right)\hbox{d}\kappa, \] 
and we remark that this is a monotonically increasing
function of $a$ \cite{kay2}.

To summarize, the overall effect of the new theory is to replace the
`would-be' Schr\"odinger wave function $\psi$ (or more precisely the
would-be density operator $\rho_0=|\psi\rangle\langle\psi|$) by the
physical density operator $\rho$ of (\ref{1.1}).

Moreover, in \cite{kay2}, an asymptotic limit of $D$ was obtained for
the case where $a \ll R$,
\begin{equation}
\label{1.2}
D(a) \simeq \alpha a^2 + O(\alpha^2 a^4\ln(a/R)),
\end{equation}
where (by default we shall assume Planck units where $G=c=\hbar=k=1$) 
$\alpha = 9M^2/R^2$. We
refer to (\ref{1.2}) as the \textit{Gaussian} asymptotic regime.   

We remark that we expect this Gaussian regime to always give a good
approximation for a ball with a mass around or bigger than the Planck mass,
since then, whenever $a$ fails to be very much less than $R$, 
$e^{-D(a)}$ will anyway be very small.

We also remark that, as long as the ball in question is far from being
on the verge of collapsing to form a black hole (which we anyway expect
it to need to be for our Newtonian approximation to be valid) then the
constant $\alpha$ in the Gaussian regime (\ref{1.2}) will (in Planck
units) be very much less than one.

An asymptotic limit was also obtained in the case where $a \gg R$,
\begin{equation} \label{1.3} D(a)\simeq 24M^2\ln
\left(\frac{a}{R}\right) + O(1). \end{equation} in view of which we
refer to this limit as the \textit{logarithmic} asymptotic regime.

We shall sometimes study a model one-dimensional analogue of (\ref{1.1})
which is taken to be exactly Gaussian, so that the relation between the
physical and would-be density operator is given by
\begin{equation}
\label{Gauss}
\rho (x,x')=\rho_0 (x,x')e^{-\kappa (x-x')^2},
\end{equation}
where
\[
\rho_0 (x,x') = \psi (x) \psi^{\ast} (x').
\]
where $\psi$ is taken to be a `would-be' wavefunction on the line and
$\kappa$ is a positive constant.  We shall call this the
\textit{one-dimensional Gaussian model}.

In view of (\ref{1.2}) and the subsequent remark, we would expect that,
if we identify $\kappa$ in (\ref{Gauss}) with $\alpha=9M^2/R^2$, then,
for $M$ around or bigger than the Planck mass, this one-dimensional
Gaussian model will give a good approximate description of the relation
between the would-be wavefunction and the physical density operator
describing the centre-of-mass degree of freedom of a uniform mass bead
(i.e. a uniform mass ball with a narrow straight tube bored through its
middle) of mass $M$ and radius $R$ constrained to slide on a straight
wire, where $x$ is taken to denote the position of the centre-of-mass of
the bead  along the wire relative to some choice of origin.

We note that if such a bead is constrained to move between
hard stops, say a distance $2(R+\delta)$ apart -- so that, calling the
position of the centre of the bead when it is half-way between
the stops `the origin', any would-be wave function, $\psi(x)$ for the
centre-of-mass motion is required to vanish at $x=\pm\delta$ --
then the Gaussian model would, again in view of (\ref{1.2}), be expected
to give a good approximation for any mass $M$ provided $\delta \ll R$.

\cite{kay2} also generalized the formula (\ref{1.1}) to the many-body
wave function (for the centre of masses) of a system of many balls.  
The general formula, for a system of $N$ balls takes the
form  
\begin{equation}
\label{1.4}
\fl \rho({\bf x}_1,\dots , {\bf x}_N ; {\bf x}_1',\dots ,{\bf x}_N') = 
\rho_0({\bf x}_1,\dots ,{\bf x}_N ; {\bf x}_1',\dots ,{\bf x}_N')
e^{-D({\bf x}_1,\dots ,{\bf x}_N ; {\bf x}_1',\dots ,{\bf x}_N')},
\end{equation}
where $\rho_0(x_1,..,x_N;x_1',...,x_N')$ takes the form $\Psi
(x_1,..,x_N)\Psi^\ast(x_1',...,x_N')$ for some many--real-variable wave
function $\Psi$ and $D$ is a  certain function which depends on  ${\bf
x}_1\dots {\bf x}_N$ and ${\bf x}_1'\dots {\bf x}_N'$  only through the
distances $|{\bf x}_I-{\bf x}_J|$, $|{\bf x}_I'-{\bf x}_J|$, $|{\bf
x}_I-{\bf x}_J'|$ and $|{\bf x}_I'-{\bf x}_J'|$ ($I, J= 1\dots N$) and
which vanishes whenever ${\bf x}_I={\bf x}_I'$ for all $I= 1 \dots N$.

In much of this paper, we shall focus our attention on states involving
a single ball in 3 dimensional space or on our even simpler
one-dimensional Gaussian model but we note that in  the generalization
to the case of $N$ balls of equal masses and radii,  there is a
generalization of the Gaussian asymptotic regime in the leading term of
which $D$ takes the explicit form  $<$\ref{Note:NGaussLog}$>$
\[
D({\bf x}_1,\dots ,{\bf x}_N ; {\bf x}_1',\dots ,{\bf x}_N')=
\alpha\left({\bf x}_1+{\bf x}_2+...
+{\bf x}_N-{\bf x}_1'-{\bf x}_2'-...-{\bf x}_N'\right)^2.
\]
This leads us to define an $N$-body generalization of
the one-dimensional Gaussian model by
\[
\fl\rho(x_1,..,x_N;x_1',...,x_N')= \rho_0(x_1,..,x_N;x_1',...,x_N') \nonumber
e^{-\kappa \left(x_1+x_2+...+x_n-{x_1}'-{x_2}'-...-{x_N}'\right)^2}
\]
where again $\rho_0(x_1,..,x_N;x_1',...,x_N')$ takes the form $\Psi
(x_1,..,x_N)\Psi^\ast(x_1',...,x_N')$ for some many-real-variable
would-be wave function $\Psi$. We shall refer to this latter  for one of
the results obtained in the Appendix (Section \ref{Sect:App}).    (Note
$<$\ref{Note:NGaussLog}$>$ also includes the generalization to the
many-ball case of the logarithmic asymptotic regime.)

\cite{kay2} further proposed a Schr\"odinger picture time-evolution rule
appropriate to this Newtonian approximation,  according to which (in the
single-ball case) the time-evolving physically relevant density operator
is obtained by adopting the formula (\ref{1.1}) at each moment in time,
$t$, so that
\begin{equation}
\label{1.5}
\rho(t;{\bf x}, {\bf x}') = \rho_0(t;{\bf x}, {\bf x}')
e^{-D({\bf x}, {\bf x}')},
\end{equation}
where $\rho_0(t; {\bf x}, {\bf x}')=\psi_t({\bf x})\psi_t^*({\bf x}')$
where the would-be wave function $\psi_t$ evolves exactly as in ordinary
non-relativistic quantum mechanics, for some choice of Hamiltonian,
$H$, so that $\psi_t({\bf x})=\langle {\bf x}|e^{-iHt}|\psi_0\rangle$  
$<$\ref{Note:timev}$>$. (Similarly, in the many-ball case and the
Gaussian model etc., we adopt formula (\ref{1.4}), respectively
(\ref{Gauss}) etc. at each moment in time, again assuming that the
relevant would-be wave functions evolve in time according to some choice
of Hamiltonian which we again call $H$.) 

For a Schr\"odinger Hamiltonian of form $H=-\nabla^2/2M + V$, it may
easily be checked that this amounts to the statement that $\rho(t; {\bf
x}, {\bf x}')$ evolves according to the master equation
\begin{equation}
\label{master}
\dot\rho={i\over 2M}\left (\nabla^2-
\nabla'^2 + V({\bf x})-V({\bf x}')
\right )\rho -{1\over M}\nabla_a' D(x-x')\left (-i\nabla_a - 
i\nabla_a' \right )\rho.
\end{equation}
We remark that it is easy to see that the analogous master equation for
the one-dimensional Gaussian model can be written in the simple operator
form 
\begin{equation}
\label{gaussmaster}
\dot\rho=-i[H,\rho]-{2\kappa\over M}[x,[p,\rho]].
\end{equation}

While these master equations can be shown  \textit{not} to have
`Gorini-Kossakowski-Sudarshan/Lindblad' (cf. \cite{GKS}, 
\cite{Lindblad}) form (we shall abbreviate this elsewhere to `GKS/L
form') and even not to arise  from a time-indexed family (i.e. not
necessarily semigroup)  of `completely positive maps' acting on an
initial $\rho$ (see  $<$\ref{Note:GKS/L}$>$ for an explanation of all
this) they clearly do have solutions, specified by the formulae
(\ref{1.5}), (\ref{Gauss}),  which consist of a density operator
$<$\ref{Note:MDM}$>$ $\rho_t$ at each instant of time and (in either
case) the map from $\rho_t$ to $\rho_{t+s}$ arises in the form
$\rho_{t+s}=\mu(s)\rho_t$ where the operators (on the space of density
operators on our matter Hilbert space) $\mu(s)$ clearly form a
one-parameter group (not just semi-group).  
 
\subsection{Purpose of present paper and summary of results}
\label{Sect:purpose}

The project which gave rise to the present paper had three purposes:
Firstly to explore the prospects for experimentally testing the theory
of \cite{kay1}, secondly to explore whether/how the theory  may form a
basis for a solution to some of the conceptual problems of quantum
mechanics (i.e. of the `measurement problem'), and, thirdly and lastly,
to explore in more detail what the theory predicts about the
time-dependence of entropy.  

In the present paper, as a first step, we shall begin to explore all
three of these issues, proceeding as if, for the relevant systems, the
Newtonian approximation of \cite{kay2} applies.  It was argued in
\cite{kay2} that this assumption is unjustified for physically realistic
laboratory-sized systems, at least as far as questions of entropy are
concerned -- the problem being that, while such systems will satisfy the
necessary approximations in the sense that the relevant speeds are slow
enough and the relevant mass distributions very far away  from
black-hole collapse etc., it is \textit{not} expected (for the reasons
listed in \cite{kay2} and discussed further in Section  \ref{Sect:discussion})
to be appropriate  to treat them as \textit{closed}.  (They should
rather be treated as open systems as indicated in  Note
$<$\ref{Note:open}$>$.)

In this article, the systems we treat will nevertheless be treated as if
they \textit{were} closed.  Thus considerable caution should be
exercised in assigning a direct physical interpretation to our results. 
However, we shall argue in the later parts of our Discussion section
(Section \ref{Sect:discussion}) that, as far as our first two 
questions (experiments and
measurement problem) are concerned, the answers we obtain will, when
properly interpreted as we explain in Section \ref{Sect:expt}, survive
when we drop this assumption and that, as far as our third question
(entropy) is concerned, the results we obtain will at least survive as a
part -- albeit perhaps a small part -- of the full results for actual
(presumably partly relativistic) closed systems and also be of interest
as providing  `toy models'  for these full results. 

Proceeding then as if the relevant systems were closed and our Newtonian
theory of \cite{kay2} were exact, we amplify on each of these issues in
turn: \cite{kay2} pointed out that, as far as the mathematical formulae
are concerned, the Newtonian theory of \cite{kay2} leads to large
effects in the `amount of decoherence'  already for `small macroscopic'
systems.  Thus e.g. a ball with 10 times the Planck mass with a would-be
wavefunction for its centre-of-mass a Schr\"odinger Cat-like
superposition  \begin{equation} \label{catintro} \psi=c_1\psi_1 +
c_2\psi_2 \end{equation} (cf. (\ref{cat}) in Note $<$\ref{Note:D}$>$)
where $\psi_1$ and $\psi_2$ are sharply localized around two distinct
locations separated by a distance $a$ equal to one tenth of the ball's
radius, will, by (\ref{1.2}), have a decoherence exponent, $D(a)$, of
approximately $9$, and thus have a physical density operator very close
to that for a statistical mixture  \begin{equation} \label{decocatintro}
\rho=|c_1|^2|\psi_1\rangle\langle\psi_1|  
+|c_2|^2|\psi_2\rangle\langle\psi_2| \end{equation} of states localized
at the two locations.  More precisely $\rho$ will be as in
(\ref{decocat}) with $\langle g_1|g_2\rangle$ approximately equal to
$e^{-9}$. It thus may seem at first sight plausible that, with suitable
experimental setups, differences from the predictions of standard
quantum mechanics should be experimentally detectable in the laboratory.
Moreover, there have been several proposals (the proposals we mention
below are \cite{Folmanetal}, \cite{Marshalletal}) for such experimental
setups, albeit the theories which the authors of these setups had in
mind were somewhat different from the theory of \cite{kay1},
\cite{kay2}.  In order to explore what the prospects are for such
experimental tests, we need of course to take a view about how our
mathematical formalism relates to experiment.  To this end, we have
explored the consequences of adopting what we shall call here the
\textit{naive pragmatic interpretation} according to which the usual
sorts of `observables', e.g. (for the single ball model) position (${\bf
x}$) and momentum (${\bf p}$) and functions of these, parity ($\cal P$)
and so on, are represented by the usual self-adjoint operators and the
expectation value of the observable represented by the self-adjoint
operator $A$ obtained in the state $\rho$ is given by the usual formula 
\begin{equation}
\label{1.6}
\fl\hbox{\textit{The expectation value, $\langle A\rangle_\rho$, of 
$A$ in the state $\rho$ is equal to $\tr(\rho A)$.}}
\end{equation}  
As we report in Section \ref{Sect:naive} we find that, with this
naive pragmatic interpretation, the
expectation values of ${\bf x}$ and ${\bf p}$ and also of
arbitrary functions of ${\bf x}$ are unchanged,
\begin{equation}
\label{xp}
\tr(\rho {\bf x})=\tr(\rho_0 {\bf x}), \quad \tr(\rho {\bf
p})=\tr(\rho_0 {\bf
p}), \quad \tr(\rho f({\bf x}))=\tr(\rho_0 f({\bf x})),
\end{equation}
but that expectation values for  ${\bf p^2}$ and parity, $\cal P$, acquire extra
terms (see (\ref{psquared}) and (\ref{tracerhoP})). 
In Section \ref{Sect:modunc}, we explore the change in the naive 
expectation values for ${\bf p}^2$ further and point out that these lead
to a non-zero state-independent minimum value for the expectation value of
${\bf p}^2$ and, related to this, to a modification of the usual ${\bf
x}$--${\bf p}$ Heisenberg uncertainty relation. (Section
\ref{Sect:modunc} stands to one side of the main line of the paper. 
The results are expected to be of interest in their own right but
are not needed or referred to in subsequent sections.)
 
We shall however (i.e. in spite of the alterations in naive expectation 
values of some observables from the those of standard quantum mechanics)
argue that our naive pragmatic interpretation is too
naive.  As we shall discuss in detail in Section \ref{Sect:expt}, if one
considers an experimental setup e.g. to determine the deviation of the
expectation value of  ${\bf p}^2$ for a ball in a particular physical
state from its expectation value according to standard quantum mechanics
in the corresponding would-be wave function, one would presumably need
to introduce at least a second `probe' particle and a measurement of
${\bf p}^2$ for the original ball might then e.g. amount
to some sort of position measurement involving the probe particle. 
However, as long as our probe particle is also non-relativistic
$<$\ref{Note:photon}$>$, the two-body versions of our formulae
(\ref{xp}) are assumed to apply and one can easily see from these 
that they entail, when formula (\ref{1.6}) is applied at the two-body
level, that 

\smallskip 

\noindent
\textit{Any sort of position measurement on the relevant physical density
operator will always give the same result as that predicted, according
to standard quantum  mechanics, for the same measurement on the
corresponding would-be wave function.}  

In the case discussed above, the relevant physical density operator is
the two-body density operator replacing the time-evolute of the tensor
product of the would-be wave function for the ball and the initial
would-be wave function for the  probe particle.  We will prove a
generalization of this result for $N$-bodies in Section \ref{Sect:expt}
and we shall refer to this latter generalization as our \textit{Position
Measurement Theorem}.

One might of course think of some cleverer setup in which the
measurement on the probe particle is also not a sort of position
measurement, but we shall advocate the view, that for a measurement to
have been convincingly made, then, sufficiently far along the
Heisenberg-von Neumann chain of measurements (see Chapter 6 in
\cite{vonN}) it eventually ends up, or can be regarded as ending up, as
a position measurement -- traditionally one talks of the position of a
`pointer' on a `dial'.  We shall call this view the  \textit{corrected
pragmatic interpretation}; it is still `pragmatic' insofar as it still
applies the formula (\ref{1.6}) at {\it some} stage along the
Heisenberg-von Neumann chain. Adopting this interpretation, the general
$N$-body argument we mentioned above applies and we thus have the rather
surprising conclusion:

\medskip

\noindent
\textit{On our corrected pragmatic interpretation, the
experimental predictions of the Newtonian approximate theory of \cite{kay2} are
identical with those of standard quantum mechanics!}
$$\quad\eqno{\rm (A)}$$

\medskip

Of course we expect that, in the fully relativistic strong-field regime,
the hypothesis of \cite{kay1} will predict (large) differences but we
expect them to depend on the post-Newtonian aspects of the theory. In
Section \ref{Sect:expt} (see especially Note $<$\ref{Note:graviton}$>$) 
we give an estimate which suggests that these
post-Newtonian corrections will, however, be completely negligible for
the sort of laboratory experiments which are currently under
consideration.

Of particular interest in connection with our conclusion (A) above is
the photon-interferometer-based experiment \cite{PenroseMP2000},
\cite{PenroseLSHM} proposed by Roger Penrose to detect modifications to
the predictions of standard quantum mechanics predicted by theories of
`collapse-model' type such as that of Ghirardi-Rimini-Weber \cite{GRW}
(elsewhere in the paper, we refer to this as the `GRW' model) and 
others and, in particular, by the  gravity-induced quantum state
reduction proposal \cite{PenroseGRG} \cite{PenroseMP2000}
\cite{PenroseLSHM} of Penrose himself.  At the heart of the experiment
is a macroscopic object (crystal) attached to a movable interferometer
mirror which, by the arrangement of the experiment would get put,
according to a standard quantum-mechanics analysis, into a
superposition, one branch of which remains at rest, while the other
oscillates between two distinct locations. If (e.g. because of Penrose's
gravity-induced quantum state reduction) the crystal decoheres as a
result of visiting (i.e. in one of the branches of the superposition)
the distinct location during its oscillation cycle, a superposition of
the photon state in the two arms of the interferometer will (by the
design of the experiment -- see Section \ref{Sect:Penrose}) get
converted into a partly decohered mixture thereby changing the predicted
exit-paths of a certain fraction of the photons travelling through the
interferometer from what would be predicted by standard quantum
mechanics, and this change can then be detected with suitable photon
detectors.  

According to Kay's theory \cite{kay2} such a crystal will (if the size
is sufficient and the distinct location suitably distant) in some ways
similarly to what is predicted by Penrose's quantum state reduction, be
predicted to decohere each time it visits its distinct location and yet,
according to our conclusion (A) (and assuming that it
applies when some of the particles involved are photons
$<$\ref{Note:photon}$>$) what
is detected at the detector must be identical to what would be predicted
by standard quantum mechanics.
 
In Subsection \ref{Sect:Penrose} we explain how this difference in
predictions between the theory of Kay and the quantum state reduction of Penrose
may be traced to the following qualitative difference between their
proposed decoherence mechanisms.  On the Penrose proposal it is (at
least implicitly) assumed that once the crystal has decohered, then it
cannot later recohere.  In the Kay theory of \cite{kay2}, on the other
hand, the superposition involving the crystal is understood to decohere
when, in one of the branches of its superposition, the crystal is in its
distinct location, but to recohere each time the crystal returns to
being in the same location in both branches.

Turning to the second of our purposes, which, we recall, was to explore
how the theory of \cite{kay1}  may form a basis for a solution to some
of the conceptual problems of quantum mechanics, we begin by remarking
that while the \textit{corrected pragmatic interpretation} described
above seems the best we can do at the moment to interpret our theory and
make experimental predictions, it is still unsatisfactory in that, like
the standard interpretation of standard quantum mechanics, it makes
essential reference to `observables' and hence implicitly to `observers'
and, as e.g. John Bell eloquently argued (see e.g. \cite{Bellpolem}) 
one might hope that our current understanding of quantum mechanics will
one day be superceded by a theory which is more objective in nature. 
Indeed many authors have asked for  a modification of quantum mechanics
(or maybe for a changed interpretation of standard quantum mechanics) in
which there is an objective notion of `events' which `happen'.   For
theories which are couched in terms of a time-evolving density operator
$\rho(t)$, as in either the full hypothesis of \cite{kay1} or the
Newtonian approximate theory of \cite{kay2} being considered here (but
also `collapse models' which we mentioned above in connection with the
Penrose experiment)) a natural proposal would seem to be
$<$\ref{Note:asfarasaware}$>$ $<$\ref{Note:unravel}$>$ 
$<$\ref{Note:objprob}$>$: 

\medskip

\noindent
\textit{The set of possible events which can happen at a given time, $t$,
is to be identified with the set of spectral subspaces of $\rho(t)$
at that time.  The probability with which a given such event occurs is
to be identified with $m\lambda$ where $\lambda$ is the eigenvalue of
$\rho(t)$ belonging to that subspace and $m$ is its multiplicity (i.e.
the dimension of the subspace).}
$$\hbox{\quad}\eqno{\rm (B)}$$

As we will discuss further in Section \ref{Sect:events}, this proposed
interpretation would, for example, interpret  the state
(\ref{decocatintro}) in terms of two possible events, namely the
subspace spanned by $\psi_1$ and the subspace spanned by $\psi_2$, and
assign to these events the probabilities $|c_1|^2$, $|c_2|^2$
respectively. We shall also explore, in Section \ref{Sect:events}, what
this proposed interpretation entails for the Gaussian model
(\ref{Gauss}) and shall actually simplify this further by studying that
only to first order in $\kappa$.
 
In order to do this, we obtain, in the Appendix, the diagonalization of the
density operator of the one-dimensional Gaussian model to first order in
$\kappa$.  A typical result is that, for a given would-be wave function $\psi$
which is an even function or an odd function, the density operator 
$\rho$ in (\ref{Gauss}) takes, to first order in $\kappa$, the form
\[
\rho=\lambda_1|\phi_1\rangle\langle\phi_1|+\lambda_2|\phi_2\rangle\langle\phi_2|
\]
where $\lambda_1=1-2\kappa\langle \psi| x^2\psi\rangle$ is (for small $\kappa$)
close to 1 and $\phi_1$ (see (\ref{gaussevents}))
resembles $\psi$ while 
$\lambda_2= 2\kappa\langle\psi|x^2\psi\rangle$ is small and 
$\phi_2$ (equal to a normalization constant times $x\psi$) has the
opposite parity to $\psi$.
According to our proposal, this would be interpreted to mean that,
for such an (even or odd) would-be symmetric quantum wavefunction on the real
line, there are two events, one with probability $\lambda_1$ which is 
identified with the subspace spanned by $\phi_1$ and one with
probability $\lambda_2$  which is identified with the subspace spanned
by $\phi_2$.  

Concerning our third purpose, which is to explore what the theory of
\cite{kay1} has to say  about the time-evolution of entropy, we make a
small beginning in Section \ref{Sect:entropy} by exploring how the
von-Neumann entropy $S$ and also the entropy-like quantity
$S_1=1-\tr\rho^2$ vary in time, mainly treating our one-dimensional
Gaussian model restricted further to would-be wave functions  which are
either even or odd and to the regime where working to first order in
$\kappa$ suffices so that we may utilize the diagonalization results
obtained in the Appendix.   As is apparent from the formulae quoted
above for the eigenvalues $\lambda_1$ and $\lambda_2$, the values of $S$
and $S_1$ for this model when the system is in a given would-be state,
$\psi$, at a given time $t$, will be correlated with the value of the
standard-quantum-mechanical quantity  $\langle\psi(t)|x^2\psi(t)\rangle$ at
that time, which, in standard quantum mechanics, is of course a measure
of the `spreading of the (would-be) wave-packet'.   This leads, if, as
is the case for a free Hamiltonian, the Hamiltonian governing the
would-be dynamics leads to spreading of the wave packet at times on
either side of a time of minimum spread, to a `two-sided entropy
increase' result.  On the other hand, for a would-be Hamiltonian, say,
of harmonic-oscillator type, the would-be wavefunction will be periodic
in time, and therefore the amount of wavepacket-spreading and therefore
also the values of $S$ and $S_1$ will be periodic in time, indicating a
continual process of decoherence and recoherence.  We point out that
this is analogous to the decoherence-followed-by-recoherence mentioned
above in our discussion of the Penrose experiment. We also compare and
contrast this Hamiltonian-dependent behaviour of $S$ and $S_1$ with the
typical sort of monotonic entropy-increase result that one finds for
density operators which satisfy master equations of GKS/L form which are
usually the relevant master equations in collapse models.  Specifically,
we compare and contrast the results mentioned above which are relevant
to the master equation (\ref{gaussmaster}) with an easily-derived result
on the Hamiltonian-independent monotonic increase of entropy for the 
Barchielli-Lanz-Prosperi \cite{BLP} master equation (\ref{BLP})
\[
\dot\rho=-i[H,\rho]-c[x, [x, \rho]] 
\] 
($c$ a positive constant) which may be regarded as a prototype 
master equation of GKS/L form. (Elsewhere in this paper, we shall refer 
to this as the `BLP' equation.)

The main part of the paper ends, in section \ref{Sect:discussion}, with a
discussion of various issues arising from the earlier sections.  

We shall summarize here some of the main overall conclusions which we
draw, in Section \ref{Sect:discussion} by comparing and contrasting what
we have learned about the theory of \cite{kay2} with the properties of
`collapse models' such as GRW \cite{GRW}.  First, we conclude that, for
ordinary laboratory-sized quantum systems, the cautionary remarks
mentioned at the start of this subsection can be ignored as far as
quantum mechanical measurements are concerned and  (in view of our
Position Measurement Theorem and our `corrected pragmatic
interpretation') our theory really does have identical experimental
predictions to standard quantum mechanics (although there will be
deviations when relativistic effects are taken into account) while, at
the same time, being free from macroscopic Schr\"odinger Cat-like 
superpositions.  This is to be compared and contrasted with collapse
models which predict small deviations from the predictions of standard
(non-relativistic) quantum mechanics while also being free from
Schr\"odinger Cat-like superpositions.   Another feature of collapse
models is that they give rise to an increasing amount of decoherence as
time increases.  For our theory, we give, in Section
\ref{Sect:discussion},  plausibility arguments for a two-sided
entropy-increase result, of the sort discussed in Section
\ref{Sect:entropy} and mentioned above, for a wide variety of `generic'
model closed systems within the Newtonian framework of \cite{kay2} and
we also argue for a plausible post-Newtonian extension of such results.
Concerning the two-sidedness, we advocate, in Section
\ref{Sect:discussion}, the natural physical interpretation that the
negative-times should be discarded as physically irrelevant and the
time-zero density operator regarded as the `initial state'.  (If the
closed system is taken as modelling the universe, then this will model
the initial physically relevant density operator of the universe.)  In
this sense, such results would amount to proofs, for appropriate models,
of the Second Law.  As far as decoherence is concerned, such Second-Law
results can be interpreted as entailing  an increasing amount of
decoherence as time increases.  However, we do not interpret these
results as applying to actual ordinary (non-relativistic) systems of 
laboratory size because indeed we will argue in Section
\ref{Sect:discussion}  that, as far as questions of entropy are
concerned, actual closed systems must be partly relativistic and indeed
may well be of galaxy-size! Rather, we argue that actual small ordinary
laboratory-sized systems need to be regarded as \textit{open} systems 
(See Note $<$\ref{Note:open}$>$) and that the correct mechanism behind
their decoherence is essentially the same as that advocated on the
traditional environment-induced decoherence paradigm. (The entropy, i.e.
amount of decoherence, will then increase or decrease with time
according to the model and state etc.)   (We have omitted here a
comparison of approaches to `events' in the  two theories.  For this, we
refer to Sections \ref{Sect:events} and \ref{Sect:discussion} and especially
Note $<$\ref{Note:unravel}$>$.)  

Our overall conclusion from this comparison is that our approximate
Newtonian theory of \cite{kay2}  does a similar job to the job which
`collapse models' such as GRW \cite{GRW} do -- albeit in a slightly
different and partly unexpected and surprising way but with the
advantage that, while collapse models are ad hoc, our Newtonian
approximate theory is part and parcel of a general hypothesis (i.e. the
hypothesis of \cite{kay1})  which also resolves a number of other
puzzles.   

This paper arose out of an ongoing research project of one of the
authors (BSK).  While both authors were involved with the new results
presented in the paper, the main involvement of VA was with Sections
\ref{Sect:naive}, \ref{Sect:modunc}, and \ref{Sect:App} with Section
\ref{Sect:modunc} and Subsections \ref{Sect:manydiag} and \ref{Sect:3D}
being mainly due to VA. Some further material related to these sections
can be found in \cite{Abyaneh}.  Some parts of the rest of the paper
(including the last paragraph of Subsection \ref{Sect:purpose}, the
proposed interpretation in terms of events in Sections \ref{Sect:intro}
and \ref{Sect:events} and the \textit{Position Measurement Theorem}
discussed in Sections \ref{Sect:intro} and \ref{Sect:expt}, 
including the discussion of the Penrose experiment in Subsection
\ref{Sect:Penrose})) are mainly  due to BSK, as are Sections
\ref{Sect:entropy} and  \ref{Sect:discussion}.  

The paper concludes with an extensive Notes section (Section
\ref{Sect:notes}, due to BSK) which was needed to put the results of the
main body of the paper in context and clarify their significance.  It
both collects together some relevant background material and also
includes some new material which is expected to be of interest in its
own right.

In particular, aside from their importance as relevant background to the
new results  reported in this paper, the first paragraph of  Subsection
\ref{Sect:background} together with Notes $<$\ref{Note:puzzles}$>$,
$<$\ref{Note:tradblack}$>$, $<$\ref{Note:entangle}$>$,
$<$\ref{Note:resolutions}$>$ and $<$\ref{Note:open}$>$ will, we hope, be
of interest as constituting a  self-contained, updated (i.e. in the
light of more recent work of BSK)  re-statement of the hypothesis, which
was first made in \cite{kay1},  for a resolution of the puzzles listed
in Note $<$\ref{Note:puzzles}$>$.

\section{Naive pragmatic interpretation: trace formulae}
\label{Sect:naive}

It is easy to see from (\ref{1.1}) that, if $A$ is any function, $f({\bf
x})$, of position and also if $A$ is the momentum operator ${\bf p}$,
then $\tr(\rho A)$ is equal to $\tr(\rho_0 A)$.  To see the former,
notice that 
\[ 
\tr(\rho f({\bf x})) =\int  f({\bf x}) \rho({\bf x}, {\bf
x})d^3{\bf x}\]
and that, by virtue of $D({\bf x}, {\bf x})=0$, 
\[\rho({\bf x}, {\bf x})=\rho_0({\bf x}, {\bf x}).\]
For the latter, we combine
\[ \tr(\rho {p_a}) =\int -i{\partial\over\partial x^a}\rho({\bf x}, {\bf
x}')|_{{\bf x}={\bf x}'} d^3{\bf x'}
\]
and
\[ 
\fl -i{\partial\over\partial x^a}\rho({\bf x}, {\bf
x'})=-i\left ({\partial\over\partial x^a}\rho_0({\bf x}, {\bf
x'})\right )e^{-D({\bf x}, {\bf x'})}-2i\alpha
\rho_0({\bf x}, {\bf x'})(x^a- x'^a)e^{-D({\bf x}, {\bf x'})}\]
\[ +
\hbox{higher order terms which vanish when ${\bf x}={\bf x}'$}
\]
(here we used (\ref{1.2})) and notice that all but the first term in the
second equation will vanish when ${\bf x}={\bf x}'$.

On the other hand, one easily sees (`a' here refers to a single index
and is not summed over) from 
\[ 
\fl -{\partial^2\over{\partial x^a}^2}\rho({\bf x}, {\bf x}')=
-\left ({\partial^2\over{\partial x^a}^2}\rho_0({\bf x}, {\bf x}')\right )
e^{-D({\bf x}, {\bf x}')}-4\alpha\left ( 
{\partial\over\partial x^a}\rho_0({\bf x},
{\bf x}')\right )(x^a- x'^a)e^{-D({\bf x}, {\bf x})} 
\]
\[
+ 2\alpha\rho_0 e^{-D({\bf x}, {\bf
x})} + \hbox{higher order terms which vanish when ${\bf x}={\bf x}'$}
\]
that, for each single Cartesian component of momentum squared,
\begin{equation}
\label{compsquared}
\tr(\rho p_a^2)= \tr(\rho_0 p_a^2)+ 2\alpha,
\end{equation}
whereupon, summing over $a$, we clearly have
\begin{equation}
\label{psquared}
\tr(\rho {\bf p}^2)= \tr(\rho_0 {\bf p}^2)+ 6\alpha.
\end{equation}

It is interesting to note that all the above results remain unchanged if
one replaces the term $e^{-D({\bf x}, {\bf x}')}$  in (\ref{1.1}) by its
Gaussian asymptotic form $e^{-\alpha ({\bf x}-{\bf x}')^2}$ or even if
one replaces the latter by its `first-order in $\alpha$' form
$1-\alpha ({\bf x}-{\bf x'})^2$, albeit the resulting expression for
$\rho$ is then not to be taken seriously except when ${\bf x}$ and ${\bf
x'}$ are sufficiently nearby for $\alpha({\bf x}-{\bf x}')^2$ to be
small compared to 1. In this spirit, we write 
\[ 
\rho_{\hbox{\small{approx}}}({\bf x}, {\bf x}')=\rho_0({\bf x}, {\bf x}')
\left(1-\alpha({\bf x}-{\bf x'})^2\right ) 
\]
i.e.
\[\rho_{\hbox{\small{approx}}}=\rho_0 -\alpha x^ax^a\rho_0 + 
2\alpha x^a\rho_0 x^a -
\alpha\rho_0 x^a x^a=\rho_0 - \alpha [x^a, [x^a, \rho_0]]\]
(summed) whereupon we have 
\begin{equation}
\label{doublecom}
\fl \tr(\rho_{\hbox{\small{approx}}} A)=\tr(\rho_0 A)-\alpha \tr ([x^a, [x^a,
\rho_0]] A) =\tr (\rho_0 A) -\alpha \tr (\rho_0 [x^a, [x^a, A]])
\end{equation}
(summed) where we have used the easy standard trace identity,
$\tr(A[B,C])=\tr([A,B]C)$, etc. in the last equality.
One easily checks that this latter equation reproduces the exact
expectation values for $f({\bf x})$, ${\bf p}$ and $p_a^2$ (unsummed)
(and ${\bf p}^2$) found above.

On the `naive pragmatic interpretation' discussed in the introduction,
the equation $\tr(\rho f({\bf x}))=\tr(\rho_0 f({\bf x}))$ predicts e.g.
an identical diffraction pattern to  that of standard quantum mechanics
e.g. for a double-slit experiment  involving a ball of ordinary  matter
even as heavy as or heavier than the Planck mass (cf. the experiments of
Zeilinger et al \cite{Zeilingeretal} albeit for smaller masses).  
On the other hand, if we
consider the wavefunction for, say, the ground state of such a ball in a
spherical or cubical box etc., then the results (\ref{compsquared}),
(\ref{psquared}) would seem to predict a detectable difference in the
expectation value of the squared momentum.  

One can also calculate $\tr (\rho {\cal P})$ where $\cal P$ is the parity
operator.
\[ 
\tr(\rho {\cal P})= \int \rho_0({-\bf x}, {\bf x})e^{-D(-{\bf
x},{\bf x})} d^3{\bf x}
\]
Assuming the above `first-order in $\alpha$' approximation
to the Gaussian approximation held (this would be the case e.g. for the
centre-of-mass wavefunction of a ball with mass much bigger than the
Planck mass and radius $R$ confined to a cubical or spherical box of
diameter not much bigger than $2R$) we would have, to first order in 
$\alpha$
\begin{equation}
\label{tracerhoP} 
\tr (\rho {\cal P}) = \tr (\rho_0 {\cal P}) - \alpha \tr (\rho_0 [x^a, [x^a,
{\cal P}]])
=\tr (\rho_0 {\cal P}) -4\alpha \langle\psi|{\bf x}^2 {\cal P}\psi\rangle.
\end{equation}
We remark 
that, if the would-be wave function $\psi({\bf x})$ has even
parity, then this is 
$1 -4\alpha \langle\psi|{\bf x}^2\psi\rangle$
i.e. $1$ minus 4$\alpha$ times the squared uncertainty
in $x$ of the would-be wave function and, if 
$\psi({\bf x})$ has odd
parity, then it is 
$-1 +4\alpha \langle\psi|{\bf x}^2\psi\rangle$
i.e. $-1$ plus 4$\alpha$ times the squared uncertainty
in $x$ of the would-be wave function.  Also, the counterpart to this
result for our one-dimensional Gaussian model (\ref{Gauss}), is that,
for a would-be wave function $\psi(x)$ on the line
\begin{equation}
\label{tracerhoPline} 
\tr (\rho {\cal P}) = \tr (\rho_0 {\cal P}) - \kappa \tr (\rho_0 [x, [x,
{\cal P}]])
=\tr (\rho_0 {\cal P}) -4\kappa \langle\psi|x^2 {\cal P}\psi\rangle.
\end{equation}
This would be appropriate e.g. to the bead-on-a-wire constrained to move
between sufficiently close-together hard stops as discussed in 
Subsection \ref{Sect:background}.  Moreover, if the wave function is 
even, we'll have 
\begin{equation}
\label{tracerhoPeven}
\tr (\rho P)=
1 -4\alpha \langle\psi|x^2\psi\rangle 
\end{equation}
(and if it is odd, 
\begin{equation}
\label{tracerhoPodd}
\tr (\rho P)=
-1 +4\alpha \langle\psi|x^2\psi\rangle).
\end{equation}

We will discuss the significance of (\ref{tracerhoP}), (\ref{tracerhoPline}),
(\ref{tracerhoPeven}) and (\ref{tracerhoPodd}) further in 
Section \ref{Sect:events}.

\section{Minimum $\Delta {\bf p}^2$ and modified uncertainty relations}
\label{Sect:modunc}

(\textit{Note} The results of this section are not referred to or needed
in the remainder of the paper.  Note also that, throughout this section,
we shall not assume that a repeated index is summed over.)

Continuing to adopt the naive pragmatic interpretation, it is
interesting to note that Eq. (\ref{compsquared}) implies that there is a
minimum value (i.e. $2\alpha=18M^2/R^2$) for the expectation value 
$\langle p_a^2\rangle=\tr(\rho p_a^2)$  of  the square of each component of
the momentum in a given physical
state for the centre of mass motion of one of our balls of mass $M$ and
radius $R$.  Given that $\tr(\rho p_a)=\tr(\rho_0 p_a)$, and defining
the squared uncertainty, $(\Delta p_a)^2$, in $p_a$ in the usual way by 
\[(\Delta p_a)^2=\tr(\rho p_a^2)-\tr(\rho p_a)^2,\]
we have
\begin{equation}
\label{uncertainty}
(\Delta p_a)^2=\tr(\rho_0 p_a^2) -\tr(\rho_0 p_a)^2 +
2\alpha=(\Delta_0 p_a)^2 + 2\alpha
\end{equation}
where $\Delta_0 p_a$ is the usual uncertainty in $p_a$ of the
corresponding would-be wave function.  So the squared uncertainty in
each component of momentum also has a minimum possible value of
$2\alpha$.

To get some idea of the orders of magnitude involved, we can convert the
minimum expectation value, $6\alpha$, of ${\bf p}^2$ into a `maximum
wavelength'  (cf. the de Broglie relation `$\lambda=2\pi\hbar/|{\bf
p}|$'), so that
\[
\lambda= 2\pi\langle{\bf
p^2}\rangle^{-1/2}={2\pi\over\sqrt{6\alpha}}={2\pi R\over 3\sqrt{6}M}.
\]
In any system of units, this is
$(2\pi/3\sqrt{6})(M_{\hbox{\small{Planck}}}/M)R$. Thus, for the
centre-of-mass motion of a uniform density ball of mass around the
Planck mass (i.e. around $10^{-5}$ g) $\lambda$ will be of the same
order of magnitude as the ball's radius. For a proton, modelled as a
uniform density ball with the proton mass and a radius around $10^{-13}$
cm, $\lambda$ will be of the order of 10 kilometres!  

In view of $\tr(\rho f({\bf x}))=\tr(\rho_0 f({\bf x}))$ (and with
obvious definitions) the squared
uncertainty $(\Delta x_b)^2$ in a given physical state for a given component
of position will equal the usual squared uncertainty $(\Delta_0 x_b)^2$
in the corresponding would-be wave function.  Combining this with 
(\ref{uncertainty}) leads to the replacement of the usual 
Heisenberg uncertainty relation
\[(\Delta_0 x_a)^2(\Delta_0 p_b)^2\ge {1\over 4}\delta_{ab}\]
by
\begin{equation}
\label{modHeis}
(\Delta x_a)^2(\Delta p_b)^2 =
(\Delta_0 x_a)^2\left ((\Delta_0 p_b)^2 + 2\alpha\right) \geq
\frac{1}{4}\delta_{ab} + 2\alpha \left ( \Delta x_a \right )^2.
\end{equation}

We remark that if we were to take this relation to be fundamental then we
could recover the lower bound $\Delta p_a \ge 2\alpha$ from it.

It is interesting to notice that, in the past few years, and partly
motivated by string theory, many authors  have proposed modified
uncertainty relations along the schematic lines
\[
\Delta x\Delta p \gtrapprox \frac{1}{2}+c\left (\Delta
p \right )^2 
\]
where $c$ is a constant.  Sometimes these are thought of as applying to
spacetime coordinates but sometimes to the coordinates of a
non-relativistic particle, see e.g. \cite{moduncert}. In the latter
case, these resemble (\ref{modHeis}) -- and even more closely, in the
case $a=b$, its approximation to first order in $\alpha$: 
\[
\Delta x_a \Delta p_a \ge \frac{1}{2}+ \alpha\left (\Delta
x_a \right )^2 
\]
\textit{except that} 
the roles of $\Delta x_a$ and $\Delta p_a$ are interchanged, so that, 
e.g. in \cite{moduncert} and other related references, one deduces a lower
bound on $\Delta x$ rather than on $\Delta p$.

\section{The experimental indistinguishability from standard quantum
mechanics and the corrected pragmatic interpretation}
\label{Sect:expt}

In this section, we elucidate further and prove, the  \textit{Position
Measurement Theorem} which was outlined in the introduction, according
to which any sort of position measurement in the Newtonian approximate
theory of \cite{kay2} based on the hypothesis of \cite{kay1} will, under
the naive pragmatic interpretation, have an outcome indistinguishable
from that predicted by standard quantum mechanics.  We also amplify on
what was said in the introduction about the \textit{corrected pragmatic
interpretation} and on the conclusion that, with this interpretation,
the Newtonian approximation of \cite{kay2} is experimentally
indistinguishable from standard quantum mechanics.  We also discuss in
detail in subsection \ref{Sect:Penrose} how this works out for the class
of experiments recently proposed by Roger Penrose which we mentioned in
the Introduction.

Consider first our example (introduced in Section \ref{Sect:intro} where
we pointed out that it is
approximately described -- when $\delta \ll R$ -- by the one-dimensional
Gaussian model, also
introduced there) involving a uniform mass-density bead of mass $M$ and
radius $R$ free to slide on a straight wire which terminates at stops a distance
$2(R+\delta)$ apart.
Calling the direction of the wire `the $x$-axis'
and the position of the centre of the bead when it is half-way between the
stops `the origin', the would-be wave function, $\psi(x)$,
of the centre-of-mass degree of freedom of the bead in its ground state
will, of course, be given by
\begin{equation}
\label{beadground}
\psi(x)=\delta^{-1/2}\cos(\pi x/2\delta)
\end{equation}
and thus according to standard quantum mechanics, its momentum in the
$x$-direction will take one of the two values $\pm \pi/2\delta$ with
equal probabilities and hence the expectation value, $\langle
p^2\rangle$, will be $\pi^2/4\delta^2$, whereas, by (\ref{compsquared}),
according to our naive pragmatic interpretation, it will be
approximately 
\begin{equation}
\label{beadpsquared}
\langle p^2\rangle =\pi^2/(4\delta^2) + 18M^2/R^2.
\end{equation}

In standard quantum mechanics, one way to experimentally determine the 
expectation value of $p^2$ (or indeed to obtain the full statistical
distribution of $p$ values) of the state of such a system might be to
rigidly  attach a light flat mirror (with diameter much larger than
$\delta$) to the bead, perpendicular to the $x$ direction and, after
each time resetting the bead to be in the state under investigation,
repeatedly to reflect off the mirror a small probe particle (with mass
much less than $M$) whose quantum mechanical wavepacket is each time
prepared to be an approximately monochromatic pencil-shaped wavepacket
with diameter much bigger than $\delta$, aimed at the mirror along a
line at an angle of $-\pi/4$ to the $x$-axis (see Figure 1),
with approximate momentum $\bf P$ pointing in the direction of the
pencil and say  $<$\ref{Note:pencil}$>$ much larger in magnitude than
$\pi/(2\delta)$.  If the mirror was fixed rigidly, this pencil would of
course turn through a right-angle when it reflects off the mirror but
because the mirror is fixed to the bead, its (standard
quantum-mechanical) state will be a superposition (or more generally a
mixture of superpositions) of wave functions
$\psi_n(x)=\delta^{-1/2}\sin(n\pi x/2\delta)$ each of which, in its
turn, is of course a superposition of states with momenta $\pm p_n=\pm
n\pi/2\delta$.  Therefore (see Note $<$\ref{Note:pencil}$>$) the probe
particle would be predicted to emerge in one of the (approximate)
directions  $+\pi/4 \pm \sqrt 2 p_n/P$ with a probability depending on
the state in the usual way.  

\begin{figure}[htbp]
\centering		
\includegraphics[width=0.60\textwidth]{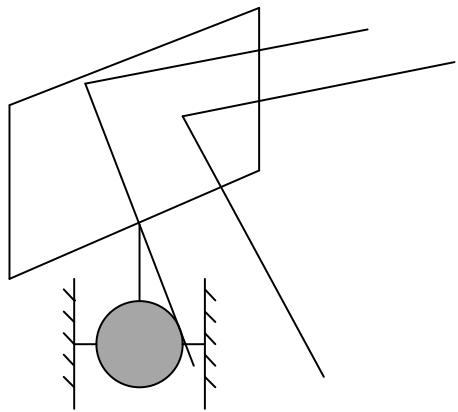}
\caption{Bead on wire with stops fitted with mirror reflecting a pencil
of light}
\end{figure}

By observing the intensity distribution of probe particles on a suitable
screen one could thus infer the probability distribution of the state of
the bead and, in particular, deduce the value of $\langle p^2\rangle$.  
In particular, if the mirror were in its ground state (\ref{beadground}), one
would have probability $1/2$ of the probe particle emerging in each of the
(approximate) directions $+\pi/4 \pm \sqrt 2 p_1/P$ (two spots on the screen
of equal intensity) and infer $<p^2>=p_1^2=\pi^2/(2\delta)^2$ 
while, in any other state, other directions would be
populated (two more widely spaced spots or more than two spots on the
screen etc.) and the value inferred
for $\langle p^2\rangle$ would be the appropriate standard quantum mechanical
value.

However, we will now argue that, on the Newtonian approximation of
\cite{kay2}, according to our naive pragmatic interpretation -- applied
now to a model closed system consisting of our
bead-with-mirror plus probe-particle system -- the
result of such an experiment will always be identical with the result
predicted by standard quantum mechanics, i.e. the result which,
according to standard quantum mechanics, would lead one to infer that
the value of $\langle p^2\rangle$ is the standard quantum mechanical
value.  Thus, in particular, if we always reset the bead to be in its
ground state, the value inferred for $\langle p^2\rangle$ will be
$\pi^2/4\delta^2$ and not the value given in  (\ref{beadpsquared}).

To see this, it is convenient to adopt a `time-dependent' approach to
the reflection of our probe particles from the mirror.  Denoting the
ground state of the bead by $\psi(x)$ and the would-be `in' wave
function for a wave-packet description of the state of one of our
probe particles at a time well before it hits the mirror by
$\phi_{\hbox{\small{in}}}({\bf y})$, so that the total  `in' wave
function  of the bead-probe-particle system is given by
\[
\Phi_{\hbox{\small{in}}}(x, {\bf y})=\psi(x)\phi_{\hbox{\small{in}}}({\bf y}),
\]
let us denote by $\Phi_{\hbox{\small{out}}}(x, {\bf y})$ the time-evolute,
according to the appropriate standard Schr\"odinger time-evolution, of
$\Phi_{\hbox{\small{in}}}$ at the time of observation of the probe particle on
the
screen and denote by $\rho^0_{\hbox{\small{in}}}$ the projector
$|\Phi_{\hbox{\small{in}}}\rangle\langle\Phi_{\hbox{\small{in}}}|$ and by
$\rho^0_{\hbox{\small{out}}}$
the projector $|\Phi_{\hbox{\small{out}}}\rangle\langle
\Phi_{\hbox{\small{out}}}|$.
Then the Newtonian approximation of \cite{kay2} tells us that the 
total physical density operators
$\rho_{\hbox{\small{in}}}$ and $\rho_{\hbox{\small{out}}}$ at the 
two times will be given by (cf. (\ref{1.4}))
\begin{equation}
\label{exptrhoin}
\rho_{\hbox{\small{in}}}(x, {\bf y}; x', {\bf y}')=
\rho^0_{\hbox{\small{in}}}(x, {\bf y}; x', {\bf y}')
\exp(-D(x, {\bf y}; x', {\bf y}'))
\end{equation}
and 
\[
\rho_{\hbox{\small{out}}}(x, {\bf y}; x', {\bf y}')=
\rho^0_{\hbox{\small{out}}}(x, {\bf y}; x', {\bf y}')
\exp(-D(x, {\bf y}; x', {\bf y}'))
\]
where, we remark, amongst its several
properties, the only relevant property of $D$ for the present argument
will be that 
\begin{equation}
\label{Ddiag}
D(x, {\bf y}; x, {\bf y})=0.
\end{equation}
In the above  thought-experiment, the result of any measurement of the
statistical distribution of probe particles on the screen will, according
to the naive pragmatic interpretation, arise from a formula of the form
$\tr(f({\bf y})\rho_{\hbox{\small{out}}})$ where $f({\bf y})$ is the same
function of the position of the probe particle that would be taken to
represent the relevant observable in standard quantum mechanics.  We have
\[
\tr(f({\bf y})\rho_{\hbox{\small{out}}})=
\int\int f({\bf y})\rho_{\hbox{\small{out}}}(x, {\bf y}; x, {\bf
y}) dx d^3 y 
\]
But in view of (\ref{exptrhoin}) and (\ref{Ddiag}) this is the same
thing as
\[
\int\int f({\bf y})\rho^0_{\hbox{\small{out}}}(x, {\bf y}; x, {\bf
y}) dx d^3 y
\]
which is of course the same result which would be predicted for the
same measurement in standard quantum mechanics.  In particular, the
statistical distribution of probe particles observed on the screen when we
perform our experiment with the mirror in its ground state will be
identical with that predicted according to standard quantum mechanics,
which is what we set out to show.

One might of course choose to do a different sort of experiment to
determine $\langle p^2\rangle$ of the state of our mirror -- perhaps
involving photons as probe particles and making use of optical devices
such as interferometers etc.   There is then, admittedly, some
uncertainty as to what the predictions of the theory in \cite{kay1}
would be because of the relativistic nature of the photon and hence the
inapplicability of the Newtonian approximation of \cite{kay2}.  But
assuming (we shall return to this below) photons can, for this purpose,
be treated as if they were non-relativistic particles (perhaps with a
tiny mass) then e.g. an experimental arrangement which results in a
detection of which of two optical channels a probe particle or photon
goes down or an experimental arrangement which results in the
examination of an interference pattern on a screen would, by these
arguments, both give the identical result to that predicted by standard
quantum mechanics, since both sorts of measurements are types of
position measurement.

We remark that similar considerations lead us to conclude that the
results of experiments along the lines of those of Folman et al
\cite{Folmanetal},  to detect changes in parity of certain quantum
mechanical states due to possible deviations from standard quantum
mechanical laws, will, on the theory of \cite{kay2} (and continuing to
assume our reasoning continues to apply when some of the particles are
photons) be identical  to the results predicted by standard quantum
mechanics, in spite of the fact that our naive pragmatic interpretation
predicts changes in the expectation value of parity as in our formula
(\ref{tracerhoP}).  The reason for this is again that the Folman et al
experiment ultimately involves a type of position measurement (again of
a photon).

In fact quite generally, for any sort of experiment, as long as the
result is obtained by a position measurement of some sort,  to the
extent that the approximation of \cite{kay2} is valid, if we apply the
theory of \cite{kay2} to the closed total system consisting of the
system of interest together with any relevant  probe particles or
`pointer variables' etc. and adopt our naive pragmatic interpretation
for this closed total system, then an adaptation of the above proof in
the case of the bead example shows that the result of our measurement
will be identical to the result predicted by standard quantum mechanics.
To see this, suppose the total system consists of $N$ bodies in all (protons,
electrons etc.) and replace the variables $(x, {\bf y})$ in the above
proof by the $N$ positions of all of these. (It may be helpful to think
of $x$ as standing for, say, $n$ `system' variables ${\bf x}_1, \dots
{\bf x}_n$ and ${\bf y}$ as standing for $m=N-n$ `pointer variables'
${\bf y}_1, \dots , {\bf y}_m$ and we shall refer to this way of
thinking in the next few parenthetical remarks.) Then the above proof
goes through, thinking now of  $\rho^0_{\hbox{\small{in}}}$ as the total
would-be density operator at the time of preparation of the experiment
and $\rho^0_{\hbox{\small{out}}}$ as its Schr\"odinger time-evolute at
the time of the measurement, $f({\bf y})$ as the appropriate position
operator (i.e. an appropriate function of all the ${\bf y}_1, \dots {\bf
y}_m$) and taking the integrations to now be over the $3N$-dimensional
total configuration space.  The key point is again that the relevant
decoherence exponent (now a function of the 2N vector variables 
${\bf x}_1, \dots ,
{\bf x}_n$; ${\bf y}_1, \dots, {\bf y}_m$; ${\bf x}_1', \dots
{\bf x}_n'$; ${\bf y}_1', \dots, {\bf y}_m'$) will vanish on the diagonal
(i.e. when ${\bf x}_a={\bf x}_a'\quad\forall a\in \lbrace 1, \dots, n\rbrace$,
${\bf y}_b={\bf y}_b'  \quad \forall b\in \lbrace 1, \dots, m\rbrace$). 
This completes the proof of our general \textit{Position Measurement
Theorem} which we referred to in the Introduction.

In our \textit{corrected pragmatic interpretation} we assume that a
realistic measurement actually consists of a whole (Heisenberg-von
Neumann) chain of measurements -- each of which is a measurement in  the
sense we have been discussing up to now -- as explained in the very
well-known discussion in Chapter 6 of von Neumann's book \cite{vonN}. 
Thus, for example, in the bead/mirror-probe-particle example with which
we opened this section, the dynamics of the bead/mirror-probe-particle
system are not the end of the story:  the measurement of the position of
the probe particle on the screen can in turn be analyzed by including
the physics behind the formation of our spots on the screen.  Let us
suppose, for example that the spots consist of black grains of metallic
silver formed when our probe particle hits a silver-iodide-coated
screen.  Then we would incorporate the screen too in the total quantum
system and include suitable terms in the total Hamiltonian to describe
the formation of the silver granules.  Then one could perform yet
another step of this sort to incorporate the physical mechanism by which
light is used to measure the position of the silver-granule-spots and
then by which the eye records the state of the light etc. etc.  However
far down such a chain one goes, our above \textit{Position Measurement
Theorem} will ensure that one will obtain the same result which is
predicted by standard quantum mechanics, provided only that, at the
stage of the chain at which one chooses to stop and apply the rule
(\ref{1.6}), the measurement is a sort of position measurement.  We
could then simply declare $<$\ref{Note:Bellink}$>$ that the only
realistic measurements are those where one does stop at a point of the
von Neumann chain which consists of a sort of position measurement. 
Alternatively and again following a well-known line of argumentation
(cf. e.g. \cite{DLP}) and cf. also the interesting recent paper
\cite{Sewell} of Geoffrey Sewell) one could argue that, whatever may be
the cause of the irreversibility, it is, in practice, only at a stage of
the chain after something macroscopic and in practice irreversible has
happened (such as the formation of our silver granules) that a
measurement has really been made $<$\ref{Note:undo}$>$ and thereafter it
is, in practice, immaterial whether the subsequent stages of the chain
are analysed with quantum or with classical physics and, in view of
that, it cannot make any difference if, insisting on still analysing
quantum mechanically things which might be analysed classically, we
decide to choose a stage of the chain in which the measurement is a sort
of position measurement -- e.g. a pointer reading on a dial.

Our \textit{corrected pragmatic interpretation} consists in adopting
either of these points of view; it doesn't matter which.  All that
matters is that, for one reason or another, we take the view that a
realistic measurement consists of a sort of position measurement at some point
of the Heisenberg-von Neumann chain. With our above \textit{Position
Measurement Theorem} we then immediately conclude (as already stated in the
introduction)

\medskip

\noindent
\textit{On our corrected pragmatic interpretation, the
experimental predictions of the Newtonian approximation of \cite{kay2} are
identical with those of standard quantum mechanics!}
$$\hbox{\quad}\eqno{\rm (A)}$$

Returning momentarily to our bead/mirror-probe-particle example, we remark
that, actually, if the probe particle is
sufficiently light (but we continue to use the non-relativistic theory
of \cite{kay2}) then $D(x,{\bf y}; x' {\bf y}')$ will be 
well-approximated by a function, say $d$, of $x$ and $x'$ only, so that
\[
\label{halfdeco}
\rho_{\hbox{\small{out}}}(x, {\bf y}; x', {\bf y}')\simeq
\rho^0_{\hbox{\small{out}}}(x, {\bf y}; x', {\bf y}')\exp(-d(x; x')).
\]
Thus, for any observable
$A$ which refers to the  probe particle so that, formally, it can be
represented by a kernel $A({\bf y}, {\bf y'})$ which is only a function
of ${\bf y}$ and ${\bf y'}$, we will have 
\[
\tr(A\rho_{\hbox{\small{out}}})=
\int\int\int A({\bf y}, {\bf y'})\rho_{\hbox{\small{out}}}(x, {\bf y'};
x, {\bf y}) dx d^3 y d^3 y' 
\]
which by (\ref{halfdeco}) is approximately equal to
\[
\int\int\int A({\bf y}, {\bf y'})\rho_{\hbox{\small{out}}}
(x, {\bf y'}; x, {\bf y}) dx d^3 y d^3 y'
\]
which, in view of $d(x,x)=0$, is equal to
\[
\int\int\int A({\bf y}, {\bf y'})\rho^0_{\hbox{\small{out}}}
(x, {\bf y'}; x, {\bf y}) dx d^3 y d^3 y'
\]
i.e. to
\[
=\tr(A\rho^0_{\hbox{\small{out}}}).
\]

This result will obviously generalize so as to conclude that, on our
naive pragmatic interpretation,  the results of \textit{any} sort of
measurement on a probe particle at some stage of a Heisenberg-von
Neumann chain of measurement, will, if the probe particle is
sufficiently light, be \textit{approximately} the same as the results
predicted by standard quantum mechanics.

However, we wouldn't assign as much importance to this result as to our
\textit{Position Measurement Theorem} because it is only approximate and also
because, in some given measurement chain, it may only apply e.g. at a
stage of the measurement chain before classicality and irreversbility
have set in.   In such a case, we would still rather rely on our
Position Measurement Theorem, applied at a later stage of the
measurement chain where the measurement is, e.g. on a macroscopic
pointer (which may well not be so light that one could make the above
approximation) in order to justify the statement (A).

As we have mentioned, all our results assume that the entire chain of
measurements involves only non-relativistic particles, whereas, in
practice (an example is the Penrose experiment we shall discuss in
detail below) photons are often involved at some stage(s) of the chain. 
This is a gap which needs to be filled.  However, we would expect that,
provided (in the lab frame) the relevant photon states are not too
energetic and provided they are sufficiently far from being on the verge
of black-hole collapse then they may be treated as if they are
non-relativistic particles with a tiny mass.  In particular, we shall
assume this to be the case for the photons in the Penrose experiment
which we discuss next.

Of particular interest to us are interferometer-like experiments along
the lines of those suggested by Penrose \cite{PenroseMP2000},
\cite{PenroseLSHM} to detect the `collapse of the wavepacket' of certain
`Schr\"odinger Cat-like' macroscopic quantum superpositions as predicted
by certain theoretical proposals for modifications of standard quantum
mechanics, notably collapse-model proposals such as that of GRW \cite{GRW}
and the proposal of Penrose himself \cite{PenroseGRG} concerning his
gravity-induced `quantum state reduction'.  In one of its simpler
versions $<$\ref{Note:felix}$>$, the setup proposed by Penrose is
reproduced here in Figure 2.  

\begin{figure}[htbp]
\centering		
\includegraphics[width=0.60\textwidth]{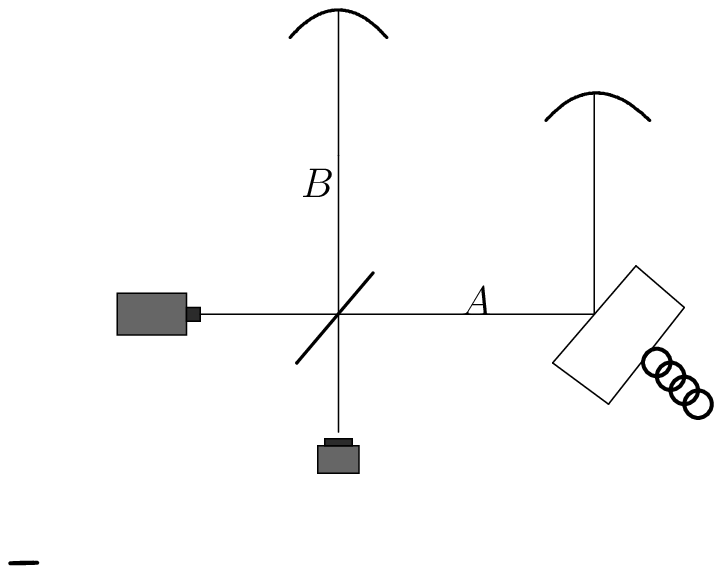}
\caption{Diagram of the Penrose experiment}
\end{figure}

As described in \cite{PenroseMP2000}, a photon, emitted from a
source,  is passed through a beam-splitter and emerges in a
superposition of two components with paths at an angle to one another,
one, say `component $B$', of which is reflected directly back towards
the beam splitter by a fixed mirror,  whereas the other, say `component
$A$', while eventually  reflected back by a second fixed mirror, also
reflects off a movable mirror of macroscopic dimensions and mass, both
on its way towards and on its way back from that fixed mirror. The
movable mirror is attached to an oscillator and (in a classical
description) initially at rest.  The mass of the movable mirror, the
spring constant of the oscillator, the frequency of the photon and the
spacing of the movable mirror and the second fixed mirror are tuned so
that (staying with a  classical description) 

\medskip

\noindent
\textit{in component `A' of the superposition (in which the photon hits the
movable mirror) its first encounter will set the movable mirror into
motion, while its second encounter will occur at the movable mirror's
initial location and bring it back to rest there -- the movable mirror
swinging to and from a macroscopically different location in between.}
$$
\hbox{ }\eqno{(\alpha)}
$$

\medskip

The positions of all the mirrors are also tuned so that, according to a
standard quantum mechanical analysis, 

\medskip

\noindent
\textit{both components of the
photon would arrive back at the beam splitter at the same time and with
relative phases such that they reassemble into a state which travels
back towards the photon source, with no photons detected at the detector
(see Figure 2) placed at the appropriate angle to that direction to catch
any photons which might have emerged from the beam-splitter in its
alternative direction.}

\medskip

This standard quantum mechanical prediction depends however on the laws
of quantum mechanics applying even though, at an intermediate stage of
the experiment (i.e. between the two reflections of component $A$ of the
photon with the movable mirror) the total quantum state of the system is
a superposition involving two macroscopically different positions of a
macroscopic object, i.e. of the movable mirror.  If a `wave-function
collapse' of this superposition occurs, e.g. as predicted by Penrose's
1996 proposal \cite{PenroseGRG}, then one would predict that, in
consequence, the coherence of the superposition of the two components of
the photon beam as they reconverge on the beam-splitter will be lost and
therefore that photons will arrive at the source and at the detector
with equal probabilities.

In this way, the Penrose experiment should be able to distinguish
between standard quantum mechanics and the gravitational-induced quantum
state reduction proposal of Penrose 1996 \cite{PenroseGRG} (or between
standard quantum mechanics and one or other collapse model).

One might at first sight expect that Kay's 1998 theory (i.e. the 
Newtonian approximate theory \cite{kay2} based on Kay's hypothesis
\cite{kay1}) would also predict a similar non-standard result for this
Penrose experiment (for suitably large crystals etc.).  After all, as we
have seen, for suitably large crystals etc., that theory too predicts 
the decoherence of superpositions involving macroscopically different
configurations.  However, our general result above tells us (assuming,
as we have discussed above, that the result still applies when the probe
particles whose position is being measured are photons) that this can't
be. Just as for any other experiment which ends in a position
measurement, the prediction must be identical with the result predicted
by standard quantum mechanics.  This difference between the predictions
of Penrose's 1996 proposal \cite{PenroseGRG} and of Kay's 1998 theory
\cite{kay2} for the Penrose experiment may be attributed to the
following difference:  In Penrose's work \cite{PenroseGRG} it is assumed
that once a macroscopic system enters a sufficiently macroscopic
superposition (in the present case involving the movable mirror in two
macroscopically different locations after the first reflection off it by
Component $A$ of the photon) then that superposition will collapse
\textit{and remain collapsed irrespective of its future dynamics}. (The
same property holds in most collapse models.) In \cite{kay2} on the
other hand, the macroscopic superposition formed after the first
reflection of component $A$ of the photon at the movable mirror will
decohere but then, at the second reflection \textit{recohere} (but we
emphasize that this is a feature of the Newtonian approximate theory
\cite{kay2}, not of the full hypothesis  of \cite{kay1} -- see the
remarks at the end of the following subsection and in Section 
\ref{Sect:discussion}).  We explain this point more fully in the
following subsection.

\subsection{Further discussion of the Penrose experiment}
\label{Sect:Penrose}
In a mathematical description of the above standard quantum mechanical
analysis, the time-evolving Schr\"odinger wave function of the 
photon/movable-mirror-cum-oscillator system may be regarded as an
element of a total matter Hilbert space
\[
{\cal H}_{\hbox{\small{matter}}}={\cal H}_{\hbox{\small{photon}}}
\otimes {\cal H}_{\hbox{\small{mirror}}}.
\]
Suppose, in a wave packet description of the photon, the photon state at
a time immediately after the photon emerges from the beam splitter takes
the form $\frac{1}{\sqrt 2}(\gamma_A(0) + \gamma_B(0))\in
{\cal H}_{\hbox{\small{photon}}}$ where $\gamma_A(0)$, $\gamma_B(0)$ represent
the $A$, $B$ components respectively of the photon superposition at that
time, and suppose that the initial state, $\mu(0)\in{\cal
H}_{\hbox{\small{mirror}}}$, of the movable-mirror-cum-oscillator is its
ground state, then we might interpret our statement ($\alpha$) by the
statement that $\gamma_A(0)\otimes \mu(0)\in  
{\cal H}_{\hbox{\small{matter}}}$ will have evolved, by an intermediate time,
$t_m$ (the suffix `m' here stands for `mid') shortly after the first
reflection of the photon off the movable mirror, say, to a state
$\gamma_A(t_m)\otimes \mu(t_m)$  $<$\ref{Note:assure}$>$  where
$\mu(t_m)$, being a quantum description of the movable mirror in its
macroscopically different location, will approximately satisfy 
$\langle\mu(t_m)|\mu(0)\rangle=0$, but that by the later time, $t_f$
(`f' here standing for `final'), when the photon has undergone its
second reflection off the movable mirror and is about to re-enter the
beam splitter in the return direction, it will have further evolved to
$\gamma_A(t_f)\otimes \mu(t_f)$ where, as the quantum counterpart to the
classical statement that the movable mirror will have returned to its
rest position, $\mu(t_f)$ will be taken to be equal to (a phase times)
the ground state $\mu(0)$.  On the other hand, Component $B$ of the
photon doesn't interact with the movable mirror at all, so the vector 
$\gamma_B(0)\otimes \mu(0)\in {\cal H}_{\hbox{\small{total}}}$ will evolve at
any time into (a phase times) $\gamma_B(t)\otimes \mu(0)$ say. 

Putting all this together, standard quantum mechanics therefore would
predict that the total state of the photon/movable-mirror-cum-oscillator
system at time $t_f$ just before (both components of) the photon
re-enter(s) the beam splitter will be $<$\ref{Note:phase}$>$
\begin{equation}
\label{photonmirror}
\Psi^{\hbox{\small{standard}}}_{\hbox{\small{matter}}}(t_f)=\frac{1}{\sqrt
2}(\gamma_A(t_f)+
\gamma_B(t_f))\otimes \mu(0)
\end{equation}
so that the partial state of the photon itself (i.e. after tracing
$|\Psi^{\hbox{\small{standard}}}_{\hbox{\small{matter}}}\rangle
\langle\Psi^{\hbox{\small{standard}}}_{\hbox{\small{matter}}}|$ over
${\cal H}_{\hbox{\small{mirror}}}$) will be the pure state with density operator
\begin{equation}
\label{photon}
\rho^{\hbox{\small{standard}}}_{\hbox{\small{photon}}}(t_f)
=\frac{1}{2}|(\gamma_A(t_f)
+\gamma_B(t_f))\rangle
\langle(\gamma_A(t_f)+\gamma_B(t_f))|
\end{equation}
and we assume that this is the photon state which passes entirely
through the beam splitter in the direction of the source, leading to zero
detection probability at the detector. 

Now, what we understand to be envisaged by Penrose \cite{PenroseMP2000},
\cite{PenroseLSHM} $<$\ref{Note:Penrose}$>$, is that, on his 1996
proposal \cite{PenroseGRG},  the initial total density operator of the 
photon/movable-mirror-cum-oscillator system
\[
\rho(0)^{\hbox{\small{Penrose}}}_{\hbox{\small{matter}}}=\frac{1}{2}|
(\gamma_A(0)+\gamma_B(0))\rangle
\langle (\gamma_A(0)+\gamma_B(0))|
\otimes |\mu(0)\rangle\langle\mu(0)|
\]
will (because of gravitational-induced quantum state reduction of the
macroscopic movable-mirror-cum-oscillator superposition according to his
proposal) evolve 
by time $t_m$ into a total density operator close to
\begin{equation}
\label{rhot1}
\fl\rho^{\hbox{\small{Penrose}}}_{\hbox{\small{matter}}}(t_m)
=\frac{1}{2}|\gamma_A(t_m)\rangle
\langle \gamma_A(t_m)|
\otimes |\mu(t_m)\rangle\langle\mu(t_m)|
+ \frac{1}{2}|\gamma_B(t_m)\rangle
\langle\gamma_B(t_m)|
\otimes |\mu(0)\rangle\langle\mu(0)|
\end{equation}
and, because it is (at least implicitly) assumed that a reduced wave
function cannot `dereduce' it is assumed that this will evolve by
$t_f$ into
\[
\rho^{\hbox{\small{Penrose}}}_{\hbox{\small{matter}}}(t_f)
=\frac{1}{2}|\gamma_A(t_f)\rangle
\langle \gamma_A(t_f)|
\otimes |\mu(0)\rangle\langle\mu(0)|
+ \frac{1}{2}|\gamma_B(t_f)\rangle
\langle\gamma_B(t_f)|
\otimes |\mu(0)\rangle\langle\mu(0)|.
\]
and therefore that the partial state of the photon itself as it
re-enters the beam-splitter will be
\[
\rho^{\hbox{\small{Penrose}}}_{\hbox{\small{photon}}}(t_m)
=\frac{1}{2}|\gamma_A(t_f)\rangle
\langle \gamma_A(t_f)|
+ \frac{1}{2}|\gamma_B(t_f)\rangle
\langle\gamma_B(t_f)|
\]
which is expected to predict equal probabilities of detecting a photon at the
source and at the detector.

On the other hand, the theory of \cite{kay1}, under the approximation of
\cite{kay2} would (see Notes $<$\ref{Note:D}$>$ and $<$\ref{Note:timev}$>$)
require the replacement of the initial density operator
$|\Psi^{\hbox{\small{standard}}}_{\hbox{\small{matter}}}\rangle\langle
\Psi^{\hbox{\small{standard}}}_{\hbox{\small{matter}}}|$,
$\Psi^{\hbox{\small{standard}}}_{\hbox{\small{matter}}}$ 
as in (\ref{photonmirror}) on
${\cal H}_{\hbox{\small{matter}}}={\cal H}_{\hbox{\small{photon}}}
\otimes {\cal H}_{\hbox{\small{mirror}}}$
by the
partial trace, $\rho^{\hbox{\small{Kay}}}_{\hbox{\small{matter}}}$ of 
$|\Phi^{\hbox{\small{Kay}}}_{\hbox{\small{total}}}\rangle\langle
\Phi^{\hbox{\small{Kay}}}_{\hbox{\small{total}}}|$
over ${\cal H}_{\hbox{\small{gravity}}}$  where 
$\Phi^{\hbox{\small{Kay}}}_{\hbox{\small{total}}}$ is
a suitable replacement for 
$\Psi^{\hbox{\small{standard}}}_{\hbox{\small{matter}}}$ on
${\cal H}_{\hbox{\small{total}}}={\cal H}_{\hbox{\small{matter}}}\otimes
{\cal H}_{\hbox{\small{gravity}}}$
($= {\cal H}_{\hbox{\small{photon}}}\otimes {\cal H}_{\hbox{\small{mirror}}} 
\otimes {\cal H}_{\hbox{\small{gravity}}}$).
In the formalism of \cite{kay2} (cf. Notes $<$\ref{Note:D}$>$ and
$<$\ref{Note:timev}$>$) one would take (assuming as above that photons
may be treated for this purpose as non-relativistic particles)
$\Psi^{\hbox{\small{Kay}}}_{\hbox{\small{total}}}(t)$ to be given by
\[
\Psi^{\hbox{\small{Kay}}}_{\hbox{\small{total}}}(t)=\frac{1}{\sqrt 
2}(\gamma_A(t)\otimes\mu(t)\otimes g(t)+
\gamma_B(t)\otimes \mu(0)\otimes g(0))
\]
where $g(t)$ is the (non-radiative) quantum counterpart to the classical
Newtonian gravitational field of a (static) mirror at the position of
the movable mirror at time $t$ when the photon is in Component `A' of
its superposition. 
$\rho^{\hbox{\small{Kay}}}_{\hbox{\small{matter}}}(t)$ will then clearly
take the following values at times $0$, $t_m$ and $t_f$:  At time zero,
it will coincide with
$\rho^{\hbox{\small{standard}}}_{\hbox{\small{matter}}}(0)$;  at time
$t_m$ it will coincide with
$\rho^{\hbox{\small{Penrose}}}_{\hbox{\small{matter}}}(t_m)$, but at
time $t_f$, it will again coincide with  
$\rho^{\hbox{\small{standard}}}_{\hbox{\small{matter}}}(t_f)
=|\Psi^{\hbox{\small{standard}}}_{\hbox{\small{matter}}}(t_f)
\rangle\langle\Psi^{\hbox{\small{standard}}}_{\hbox{\small{matter}}}(t_f)|$
(see (\ref{photonmirror})).  In other words, between times $t_m$ and
$t_f$,  $\rho^{\hbox{\small{Kay}}}_{\hbox{\small{matter}}}$ will
`recohere' and in fact, as we know it must, will therefore result in 
identical experimental results to those predicted by standard quantum
mechanics $<$\ref{Note:allthree}$>$.

The reason for this recoherence is (cf. Note $<$\ref{Note:timev}$>$ and
also the last paragraph in Section \ref{Sect:entropy}) because
the gravitational state-vector $g(t)$ depends only on the state of the
movable mirror at the time $t$ and
not on its history and therefore, since (reverting to the classical
description which is relevant here) the movable mirror returns to its
initial position at time $t_f$, we will have $g(t_f)=g(0)$
$<$\ref{Note:furtherinsight}$>$.   Of
course, in a fully relativistic treatment, this is expected no longer to
be the case and in particular, if the motion of the movable mirror between times
$0$ and $t_f$ causes the emission of one or more `gravitons' then one
would expect  $\langle g(t_f)|g(0)\rangle=0$ (approximately) in
consequence of which the partial density operator $\rho(t_f)$ of the
photon at time $t_f$ (and hence the experimental prediction) will
coincide with $\rho^{\hbox{\small{Penrose}}}_{\hbox{\small{matter}}}(t_f)$. 
However, an order of magnitude estimate shows that,
with typical experimental values for the mass and size and frequency of
the movable mirror, the amplitude to emit a graviton will be tiny on
time-scales relevant to the experiment $<$\ref{Note:graviton}$>$.

\section{A tentative interpretation in terms of events}
\label{Sect:events}

As discussed in the Introduction around the passage labelled (B) and
discussed further in Notes $<$\ref{Note:asfarasaware}$>$,
$<$\ref{Note:unravel}$>$, $<$\ref{Note:objprob}$>$, given any candidate
for a fundamental alternative theory to standard quantum mechanics which
is couched in terms of a time-evolving density operator, $\rho(t)$, it
would seem natural to attempt an interpretation in terms of `events'
which `happen', by identifying the set of events which can happen at a
given time $t$ with the set of spectral subspaces of $\rho(t)$ and
identifying the probability with which a given such event happens with
the associated eigenvalue, $\lambda$, multiplied by its multiplicity,
$m$.  Thus, in particular, it is interesting to explore the prospects
for such an interpretation in the case of the hypothesis of \cite{kay1}
and one might hope that it would have a more fundamental status than
either of our `naive' or `corrected' pragmatic interpretations (i.e. of
the Newtonian approximate theory of \cite{kay2}).  From now on (and in
Note $<$\ref{Note:beable}$>$) we shall refer to it as our `events'
interpretation.  To make a small beginning in this, we shall begin to 
explore this question here in the context of the Newtonian approximation
of \cite{kay2}.   The caveats mentioned at the beginning of Section
\ref{Sect:purpose} should of course continue to be borne in mind. (They will
be discussed further Section \ref{Sect:discussion}). 

As we have already indicated in the Introduction and in the notes
mentioned above,  in the case of a would-be wave function for the centre
of mass of a single ball in a Schr\"odinger Cat-like state
\[
\psi=c_1\psi_1 + c_2\psi_2
\]
where $\psi_1$ and $\psi_2$ are sharply localized around two different
locations, provided the mass of the ball is much larger than the Planck mass
and the distance between the two locations a big enough fraction of the
ball radius, the physical density
operator will be well-approximated  by (\ref{decocatintro})
\[
\rho=|c_1|^2|\psi_1\rangle\langle\psi_1|  
+|c_2|^2|\psi_2\rangle\langle\psi_2|.
\]
Our events interpretation then tells us
that (except when $|c_1|^2=1/2=|c_2|^2$) 
two events are possible, one corresponding to the subspace spanned
by $\psi_1$, and one corresponding to the subspace spanned by $\psi_2$. 
Further, these events will occur with probabilities $|c_1|^2$, $|c_2|^2$
respectively.  This is arguably just what one would hope for from a
resolution to the Schr\"odinger Cat puzzle -- see Note
$<$\ref{Note:catpuzzle}$>$ for further discussion.

In the remainder of this section, we shall mainly discuss how our events
interpretation works out for a submodel of the one-dimensional Gaussian
model which was introduced in Section \ref{Sect:intro}, postponing
discussion of more general models to Section \ref{Sect:discussion}.  We
recall from Section \ref{Sect:intro} that in this model the relationship
between the would-be wave function, $\psi$ (on the line) and the
physical density operator, $\rho$, is given by \begin{equation}
\label{Gaussbis}
\rho (x,x')=\psi(x)\psi^*(x')e^{-\kappa(x-x')^2}
\end{equation}
and that the model is expected to give a good approximation to the
physical density operator for the centre-of-mass
motion of a uniform-mass-density bead of mass $M$ and radius $R$
constrained to move along a straight wire provided $M$ is around or bigger than
the Planck mass, or, in the case the wire terminates at
hard stops a distance $2(R+\delta)$ apart, for any
mass provided $\delta \ll R$.

We shall restrict ourselves further to would-be wave functions $\psi$ in
this model which are either even or odd functions of $x$.  As we show in
the Appendix, using perturbation-theoretic methods, it is very easy to 
explicitly diagonalize any $\rho$ in 
(\ref{Gaussbis}) for a $\psi$ which
satisfies this further
restriction, to first order in $\kappa$, and one finds, to this
order, that there are only two non-zero eigenvalues (each non-degenerate), 
\begin{equation}
\label{gaussprobs}
\lambda_1=1-2\kappa\langle x^2\rangle , \quad 
\lambda_2= 2\kappa\langle x^2\rangle
\end{equation}
with corresponding eigenfunctions
\begin{equation}
\label{gaussevents}
\phi_1=c_1((1+\kappa\langle x^2\rangle)\psi -\kappa x^2\psi, \quad
\phi_2=c_2 x\psi.
\end{equation}
where $c_1$ and $c_2$ are normalization constants.  Here and below,
$\langle x^2\rangle$ stands for the expectation value $\langle\psi|
x^2\psi\rangle$ of $x^2$ in the would-be state $\psi$. 

As long as $\kappa\langle x^2\rangle$ is very much less than 1, 
\begin{equation}
\label{smallpert}
\kappa\langle x^2\rangle \quad (=\kappa\langle\psi|
x^2\psi\rangle)\quad \ll 1,
\end{equation}
the eigenvalues (\ref{gaussprobs}) will be close to 1 and zero and therefore
one expects this first-order perturbation theory result to give a good
approximation to the true spectrum.

We conclude that, if the would-be wave function, $\psi$, satisfies
$\langle\psi|x^2\psi\rangle \ll 1$ and is either an even function or an odd
function of $x$ then, to a good approximation,
the possible events, according to our events interpretation, will be
the one-dimensional subspaces spanned by $\phi_1$ and $\phi_2$ in 
(\ref{gaussevents}) and their respective probabilities will be $\lambda_1$
and $\lambda_2$ in (\ref{gaussprobs}). 
To illustrate this, we have sketched, in Figure 3, 
$\phi_1$ and $\phi_2$ when the would-be wave function is 
the ground state (\ref{beadground})
\[
\psi(x)=\delta^{-1/2}\cos(\pi x/2\delta)
\]
of our bead-on-a-wire example (see Section \ref{Sect:intro}  after
equation (\ref{Gauss}) and equation (\ref{beadground}) in Section
\ref{Sect:expt} and below) with stops at $\pm(R+\delta)$.

\begin{figure}[htbp]
\centering		
\includegraphics[width=0.60\textwidth]{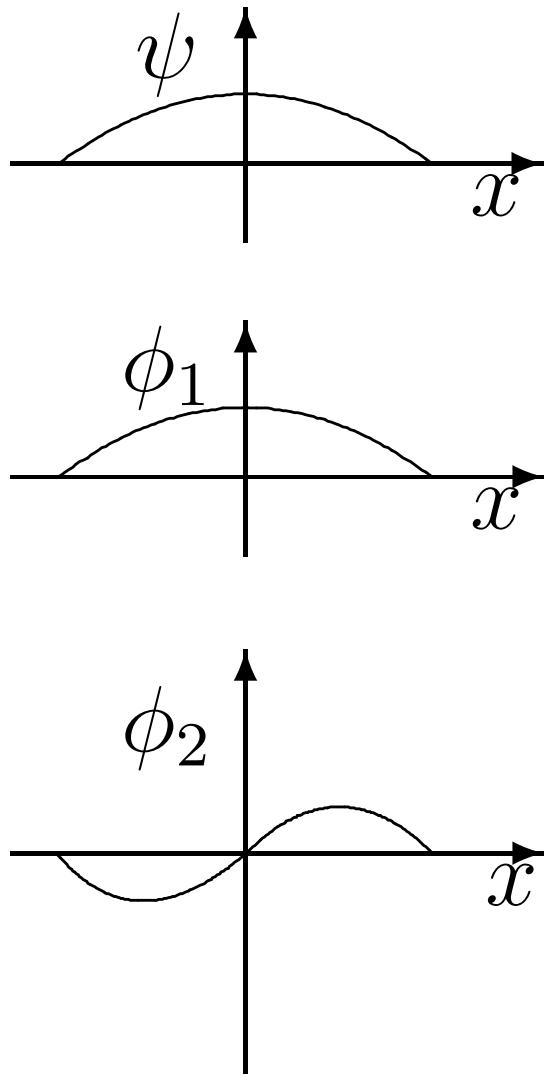}
\caption{A typical wave function for the one-dimensional Gaussian model
and its two `events'}
\end{figure}

In fact, as explained in Section \ref{Sect:intro}, as long as
$\delta \ll R$ then 
the Gaussian model will give a good approximation for the physical
density operator, $\rho$, of this
bead-on-a-wire-with-stops example whatever the mass,
while, in view of (\ref{smallwidth}) and, recalling that $\kappa=9M^2/R^2$, 
we expect the above first-order perturbation theory
results to be applicable provided $\delta \ll R/3M$ which is of course not an
additional restriction as long as $M$ is around or smaller than the Planck mass.

On the other hand, we note that for the application to our bead moving along an
infinite straight wire, then the 
condition (\ref{smallpert}) amounts, in view of $\kappa=9M^2/R^2$, to
the condition 
\begin{equation}
\label{smallwidth}
\langle\psi|x^2\psi\rangle^{1\over 2} \ll R/3M.
\end{equation}
so, since, as we saw in Section \ref{Sect:intro}, we need to assume $M$ to be 
around or greater than 1 for the
Gaussian model to be a good approximation at all, we can trust our above
first-order perturbation theory result only when the width of the
would-be wave function for the centre-of-mass of the bead is very much less than
the bead's radius.

In any case (i.e. whether or not the one-dimensional Gaussian model is a
good approximation for some bead-on-a-wire system) it is interesting to
notice (see Figure 3) that (to first order in $\kappa$) if we take, say,
an even would-be wave function $\psi$, then $\phi_1$ is even and
approximately  resembles $\psi$ and occurs with a probability,
$\lambda_1$, close to 1, but there is a small but non-zero probability,
$\lambda_2$, for $\phi_2$ to occur and this is a quite different
function from (in fact it is, of course, orthogonal to) $\psi$ and it is odd! 
Moreover, the expectation value of parity $\langle {\cal P}\rangle$, by which
we now mean the ordinary appropriately weighted sum of probabilities 
$(+1)\lambda_1+(-1)\lambda_2$ is then, by (\ref{gaussprobs}), given by
\begin{equation}
\label{paritycoincidence}
\langle {\cal P}\rangle=1 
-4\kappa \langle\psi|x^2\psi\rangle.
\end{equation}
and coincides with the formula (\ref{tracerhoPeven}) for the expectation
value of parity (albeit defined there, differently, to be $\tr (\rho
{\cal P})$!) obtained under the naive pragmatic interpretation in Section
\ref{Sect:naive}.  (And one can similarly reproduce (\ref{tracerhoPodd})
when $\psi$ is odd.) The coincidence is of course explained by the fact
that, for a would-be wave function, $\psi$, which is either even or odd,
the conventional Parity operator $\cal P$  is simultaneously diagonalizable
with $\rho$ and has eigenvalue +1 on $\phi_1$ and $-1$ on $\phi_2$ 
$<$\ref{Note:beable}$>$.  One can also show, using the results in
Subsection \ref{Sect:3D} in the Appendix, that similar coincidences
obtain, for the analogous pair of different possible meanings for the phrase 
`expectation value of parity' in the
case of the three-dimensional situation discussed in the sentence
containining equation \ref{tracerhoP} and the subsequent sentence.

\section{Entropy and its time dependence}
\label{Sect:entropy}

As soon as one knows how to diagonalize $\rho$, one can calculate its
(von Neumann) entropy, $S=-\tr(\rho\ln\rho)$, and other 
entropy-like measures of its amount of decoherence -- in particular
$S_1=1-\tr\rho^2$ (see e.g. \cite{Wehrl}) which, while less significant
physically, is often more tractable mathematically. 

Staying with the one-dimensional Gaussian model and working to lowest
order in $\kappa$ as explained  in Section \ref{Sect:events} and the
Appendix, one finds, for a given would-be wave function $\psi$,
\[
S=-2\kappa\langle x^2\rangle\ln(2\e\kappa\langle x^2\rangle),
\]
\[
S_1=4\kappa\langle x^2\rangle
\]
where, as in the previous section, $\langle x^2\rangle$ stands for
$\langle\psi|x^2\psi\rangle$.  These are expected to be good
approximations as long as $\kappa\langle x^2\rangle$ is very much less than 1.

If the would-be wave function $\psi$ evolves in time according to some
would-be Hamiltonian $H$, as discussed in the Introduction around
equation (\ref{1.5}), so that the physical density operator evolves
according to the master equation (\ref{gaussmaster})
\[
\dot\rho=-i[H,\rho]-{2\kappa\over M}[x,[p,\rho]],
\]
then, clearly, the time-evolution of the approximate expressions for $S$
and $S_1$ above is controlled by the time-function
$t\mapsto\langle\psi(t)|x^2\psi(t)\rangle$. (Below, we call this
$\langle x^2(t)\rangle$.)  If this increases,  our
approximate $S_1$ (which is just $4\kappa$ times this quantity) will
obviously increase and so will our approximate $S$ provided
$\kappa\langle x^2\rangle < 1/\e^2$.  (But we are anyway assuming that
$\kappa\langle x^2\rangle \ll 1$.) If $\psi(t)$ evolves according to the
free Hamiltonian $H=p^2/2M$, we have the easily derived and well-known
`spreading-of-the-wavepacket' result that $\langle x^2(t)\rangle$ will
take the quadratic form
\begin{equation}
\label{wavespread}
\langle x^2(t)\rangle = A + (2E/M)(t-t_{\hbox{\small{min}}})^2
\end{equation}
where $E$ is the (conserved) expectation value 
$\langle\psi(t)|H\psi(t)\rangle$ of the energy in the state $\psi(t)$.
The feature of (\ref{wavespread}) to which we wish to draw attention here is
that
$\langle x^2(t)\rangle$ has a single minimum (with value $A$) -- occurring at
$t_{\hbox{\small{min}}}$.  In consequence of this, substituting 
(\ref{wavespread}) into either
of our above approximate formulae for $S$ or $S_1$ gives rise to 
a `two-sided' entropy-increase result in the sense that, if the time
$t_{\hbox{\small{min}}}$ is taken as an appropriate `zero time' then the
entropy will only increase on either side of that (in other words, it
will decrease monotonically before the zero time and increase
monotonically after it) at least until 
$\kappa\langle x^2(t)\rangle$
ceases to be very much less than 1, at which point our approximations
and assumptions will break down. 

This two-sided entropy increase result would be applicable, for example,
to the centre-of-mass degree of freedom of our bead of mass $M$  (see Section
\ref{Sect:events}) bigger than or around the Planck mass, free to slide
along an infinite wire and with an initial wave function at
$t_{\hbox{\small{min}}}$ such that $A$ in  (\ref{wavespread}) is very
much less than $R^2/9M^2$, where $R$ is the bead radius, as long as the
times considered are sufficiently short for the wave-packet not to have
spread so much that the  condition (\ref{smallwidth}) ceases to hold. 
To give a quantitative example, consider a Planck mass bead ($\approx
2\times {10}^{-5}$ g) with radius  $R={10}^{-2}$ cm and a state in which
 $\langle x^2(t_{\hbox{\small{min}}}\rangle)=A= {10}^{-9}$ cm$^2$.  Then
${S_1}_{\hbox{\small{min}}}$  ($=4\kappa A =4\times
\frac{9M^2}{R^2}A)\approx 4\times {10}^{-4}$.  Moreover, if we assume
that we can trust (\ref{gaussprobs}) (and hence our formula for $S_1$) 
as long as $\langle x^2\rangle < 10^{-7}$ cm$^2$, say, so that
$\kappa\langle x^2\rangle < {10}^{-2}$, then the time, $t$, at which
$\langle x^2(t)\rangle$ attains this value, can be estimated (by setting
$\frac{9M^2}{R^2}\frac{2E}{M}t^2={10}^{-2}$ and taking
$E=\frac{p^2}{2M}$ with $p^2$ by an uncertainty principle estimate 
$=\frac{1}{4A}$) to be, within an order of magnitude or so, around
${10}^9$ years!  $\dots$ by which time, $S_1$ will have increased to
around $4\times {10}^{-2}$!

It is interesting to compare and contrast this sort of (two-sided) entropy
increase result (for either $S$ or $S_1$) 
with the sort of entropy increase results which one can obtain 
for the BLP master equation \cite{BLP}
\begin{equation}
\label{BLP}
\dot\rho=-i[H,\rho]-c[x, [x, \rho]]
\end{equation}
(Here $H$ is some choice of Hamiltonian and $c$ a positive constant.)
which may be regarded as a prototype example of
a master equation of GKS/L form.  (See Note $<$\ref{Note:GKS/L}$>$.)

For this BLP model, it is easy to show that, irrespective of the
Hamiltonian $H$, the quantity $S_1$ will, for positive times, increase
monotonically in time. To show this, begin by noticing that it follows
immediately from the definition $S_1 = 1 - tr(\rho^2)$ that the time
derivative, $\dot S_1 = - 2\tr\dot\rho\rho= - 2 tr(-i[H,\rho]\rho) +2c
tr([x, [x, \rho]])\rho).$

The first term here vanishes by the cyclic invariance of the trace.
The second term equals
$-2c tr([x,\rho]^2) = +2c tr([x \rho]^\dagger [x \rho])$
which is $\ge 0$. (End of proof.)

This situation differs in at least two important respects from that we 
saw above for our Gaussian model,
which, we recall, satisfies the master equation (\ref{gaussmaster})
when $H$ is taken to be the free Hamiltonian $H=p^2/2M$. To explain the
first difference, we first need to notice that the BLP master equation
(\ref{BLP}) is itself of a quite different nature  to
(\ref{gaussmaster}) in that (see Note $<$\ref{Note:GKS/L}$>$) it holds
unrestrictedly for any $\rho$ but only for positive times, whereas
(\ref{gaussmaster}) is only meaningful for $\rho$ for which 
$\rho(x,x')$ takes the special form (\ref{1.1}) but then holds both for
positive and negative times. (See Note $<$\ref{Note:MDM}$>$.)
This difference is of course reflected in the different sorts of entropy
increase results which are appropriate, with our two-sided result for
(\ref{gaussmaster}) and the above, more familiar, one-sided result for 
(\ref{BLP}).

Second, the above `one-sided' entropy-increase result for the BLP master
equation (\ref{BLP}) holds {\it irrespective} of the choice of
Hamiltonian $H$ in (\ref{BLP}).  In contrast, the  two-sided
entropy-increase result for our master equation (\ref{gaussmaster}) 
which we obtained above in the case that $H$ in (\ref{gaussmaster}) is
the free Hamiltonian $H=p^2/2M$, depended on the form of the
Hamiltonian.  Indeed, suppose one were to take, instead, for example,
the Harmonic oscillator Hamiltonian $H=p^2/2M + kx^2/2$ (describing,
say, our bead-on-a-wire when the bead is connected to the origin with a
spring with spring constant $k$) then it is obvious that, since the
would-be wave function $\psi(t)$ will be periodic in time (with period
$2\pi(M/k)^{1/2}$) then so will $S_1$ and $S$ -- these being functionals
of $\psi$.  This second difference can be traced (see Notes
$<$\ref{Note:GKS/L}$>$ and $<$\ref{Note:MDM}$>$) to the difference that
master equations of GKS/L form such as BLP arise from tracing over
`radiation' type modes which `fly away' whereas our master equation
(\ref{gaussmaster}) arises (cf. the last paragraph of Subsection
\ref{Sect:Penrose} and  Notes $<$\ref{Note:timev}$>$ and
$<$\ref{Note:furtherinsight}$>$) from tracing over modes which (in our
bead-on-wire interpretation) are  dragged around by the would-be wave
function of our bead.

The relevance of the two-sided entropy result obtained in this Section
for our Gaussian model to the general understanding of the Second Law
entailed by the hypothesis of
\cite{kay1} will be discussed in the next (Discussion) section.

\section{Discussion}
\label{Sect:discussion}

Our main purpose has been to explore further the non-relativistic
approximate theory of \cite{kay2} which, as argued in \cite{kay2} and
explained in the Introduction, is expected to be the non-relativistic
limit of any full theory of quantum gravity consistent with the
hypothesis of \cite{kay1} which, in its turn, as argued in \cite{kay1}
(see also Note $<$\ref{Note:resolutions}$>$) offers a natural resolution to
the several puzzles and problems mentioned at the start of the 
Introduction and in Note $<$\ref{Note:puzzles}$>$.
  
As we have explained in the Introduction, this approximate theory
amounts, in the most general form in which we have stated it (see
Equation (\ref{1.4}), Note $<$\ref{Note:NGaussLog}$>$, and the parenthetical
remark after Equation (\ref{1.5})) to the following  modification of
standard (many-body) non-relativistic quantum mechanics:  

One has an `underlying' many-body wave function 
$\Psi(t; {\bf x}_1,\dots , {\bf x}_N)$
which evolves in time {\it exactly} as in standard non-relativistic
quantum mechanics for the usual many-body Hamiltonian that one would
adopt for the problem of interest.  However, while one would
normally take the density operator of the system at time $t$ to be given by
the (pure) projector onto $\Psi$, which in its position-space
representation is given by
\begin{equation}
\label{rhonought}
\rho_0(t; {\bf x}_1,\dots , {\bf x}_N ;{\bf x'}_1,\dots , {\bf x'}_N)=
\Psi(t; {\bf x}_1,\dots , {\bf x}_N))\Psi(t; {\bf x'}_1,\dots , {\bf x'}_N)^* 
\end{equation}
one declares that the {\it physically relevant} density operator $\rho$
is given, in its position-space representation by 
\begin{equation}
\label{rhophysical}
\fl \rho(t; {\bf x}_1,\dots , {\bf x}_N ; {\bf x}_1',\dots ,{\bf x}_N') = 
\rho_0(t; {\bf x}_1,\dots ,{\bf x}_N ; {\bf x}_1',\dots ,{\bf x}_N')
e^{-D({\bf x}_1,\dots ,{\bf x}_N ; {\bf x}_1',\dots ,{\bf x}_N')},
\end{equation}
where the {\it decoherence exponent}  $D({\bf x}_1,\dots ,{\bf x}_N ;
{\bf x}_1',\dots ,{\bf x}_N')$ is a specific function on configuration
space (see Note $<$\ref{Note:NGaussLog}$>$) which depends both on the mass
of each body and also on a suitable radius 
(see the Introduction, Note  $<$\ref{Note:NGaussLog}$>$ and
\cite{Abyaneh-Kay}) which is ascribed to each particle and which 
satisfies the general properties stated
after Equation (\ref{1.4}) -- the most important of which for the present
discussion are 
 
\smallskip

(a) $D$ tends to be large for pairs of configurations $({\bf x}_1,\dots
, {\bf x}_N)$,  $({\bf x'}_1,\dots , {\bf x'}_N)$ which (in the
terminology of \cite{kay2}) involve `mass relocations' around or bigger
than the Planck mass.  As discussed further in \cite{kay2}, for a given
pair of configurations, the {\it mass relocation} involved may be
defined to be the minimum amount of mass that has to be moved in order
to convert the unprimed  configuration into the primed configuration (or
vice versa).

(b) $D$ vanishes on the diagonal, i.e. 
\[
D({\bf x}_1,\dots , {\bf x}_N; {\bf x}_1,\dots , {\bf x}_N)=0
\]

As was discussed in the Introduction and in \cite{kay2}, Property (a)
leads to the suppression of macroscopic (`Schr\"odinger Cat-like')
superpositions while property (b) is what lies behind our
\textit{Position Measurement Theorem} which, on our \textit{corrected
pragmatic interpretation} (see Section \ref{Sect:expt}) leads us to
conclude that the experimental predictions of our non-relativistic
approximate theory are identical to those of standard quantum mechanics.
(As we cautioned at the beginning of Subsection \ref{Sect:purpose} this
conclusion is also predicated on being able to treat the relevant small
non-relativistic systems as closed for the purposes of this discussion
of measurement.  We shall provisionally assume that it is and give
arguments which confirm this towards the end of this section.)

As we remarked in the Introduction (see before the paragraph labelled
(A) in Section \ref{Sect:purpose})
the latter conclusion might seem surprising.  After all, for large
macroscopic systems,  the difference between the mathematical quantities
$\rho$ and $\rho_0$ can be huge by any measure.  On the other hand, the
conclusions of \cite{kay2} that quantum gravitational effects could have
any sort of big effect on every-day laboratory sized quantum systems
were themselves surprising.  So  it is perhaps reassuring that we  have
now arrived at the conclusion that (in the non-relativistic regime) they
don't have any \textit{measurable} effect after all. The way in which
they manage not to have a measurable effect, however, is rather
subtle and despite the experimental indistinguishability, the theory of
\cite{kay2} possesses properties and prospects quite different from
those of standard quantum mechanics as we now discuss.

First and foremost, the theory of \cite{kay2} achieves the goal of
replacing standard quantum mechanics by a modified theory in which
macroscopic (`Schr\"odinger Cat-like')  superpositions are absent.  As far as
the achievement of this goal is concerned, it would, indeed, seem to
be an attractive rival e.g. to the `collapse models' such as that of GRW
\cite{GRW} and others, which achieve the same goal at the expense of
certain ad hoc modifications of quantum mechanics (and which also entail
certain small modifications to the experimental consequences).  

It is interesting to notice that, in the light of the present work, one
realizes that it would have been possible to achieve the same goal
simply by positing that the usual pure density operator
(\ref{rhonought}) of standard quantum mechanics  should be replaced by a
physical density operator given by (\ref{rhophysical}) with $D$
\textit{any} (ad hoc) \textit{choice} of function on 
\textit{Configuraton~space}~$\times$~\textit{Configuration~space} 
(perhaps quite unrelated to quantum gravity) designed to have the
properties (a) and (b) above (and the other general properties stated
after Equation (\ref{1.4})).   For, once one posits any such
replacement, one will be able to prove  a \textit{Position Measurement
Theorem} and to propose a \textit{corrected pragmatic interpretation}
etc.  To the best of our knowledge, however, no such proposal had
hitherto been made.  In any case, the proposal with our approximate
theory of \cite{kay2} would (in the absence of any current experimental
guidance) seem to be more satisfactory than that with any such ad hoc
choice of $D$ and also more satisfactory than the resolution achieved
with any ad hoc collapse model, not only because it is less ad hoc, but
also insofar as it is part and parcel of the hypothesis of \cite{kay1}
which offers a much more wide-ranging collection of simultaneous
resolutions to several other puzzles as explained at the outset of the
Introduction and in Notes $<$\ref{Note:puzzles}$>$,
$<$\ref{Note:tradblack}$>$, $<$\ref{Note:entangle}$>$,
$<$\ref{Note:resolutions}$>$ and $<$\ref{Note:open}$>$.

The fact that our non-relativistic approximate theory manages to resolve
the Schr\"odinger Cat puzzle (i.e. by eliminating all macroscopic 
Schr\"odinger Cat-like superpositions) while preserving (on our
corrected pragmatic interpretation) exactly the same  experimental
predictions as standard quantum mechanics might be regarded as another
advantage.  But it also of course means that, viewed in its own right,
it is only what one might call a \textit{philosophical} resolution in
that it leads to a changed ontology (i.e. with no Schr\"odinger Cat-like
macroscopic superpositions) from that of standard quantum mechanics  but
makes no experimentally distinguishable predictions.  

However, it is saved from being `merely' philosophical since the
experimental indistinguishability of course only holds for our
non-relativistic approximate theory.  As is implicit in Section
\ref{Sect:expt} (see also Note $<$\ref{Note:graviton}$>$) when
post-Newtonian corrections to our theory become significant, one would
expect experiments along the lines of the Penrose experiment to detect
differences.  We next turn to consider at more length what the
post-Newtonian corrections to our theory might be like.

As our starting point, we return to the Newtonian time-evolution rule
summarized in the first paragraphs of this section.  When some details
are stripped away,  this specifies, for a given zero-time density
operator $\rho(0)$ with position space representation as in
(\ref{rhonought}) with $t$ set to zero, a map 
\begin{equation}
\label{ttorhomap}
t\mapsto \rho(t)
\end{equation}
from the full real time-line (i.e. the full set of positive and zero and
negative times) to the space of density operators on the relevant many
body Hilbert space ${\cal H}_{\hbox{\small{matter}}}$.

This rule will arise as the solution to a suitable master
equation.  We won't write this down explicitly; it is straightforward to
obtain it by differentiating Equation (\ref{rhonought}) with respect
to time just as (\ref{master}) is obtained by 
differentiating the one-body special case (\ref{1.5}).  Instead, we will
write it schematically as
\begin{equation}
\label{Newton}
\dot\rho=-i[H,\rho]+\hbox{\textit{Newton}}\,\rho
\end{equation}
where $\hbox{\textit{Newton}}$ is the appropriate operator which acts on
the set of density operators.  It will be a `master equation of MDM
type' in the terminology introduced in Note $<$\ref{Note:MDM}$>$.

In the post-Newtonian regime, we would expect the nature of the dynamics
of the physical density operator $\rho(t)$ (i.e. of the trace over
${\cal H}_{\hbox{\small{gravity}}}$ of the time-evolving pure total
density operator as explained in Subsection \ref{Sect:background}) to be
qualitatively different (as we are about to discuss)  but we would still
expect it to admit of a description in terms of a map of the form of
(\ref{ttorhomap}) -- i.e.  from the full real time-line to the same
space of many-body density operators.  However, unlike
(\ref{rhophysical}) and its associated master equation (\ref{Newton}),
one expects a new sort of rule determining the time-evolution of
$\rho(t)$ which incorporates the back-reaction on our many-body matter
from radiated gravitons. We would not necessarily expect this new rule
to obey an exact master equation but, in suitable models, we would
expect there to be a  preferred zero-moment of time at which the entropy
is a minimum --  the many-body post-Newtonian counterpart to  the moment
($t_{\hbox{\small{min}}}$ in (\ref{wavespread})) of minimum entropy in
our one-dimensional Gaussian model example of Section \ref{Sect:entropy}
--  and, while, of course, one now expects the rule which determines
$\rho(t)$ at other times to depend on the underlying total time-zero
state and not just on the physically relevant density operator $\rho(0)$
(which is its partial trace over gravity -- see the beginning of
Subsection \ref{Sect:background}), we would expect, in view of the
considerations discussed in  Note $<$\ref{Note:GKS/L}$>$, that,
\textit{for positive times} $\rho(t)$ will \textit{approximately} obey a
master equation of schematic form
\begin{equation} 
\label{postNewton}
\dot\rho=-i[H,\rho]+\hbox{\textit{Newton}}\,\rho 
+\hbox{\textit{postNewton}}\,\rho 
\end{equation} 
where \textit{postNewton} is approximately of GKS/L form while, \textit{for
negative times} it will approximately obey a similar master  equation
but with the term \textit{postNewton} replaced by a term
\textit{antipostNewton} which is of anti-GKS/L form $<$\ref{Note:GKS/L}$>$.
The reason is that (for positive times) the post Newtonian terms will
arise by, at each time, tracing the full matter-gravity (pure) density
operator over modes of the gravitational field which are {\it radiative}
(see $<$\ref{Note:GKS/L}$>$), and, unlike the non-radiative modes which,
in virtue of being `dragged around' by the matter are responsible for
the special features of \textit{Newton} as we discussed in Subsection
\ref{Sect:Penrose},  Section \ref{Sect:entropy} and Note
$<$\ref{Note:MDM}$>$, radiative modes will, of course, be expected to
fly away from the matter.  (And there is of course a similar argument
for the expected anti-GKS/L property for negative times.)   Moreover,
while the terms \textit{postNewton} and \textit{antipostNewton} are
expected to be  negligibly tiny in comparison to \textit{Newton} in the
Newtonian regime (see Note $<$\ref{Note:graviton}$>$), we would expect
that, as one moves away  from the Newtonian regime,  they will become
comparable in size, and as one approaches an extreme relativistic
regime, they will be much more important.   So one might say that the
non-relativistic approximate theory, which is the main subject of the
present paper, is expected to just be the small non-relativistic tip of
a large relativistic iceberg!  (Of course, the possibility of describing
$\rho$ as a many-particle density operator will presumably break down 
at some point and need to be replaced e.g. by a description in terms of 
quantum fields or strings etc.) 

We next turn to discuss the issue of which systems deserve to be
considered \textit{closed} for our various purposes.   As we have
indicated towards the beginning of Subsection \ref{Sect:purpose} and
warned in several places above, it was already concluded in \cite{kay2}
that, at  least as far as the concept of entropy is concerned, ordinary
small laboratory-sized non-relativistic systems cannot be considered
closed.  We shall next recall the reasoning behind this conclusion and
shall then discuss whether or not such a conclusion should also apply to
the other matters under discussion here -- notably our conclusions about
the Penrose (and other) experiment(s) and our tentative interepretation 
in terms of events.

By a well-known discussion (see e.g. the article by Joos in
\cite{Joosetal}) based on the argument that there will always inevitably
be quantum mechanical  entanglement over widely spatially separated
subsystems, one can argue that it only really can make exact sense to
regard a subsystem of the universe as closed  and to regard its
underlying state as describable by a pure  (i.e. vector) state if  the
subsystem  is the entire universe. Indeed, at a classical level, it was
famously pointed out by Emil Borel in 1914 that the motion of a few
grams of material on the star, Sirius, would, just because of its
gravitational interaction, after a short time,  significantly change the
configuration of a gas in a box on the Earth.  Similar considerations at
a quantum level (and for not-necessarily gravitational interactions)
make it plausible that, in a quantum description, in the actual state of
the universe  there will be a considerable degree of entanglement
between a bit of gas in a box on the Earth and a bit of gas in the star
Sirius etc. (For a discussion of all this, see for example Chapter 3 by
E. Joos in \cite{Joosetal})

However, one can ask whether it is legitimate to assume a given system
is closed for the purposes of some given calcuation -- if one is content
with a good approximate (rather than an exact) answer.  We know from
experience that, for many questions, such an assumption can be
successful for laboratory-sized systems.  Otherwise standard quantum
mechanical calculations would not be anything like as successful as they
actually are.  However,  if our purpose is the explanation, on the
hypotheses of \cite{kay1}, and within the approximate Newtonian
framework of \cite{kay2} of how  e.g. a box of gas in the laboratory can
have an entropy comparable to the usual entropy one would calculate for
a thermal equilibrium state at say room temperature, then it appears,
from the preliminary model calculations reported in \cite{kay2} that
this cannot be done:  The typical entropies we calculate (see
\cite{kay2}), while non-zero, are orders of magnitude smaller.
Our conclusion from this is not that the theory of \cite{kay2} or the
calculations are wrong, but rather that  hot boxes of gas in the
laboratory cannot be regarded as closed systems.  Rather, as we discuss
further in Note $<$\ref{Note:open}$>$ (see also Figures 4 and 5) we
conclude that the entropy of such systems needs to be explained by
regarding them to be open systems and is due to tracing over a total
environment, part of which is gravitational, but the main part of which
is essentially the same matter environment which (on one standard view
about the origin of the entropy of gases in boxes -- see the last
paragraph of Note $<$\ref{Note:open}$>$) would usually be held
responsible for their entropy. 

It then becomes an important question, whether, to get physically
correct values for the entropy of such laboratory-sized systems and to
get physical thermodynamic behaviour (i.e. a `Second-Law' result to the
effect that the entropy of a closed system always increases) it could
suffice to regard as closed, say, the solar system, or whether one needs
to include our whole galaxy or whether one actually needs to go as far
as the entire universe.  We have argued in \cite{kay2} and will argue
further below that any answer to this question must involve enlargement
to a system, parts of which are  essentially relativistic in nature (the
remark in Note $<$\ref{Note:open}$>$ about total closed sytems which
contain black holes is also relevant to this conclusion).  Thus the
galaxy or the entire universe would seem to be the more likely answers. 
And if the galaxy turns out to be a valid answer, then one might
speculate that if it is true, as is currently believed likely, that
every large galaxy contains a supermassive black hole, then this fact
may play a role in this.

What we have studied in \cite{kay2} and in the present paper are small
laboratory-sized model systems on the (we have concluded above as far as
entropy is concerned) counterfactual assumption that they are closed. 
While not expecting their entropies to be physically realistic, we can
endeavour to learn lessons about the mechanisms which give rise to
entropy and about the way that entropy tends to vary with time.  In
\cite{kay2} we considered a number of laboratory-sized model static
systems and calculated their entropies on this assumption, and showed,
in particular, that these entropies, while (as we mentioned above)
small,  appear to correlate, to some extent, with their degree of
`matter clumping'. In Section \ref{Sect:entropy} here, we have augmented
these results with some first results on the time-dependence of entropy
and found, for our Gaussian model to first order in $\kappa$ (describing
our free bead-on-a-wire model) a two-sided entropy increase result
whereby there is a moment of minimum entropy while, as a result of
`spreading of the would-be wave-packet', entropy increases for times on
either side of this moment (i.e. it decreases before this moment and
increases after it).  For more `realistic' small non-relativistic
many-body systems we would expect there to be a similar two-sided
entropy-increase result in suitable circumstances, but we expect the 
mechanism behind it to involve not only spreading of the would-be wave
packet, but also, and in fact often mainly, `dynamical clumping', by
which we mean changes in time of the amount of matter clumping.  Turning to
the post-Newtonian regime, we would still expect to obtain a two-sided
entropy-increase result in suitable models, but we expect the way one
will understand mathematically the origin of this result will be
different (and this will be traceable to the different behaviour of the
non-radiative and radiative modes which will be responsible for the two
regimes -- see $<$\ref{Note:GKS/L}$>$  and $<$\ref{Note:MDM}$>$). In the
case of our Gaussian model and (one assumes) in the case of
non-relativistic  many-body systems, the demonstration of such a
two-sided entropy increase result depended/will presumably depend on a
study of the relevant single exact master equation of `MDM' type (see
after (\ref{Newton})  which describes the dynamics of $\rho(t)$. In the
post-Newtonian regime, one expects, instead, that after choosing a
suitable model for a low-entropy inital $\rho(0)$, the demonstration of
a two-sided entropy increase result will depend on a study of an appropriate
approximate master equation of form (\ref{postNewton}) for positive
times and its negative-time counterpart for negative times.  
To the extent that one can
neglect the \textit{Newton} terms in these (cf. our `tip of iceberg'
remark in the paragraph containing (\ref{postNewton})) the question 
will be whether or not the  GKS/L term \textit{postNewton} satisfies the
criterion of Benatti-Narnhofer \cite{Benatti-Narnhofer} discussed in
Note $<$\ref{Note:GKS/L}$>$ (etc. for negative times).

The interpretation of such two-sided entropy increase results will of
course be that, while, theoretically, they represent possible dynamics
for all times, positive and negative, in actual physical application,
only the positive-time behaviour will be relevant and the negative time
part should simply be discarded.  Either one has in mind that one's
closed system is modelling the whole universe in which case $\rho(0)$
will model the initial state of the universe, or, if one is interested
in a model which idealizes some subsystem of the universe as closed
(e.g. a star collapsing to a black hole in an  otherwise empty universe
as is done in Note $<$\ref{Note:resolutions}$>$) then $\rho(0)$ will
represent the initial state of that system and its low entropy can
either be thought of as an assumption of the model or alternatively, perhaps,
as ultimately  explainable outside the model as traceable back, via
processes in the wider universe, to the low entropy of the initial state
of the universe.  (Cf. the notion of `branch systems' in
\cite{Reichenbach} although note that, in view of our remarks above
about how large closed systems may have to be,
it is a delicate question if/when/how this notion is applicable here.)

Of course, the only result we have proven here is our two-sided
entropy result, obtained in Section \ref{Sect:entropy}, for our 
one-dimensional Gaussian model, interpretable in terms of a model closed
system consisting of our free non-relativistic bead on a wire.  This is
of course a thoroughly counterfactual model closed system but
nevertheless our result is of value as a first entropy-increase result
on the hypothesis of \cite{kay1}.  If that hypothesis and the extra
assumptions involved in \cite{kay2} are correct, then that model will,
as we have indicated above, be of some significance as one extreme tip
(it concerns only one single small non-relativistic bead) of a 
non-relativistic many-body tip (cf. the paragraph containing
({\ref{postNewton})) of a full relativistic iceberg which is the Second
Law of Thermodynamics!

The other side of the coin to entropy increase (on the hypothesis of
\cite{kay1}) is of course dynamical \textit{decoherence} since,  after
all,  aside from its role in thermodynamics, our $S$ can be thought of
as a measure of the `amount of decoherence' of our closed system. So, if
our expectations concerning results of Second-Law type for suitable
(say, laboratory-sized non-relativistic) models are fulfilled, then we
will be able to say that, for suitable `time-zero' initial total states,
the amount of decoherence will always  increase.  Thus, returning to our
comparison with theories of `collapse model' type, we will be able to
say that our non-relativistic approximate theory shares with `collapse
models' (i.e. in addition to eliminating Schr\"odinger Cat-like
superpositions as discussed above)  the ability to predict a continual
process of decoherence.  (The way that GRW does this is mentioned in 
Note $<$\ref{Note:GKS/L}$>$.)   However, the comparison of the two sorts
of theories has to be more complicated than that, since, as we have
argued above, small laboratory systems cannot, on our theory be
regarded, for the purposes of calculating entropy, as closed.  Regarding
them, instead, as open (as discussed in Note $<$\ref{Note:open}$>$) we
see that, instead, our theory (i.e. of \cite{kay2}) entails,  as far as
its predictions of decoherence are concerned, essentially the same
predictions as (see  Note $<$\ref{Note:resolutions}$>$) the usual
`environment induced decoherence paradigm'.   (See however the
next-but-two paragraph.)

Turning to our tentative notion of `events' defined according to the
passage labelled (B) in Section (\ref{Sect:purpose}), we have to expect
that, for related reasons to those given above, the detailed structure
of the set of events and their probabilities and of the way these change
with time will differ considerably, for a large relativistic closed
system, from the structure etc. of this set etc. when one analyses
simple non-relativistic models of closed systems as in Section
\ref{Sect:events}.  Nevertheless, the results of Section
\ref{Sect:events}  are still of interest as a model within which one can
attempt to explore what might be the mechanism by which such an
events-based interpretation might be able to supplant the stop-gap naive
and corrected pragmatic interpretations discussed in Sections
\ref{Sect:naive} and \ref{Sect:expt}.   What one hopes for here is to
have a theory which explains how quantities such as position and
momentum and parity etc. which are treated as `observables' in our naive
and corrected pragmatic interpretations \textit{emerge} in some suitable
sense from an events interpretation.   Our discussion, in Section
\ref{Sect:events} (see also Note $<$\ref{Note:beable}$>$) of the
relationship between equations (\ref{tracerhoPeven}} and
(\ref{paritycoincidence}) in the context of our one-dimensional Gaussian
model,  appears to give a promising  indication, for the example of 
`parity' of how this can happen and the challenge which remains is to
study whether an explanation along the lines indicated by that example,
of the emergence of less primitive observables in more complicated
(many-body) examples.  

We emphasize that even if a theory along such lines for how observables
emerge from our events interpretation is possible, it is not obvious
that it will be possible staying within the context of small
non-relativistic model closed systems.  After all, we argued above that
for questions related to entropy and thermodynamics, such models need to
be rejected in favour of models of large relativistic closed systems. 
However the small success of our discussion of parity encourages us to
hope that, for these different purposes, our small non-relativistic
models may suffice.  Indeed, it seems conceivable that, even though a
different, and perhaps richer and ultimately arguably more `physical',
theory of events would be had for large relativistic models of closed
systems and by treating actual ordinary laboratory-sized systems as open
subsystems of these, the theory obtained by modelling actual ordinary
(non-relativistic) laboratory-sized systems as closed may, in itself,
yield a viable events-based replacement for standard quantum mechanics.

Concomitantly, one might possibly argue that, returning to our
comparison between our theory and `collapse models', if we treat a given
small non-relativistic laboratory-sized system as if it is `closed' for
the purposes of calculating its entropy, $S$, then, while the resulting
$S$ should not be taken seriously as having anything to do with
thermodynamics or `true extent of decoherence', it  might still be an
interesting measure of `decoherence relevant to the model' and might,
under certain circumstances, always increase.

Finally, returning to our naive and pragmatic interpretations which we
discussed at the outset, we return to the question whether it is
justified, for  the results which were obtained within them (including
our Position Measurement Theorem and including our analysis of the
Penrose experiment)  to be taken to be physically relevant.  The
question is whether such matters can (unlike what we argued above for
questions of entropy and thermodynamics) be discussed in the context of
the models of small non-relativistic closed systems which we have
considered here.  Another way of putting this is to ask whether, e.g. in
our discussion of the Penrose experiment, we can ignore entanglement of
our small model closed systems with the environment.  The answer would
seem to be that it is justified since (cf. Note $<$\ref{Note:open}$>$)
the issue of entanglement of such systems with the environment within
our theory is more or less the same as what would usually be regarded as
the issue of the entanglement of such systems with the environment and
this is, of course (see especially \cite{Marshalletal}) precisely the
issue which makes the Penrose experiment so difficult (but not
impossible) to perform and which the experiment is designed to overcome.

(A brief summary-statement of some of the main conclusions from the
above discussion with an emphasis on comparing and contrasting
our Newtonian approximate theory with `collapse models' such as GRW \cite{GRW}
was given towards the end of the Introduction.)

\section{Acknowledgments} 

BSK would like to thank The Leverhulme Foundation for a Leverhulme
Fellowship (RF\&G/9/RFG/2002/0377) from October 2002 to June 2003 during
the course of which an important part of this research was done.
BSK's research was also partially supported by PPARC grant 
`Quantum Physics of Fields, Boundaries and Gravitation' (Sept 2004 to
Aug 2006).  VA thanks PPARC for a research studentship from October 2002 to
September 2005.

\section{Appendix: Approximate diagonalisation of the density operator in the
one-dimensional Gaussian and some other, related, Models}
\label{Sect:App}

Given $\rho$ as in (\ref{1.6})
\[
\rho (x,x')=\rho_0 (x,x')e^{-\kappa (x-x')^2},
\]
where
\[
\rho_0 (x,x') = \psi (x) \psi^{\ast} (x'),
\]
our aim is to find the eigenvectors and eigenvalues of $\rho$. I.e.
we wish to solve
\begin{equation}
\label{2.1} \rho | \phi \rangle =\lambda| \phi \rangle,
\end{equation}
for $|\phi\rangle$ and $\lambda$. This could be studied exactly, but we 
content ourselves here with solving the simpler problem where
$\rho$ is replaced by its first order in $\kappa$ form, which shall call 
$\rho^\kappa$,
\begin{eqnarray} \fl
\rho^\kappa (x,x') = \rho_0 (x,x') \left(1-\kappa (x-x')^2\right)
=\rho_0 (x,x')\left(1-\kappa x^2+2\kappa xx'-\kappa {x'}^2\right)
\nonumber \\
\lo =\rho_0 (x,x')+\kappa \rho_1 (x,x'), \nonumber
\end{eqnarray}
where
\[
\rho_1 (x,x')=-\kappa\psi (x)\psi^\ast (x') (x-x')^2
\]
and the resulting eigenvalue equation is solved to first order in $\kappa$.
It is useful to note, that, in Dirac notation, $\rho^\kappa$ may be written
\begin{equation} \fl
\label{2.3} \rho \simeq \rho_0 +\kappa \rho_1=| \psi \rangle
\langle \psi |+\kappa \left( -x^2| \psi \rangle \langle \psi
| +2x| \psi \rangle \langle \psi | x-| \psi \rangle
\langle \psi | x^2\right).
\end{equation}
Taking this as our starting point, we can make the following
perturbation-theoretic argument:  If $\kappa$ is small, we expect
$\rho$, and hence also $\rho^\kappa$ to have an eigenvector 
close to $|\psi \rangle$, and
that its eigenvalue will be close to unity. Thus we write
this first eigenvector as
\begin{equation}
\label{2.4} | \phi_1 \rangle =| \psi \rangle +\kappa |
\psi_\bot \rangle,
\end{equation} 
where $| \phi_1 \rangle $ is not yet normalized and we assume the
orthogonality condition,
\[
\langle \psi | \psi_\bot \rangle =0
\]
holds. As $\tr \rho=\tr \rho^\kappa=1$, and $\rho$ is  positive (we will
confirm that $\rho^\kappa$ is positive for sufficiently small $\kappa$
below)  we expect that the eigenvalue of  $| \phi_1 \rangle $ will be
slightly less than unity so we write
\begin{equation}
\label{2.5} 
\rho^\kappa | \phi_1 \rangle =(1-a\kappa ) | \phi_1
\rangle,
\end{equation}
where $a$ is to be determined. From
(\ref{2.3}), (\ref{2.4}), and (\ref{2.5}) we get
\begin{equation}
\label{2.6} 
\rho^\kappa | \phi_1 \rangle =\left( \rho_0+\kappa \rho_1
\right) \left( | \psi \rangle +  \kappa | \psi_\bot \rangle
\right)=(1-a\kappa )\left( | \psi \rangle +\kappa |
\psi_\bot\rangle \right).
\end{equation}
We can obtain $|\psi_\bot \rangle$ from (\ref{2.6}) by expanding out
the brackets to give
\[
|\psi_\bot \rangle= (a+\rho_1)|\psi\rangle,
\]
and substituting the expression for $\rho_1$, given in
(\ref{2.3}), into the above equation gives
\begin{equation}
\label{2.7} |\psi_\bot \rangle=a|\psi\rangle -x^2|\psi\rangle+2
\langle x\rangle x|\psi\rangle- \langle x^2\rangle |\psi\rangle.
\end{equation}
By only keeping terms up to order $\kappa$ in (\ref{2.6}) we get
\begin{equation}
\label{2.8} \kappa \rho_1 | \psi \rangle =-a\kappa | \psi
\rangle +\kappa | \psi_\bot \rangle
\end{equation}
Now acting on (\ref{2.8}) with $\langle \psi |$ it can be seen
that
\begin{equation}
\label{2.9} a=-\langle \psi | \rho_1 | \psi \rangle =2(\Delta
x )^2,
\end{equation}
where
\[
\Delta x=\left( \langle \psi | x^2 | \psi \rangle -\langle
\psi | x | \psi \rangle ^2 \right)^\frac{1}{2}.
\]
Note that the above is the uncertainty in position of the would-be wave
function (as we call it in Subsection \ref{Sect:background}). So we are now in
a position to write an explicit expression (not normalized) for our
first eigenvector of $\rho$ to first order in $\kappa$, 
using (\ref{2.4}), (\ref{2.7}), and
(\ref{2.9})
\begin{equation}
\label{2.10} | \phi_1 \rangle =\left(1+2\kappa (\Delta x)^2
\right) | \psi \rangle -\kappa x^2 | \psi \rangle +2\kappa
\langle x\rangle x |\psi\rangle - \kappa \langle x^2 \rangle
|\psi\rangle,
\end{equation}
and its eigenvalue, from (\ref{2.5}) and (\ref{2.9}), is
\begin{equation}
\label{2.11} \lambda_1 =1-2\kappa (\Delta x)^2.
\end{equation}
Until now we have made no assumptions about the state (i.e. would-be
wave function)
$|\psi\rangle$, but in order to have a simple form for the remaining 
eigenvectors we shall specialize from now on to the cases where
$\psi$ is either an even function of $x$ (i.e. $\psi
(x)=\psi(-x)$) or an odd function of $x$ (i.e. $\psi
(x)=-\psi(-x)$). Using this assumption in (\ref{2.10}) and
(\ref{2.11}) we have that
\[
|\phi_1\rangle= \left( 1+\kappa \langle x^2 \rangle \right)
|\psi\rangle -\kappa x^2 | \psi \rangle,
\]
 and
\[
\lambda_1=1-2\kappa \langle x^2 \rangle.
\]
Next we note that any eigenvector $|\phi\rangle$ must (to order $\kappa$)
satisfy
\begin{eqnarray} \fl
\label{2.12} \rho^\kappa | \phi\rangle =| \psi \rangle \langle
\psi | \phi\rangle -\kappa x^2| \psi \rangle \langle \psi
| \phi\rangle \nonumber \\ \lo
 +2\kappa x| \psi \rangle \langle \psi | x | \phi
\rangle -\kappa | \psi \rangle \langle \psi | x^2 | \phi
\rangle =\lambda | \phi\rangle .
\end{eqnarray}
from which it can be seen that any non-zero eigenvector of
$\rho^\kappa$, must be a linear combination of $| \psi \rangle,\ x| \psi
\rangle$ and $x^2 | \psi \rangle$.  Using this observation as a clue, it
is easy to see that there is another eigenvector, $|\phi_2\rangle$, with
a non-zero eigenvalue, and that it has the simple form
\begin{equation}
\label{2.13} | \phi_2 \rangle =x | \psi \rangle ,
\end{equation}
with an eigenvalue equal to $2\kappa \langle x^2\rangle$, which can
be seen from (\ref{2.13}) and (\ref{2.3}).

Since the sum of the eigenvalues we have found so far is 1 ($=\tr \rho
=\tr \rho^\kappa$) and on the assumption that $\rho^\kappa$, like $\rho$ is
positive, it must be that the orthogonal complement to the 
(two-dimensional) subspace spanned by the eigenvectors we
have found so far must consist entirely of eigenvectors with eigenvalue
zero (to first order in $\kappa$).  Thus, to first order in $\kappa$, we
have found all eigenvectors and eigenvalues of $\rho^\kappa$ and (to this
order) these are obviously the same as for $\rho$.

We remark that an alternative derivation of the same result, and
confirmation that, for sufficiently small $\kappa$, $\rho^\kappa$ is positive,
may be had by noting that (still with our specialization to $\psi$
either even or odd) to first order in $\kappa$,  (\ref{2.3}) may
be written
\[
\rho=|(1-\kappa x^2\psi)\rangle\langle(1-\kappa x^2\psi)|
+2\kappa |(x\psi)\rangle\langle(x\psi)|
\]
This is in the canonical form, 
\[
\rho=\sum_n b_n|\psi_n\rangle\langle\psi_n|
\]
{\it except that} the vectors
which appear, while orthogonal, are not normalized.  To remedy this,
we easily have (to first order): 
\[
||(1-\kappa x^2\psi)||=1-\kappa\langle\psi|x^2\psi\rangle=1-a\kappa
\]
(where, for convenience, we resume our abbreviation of
$\langle\psi|x^2\psi\rangle$ by $a$) so
that (again, all equalities are to first order) the normalized vector is 
\[
(1+a\kappa)(1-\kappa x^2\psi)=1+a\kappa-\kappa x^2\psi
\]
and also
\[
||x\psi||=a^{1/2}.
\]
We can now replace the vectors by their normalized forms at the
expense of introducing prefactors equal to the square of their norms.
Thus
\[ 
\rho=(1-2a\kappa)|(1+a\kappa-\kappa
x^2\psi)\rangle\langle(1+a\kappa-\kappa x^2\psi)|
+2a\kappa |(a^{-{1\over 2}}x\psi)\rangle\langle(a^{-{1\over 2}}x\psi)|,
\]
from which we can read off the eigenvectors and eigenvalues.  We see
that these are the same as those we obtained (by a rather
different route) above.

\subsection{Diagonalizing a tensor product state in one dimension}
\label{Sect:manydiag}

Consider the tensor product state \[ | \Psi \rangle =| \psi_1
\rangle \otimes | \psi_2 \rangle \otimes ... \otimes | \psi_N
\rangle, \] and the density matrix \begin{eqnarray} 
\fl
\rho(x_1,..,x_N;x_1',...,x_N')= \Psi (x_1,..,x_N)\Psi^\ast
(x_1',...,x_N') \nonumber
 e^{-\kappa \left(x_1+x_2+...+x_n-{x_1}'-{x_2}'-...-{x_N}'\right)^2}
 \\ \fl
=\Psi (x_1,..,x_N)\Psi^\ast (x_1',...,x_N') \left(1-\kappa\left[
\left( x_1+x_2+...+x_N \right)-\left( {x_1}'+{x_2}'+...+{x_N}'
\right) \right]^2 \right), \nonumber
\end{eqnarray}
to O($\kappa$). (From now on, we shall not always bother to mention that
we are working to order $\kappa$ and we shall not bother to distinguish
between $\rho$ and the counterpart to what we called above $\rho^\kappa$.)
In Dirac notation this is
\begin{eqnarray}
\label{eqn:4.4} \fl
\rho=|\Psi\rangle\langle\Psi|
-\kappa(x_1+...+x_2)^2|\Psi\rangle\langle\Psi| \nonumber \\
+2\kappa(x_1+...+x_N)|\Psi\rangle\langle\Psi|(x_1+...+x_N)-
\kappa|\Psi\rangle\langle\Psi|(x_1+...+x_N)^2.
\end{eqnarray}
From the previous example we are tempted to hope that there will
once again be only be two eigenstates of our density matrix, and
that they will have a similar form. We know that there is only one
eigenstate of the density  matrix $\rho_0=|\Psi\rangle\langle\Psi|$,
namely $|\Psi\rangle$. Thus we expect that there will be an
eigenstate close to $|\Psi\rangle$ with eigenvalue slightly less
than unity. Let us write our first eigenvector as
\begin{equation}
\label{eqn:1.7}
|\Phi_1\rangle=|\Psi\rangle+\kappa|\Psi_\bot\rangle
\end{equation}
where $\langle\Psi|\Psi_\bot\rangle=0$. Then
\begin{eqnarray}
\label{eqn:4.5} \fl \rho|\phi_1\rangle=[|\Psi\rangle\langle\Psi|
-\kappa(x_1+...+x_2)^2 |\Psi\rangle\langle\Psi|
+2\kappa(x_1+...+x_N)|\Psi\rangle\langle\Psi| (x_1+...+x_N)
\nonumber \\ - \kappa|\Psi\rangle \langle \Psi|(x_1+...+x_N)^2]
\left(|\Psi\rangle+\kappa|\Psi_\bot\rangle\right)  = (1-\kappa
a)|\phi_1\rangle \nonumber \\ =(1-\kappa
a)(|\Psi\rangle+\kappa|\Psi_\bot\rangle).
\end{eqnarray}
Equating the O$(\kappa)$ terms in the above equation yields
\begin{eqnarray}
\fl -(x_1+...+x_N)^2|\Psi\rangle +2 (x_1+...+x_N) |\Psi \rangle
\langle \Psi|(x_1+...+x_N)|\Psi\rangle
\\ -|\Psi\rangle \langle \Psi|(x_1+...+x_N)^2 |\Psi\rangle=
-a|\Psi\rangle + |\Psi_\bot\rangle. \nonumber
\end{eqnarray}
In order to proceed and to be able to find the remaining eigenvector
we assume that the wave functions $\psi_j (x)$ are either even or
all odd for all $j \in 1...N$, bearing in mind that
$|\Psi\rangle=|\psi_1\rangle \otimes...\otimes|\psi_N\rangle$. That
is to say that for each $j$ we have either $\psi_j(x)=\psi_j(-x)$ or
$\psi_j(x)=-\psi_j(-x)$. Assuming this and acting on both sides of the
above equation with $\langle\Psi|$ gives
\begin{equation}
\label{eqn:c1.6} a=2\left(\langle x_1^2 \rangle+...+\langle
x_N^2\rangle \right).
\end{equation}
Substituting (\ref{eqn:c1.6}) into (\ref{eqn:4.5}) we are able to
deduce $|\Psi_\bot\rangle$, and hence the first eigenstate (by
substituting $|\Psi_{\bot}\rangle$ into (\ref{eqn:1.7})),
\[ \fl
| \phi_1 \rangle =\left[ 1+\kappa \left( \langle
{x_1}^2\rangle+\langle {x_2}^2\rangle +...+\langle {x_N}^2 \rangle
\right) \right] | \Psi \rangle  -\kappa\left( x_1+x_2+...+x_N
\right)^2 | \Psi \rangle,
\]
along with its eigenvalue
\[
\lambda_1=1-2\kappa\left(\langle {x_1}^2 \rangle +\langle {x_2}^2
\rangle +...+\langle {x_N}^2\rangle \right).
\]
It can be shown that there is only one other eigenvector,
\[
| \phi_2 \rangle =(x_1+x_2+...+x_N) | \Psi \rangle,
\]
with eigenvalue
\[
\lambda_2 =2\kappa\left( \langle {x_1}^2 \rangle +\langle {x_2}^2
\rangle+...+\langle {x_N}^2 \rangle \right),
\]
which is similar to the previous example.

\subsection{Diagonalizing the density operator in three dimensions}
\label{Sect:3D}
Let the wave function $\psi({\bf x})$ satisfy the following symmetry
conditions
\begin{eqnarray}
\label{eqn:4.5.1} \fl
\psi(x,y,z)=\psi(-x,y,z)=\psi(-x,-y,z)=\psi(-x,-y,-z)=
\\ \psi(-x,y,-z)=\psi(x,-y,z)=\psi(x,-y,-z)=\psi(x,y,-z),
\end{eqnarray}
and be such that it can be written as a product of three independent
functions,
\[
\psi(x,y,z)=\alpha(x)\beta(y)\gamma(z).
\]
Consider the density operator
\[
 \rho({\bf x},{\bf x}')=\rho_0({{\bf x},{\bf
x}'})e^{-\kappa({\bf x}-{\bf x}')^2}\simeq\rho_0( {\bf x}, {\bf
x}')-\kappa({\bf x}-{\bf x}')^2\rho_0({\bf x},{\bf x}')
\]
where $\rho_0( {\bf x},{\bf x}')=\psi({\bf x})\psi^\ast({\bf x}')$.
We wish to find the eigenvectors and eigenvalues of $\rho$ to
$O(\kappa)$, which means that we want to find $\phi(\bf{x})$ and
$\lambda$ that satisfy
\begin{equation}
\label{eqn:4.6} \int d^3 {\bf x}'\rho({\bf x},{\bf x}')\phi({\bf
x}')=\lambda\phi({\bf x}).
\end{equation}
In Dirac notation our density operator, to first order in $\kappa$,
is
\[
\rho = |\psi\rangle \langle\psi|- \kappa{\bf x}^2
|\psi\rangle\langle\psi| +2\kappa {\bf
x}|\psi\rangle\langle\psi|-2\kappa|\psi\rangle\langle\psi|{\bf x}^2.
\]
To find the first eigenvector we use a similar perturbative argument
to the previous one to deduce that there must be an eigenvector
close to $|\psi\rangle$ with eigenvalue slightly less than unity,
i.e.
\[
\phi_1(\bf{x})=\psi(\bf{x})+\kappa\psi_\bot (\bf{x}),
\]
or in Dirac notation
\begin{equation}
\label{eqn:c1.8}
|\phi_1\rangle=|\psi\rangle+\kappa|\psi_\bot\rangle.
\end{equation}
We want to find $|\psi_\bot\rangle$ such that $|\phi_1\rangle$ is an
eigenvector of $\rho$. To do this we first write
\begin{eqnarray} \fl
\label{eqn:4.6.1} \rho|\phi_1\rangle= \left( |\psi\rangle \langle
\psi|-\kappa {\bf x}^2 |\psi\rangle\langle\psi|+2\kappa {\bf
x}|\psi\rangle\langle\psi| {\bf x}-\kappa|\psi\rangle\langle\psi|
{\bf x}^2 \right)
\left(|\psi\rangle+\kappa|\psi_\bot\rangle  \right) \nonumber \\
= (1-a\kappa)(|\psi\rangle+\kappa|\psi_\bot\rangle).
\end{eqnarray}
From this, and using the symmetry restrictions we imposed on
$|\psi\rangle$, we obtain $|\psi_\bot\rangle$,
\begin{equation}
\label{eqn:c1.9} |\psi_\bot\rangle=-({\bf x}^2+\langle {\bf
x}^2\rangle)|\psi\rangle.
\end{equation}
By acting on both sides of (\ref{eqn:4.6.1}) with $\langle\psi|$ we
find that $a=2\langle{\bf x}^2\rangle$. Thus our first eigenvalue is
\[
\lambda_1=1-2\langle{\bf x}\rangle^2
\]
and the corresponding eigenvector (obtained by substituting
(\ref{eqn:c1.9}) into (\ref{eqn:c1.8})),
\[
|\phi_1\rangle= (1-\kappa \langle{\bf x}^2\rangle)
|\psi\rangle-\kappa {\bf x}^2|\psi \rangle.
\]
We claim that the following are eigenvectors of $\rho$,
\[
\phi_2({\bf x})=x\psi({\bf x});\  \phi_2({\bf x})=y\psi({\bf x});\
\phi_3({\bf x})=z\psi({\bf x}),
\]
with respective eigenvalues
\[
\lambda_2=2\kappa\langle x^2\rangle;\ \lambda_3= 2\kappa\langle
y^2\rangle;\ \lambda_4= 2\kappa\langle z^2\rangle.
\]
To show that $\phi_2(\bf{x})$ is an eigenvector we simply show that
it satisfies (\ref{eqn:4.6}) through
\begin{eqnarray} \fl
\int d^3 {\bf x}'\rho({\bf x},{\bf x}')\phi_2({\bf x}')= \int
dx'dy'dz' \alpha(x)\beta(y) \gamma(z) \alpha^ \ast(x')\beta^\ast(y')
\gamma^\ast(z') x'\alpha^ \ast(x')\beta^\ast(y') \gamma^\ast(z')
\nonumber
\\ -\kappa\int dx'dy'dz'(x^2+y^2+z^2-2xx'-2yy'-2zz'+x'^2+y'^2+z'^2)
\nonumber
\\ \alpha(x)\beta(y) \gamma(z) \alpha^\ast(x') \beta^\ast(y') \gamma^
\ast(z')x' \alpha^\ast(x')\beta^\ast(y')\gamma^\ast(z'). \nonumber
\end{eqnarray}
The symmetry restrictions on $\psi(x,y,z)$ imply that the first term
in the above equation vanishes. Noting this and then expanding out
the second term gives
\begin{eqnarray} \fl
\int d^3 {\bf x}'\rho({\bf x},{\bf x}')\phi_2({\bf x}')
=-\kappa\psi(x,y,z)(x^2+y^2+z^2) \nonumber \\ \times \int dx'dy'dz'
\alpha^\ast(x')\beta^\ast (y')\gamma^\ast
(z')x'\alpha(x')\beta(y')\gamma(z')  -\kappa\psi(x,y,z) \nonumber \\
\times \int dx'dy'dz'\alpha^\ast (x')\beta^\ast (y')\gamma^\ast (z')
(x'^2+y'^2+z'^2)x' \alpha(x')\beta(y')\gamma(z') \nonumber
\\ +2\kappa \psi(x,y,z)\int dx'dy'dz' \alpha^\ast(x') \beta^\ast(y')
\gamma^\ast(z') (xx'+yy'+zz')x'\alpha(x')\beta(y')\gamma(z').
\nonumber
\end{eqnarray}
Once again the first two terms of the above equation vanish and we
are left with
\[
\int d^3 {\bf x}'\rho({\bf x},{\bf x}')\phi_2({\bf x}')=2\kappa
\langle x^2\rangle x \psi(x,y,z)=2\kappa \langle x^2\rangle
\phi_2({\bf x}),
\]
which shows that $\phi_2(\bf{x})$ is an eigenvector. It is easy to
show that $\phi_3({\bf x})$ and $\phi_4({\bf x})$ are eigenvectors
by a similar argument. This example, with its artificial symmetry
restrictions on the wave function $\psi ({\bf x})$, serves as an
indication that there are more eigenstates of $\rho$ when
investigating higher-dimensional systems.

\section{Notes}
\label{Sect:notes}

\begin{enumerate}

\item
\label{Note:puzzles} The problems and puzzles which 
concern quantum physics in general include:

{\begin{description}

\item [measurement]
Those usually collected under the heading `the measurement problem in
quantum mechanics'. 
 
\smallskip

\item [entropy] 
The puzzling contradiction between the traditional understanding
of the entropy of a general closed system in terms of (subjective)
`coarse-graining' and our understanding, since the work of
Bekenstein \cite{Bekenstein} and Hawking \cite{Hawkingevap} in the 1970s
that the entropy of the specific closed system consisting of a black hole
(either sitting in and radiating into, an asymptotically flat empty 
universe or in equilibrium with radiation in a suitable box)
is something objective -- namely something which is approximately equal to 
`one quarter of the area of the event horizon'. 

\smallskip

\item [second law]
The problem of supplying a microscopic explanation for
the second law of thermodynamics, i.e. a microscopic explanation for 
why the entropy of a closed system always increases.

\end{description}}

\smallskip

The problems and puzzles which relate specifically to black holes 
include:

\begin{description}

\smallskip

\item [information loss] 
The `information loss puzzle', which we shall take here to mean
the puzzle as to how it can be that, on the one hand, during the process
of stellar collapse and black-hole formation and then black-hole
evaporation,  entropy presumably actually continually increases while,
on the other hand, the underlying quantum  dynamics of the process is
presumably unitary.   We remark that what makes this a puzzle is the
traditional (but incorrect according to \cite{kay1} and the present paper)
assumption that (in this context) the physical entropy of a quantum
mechanical system should be identified  with the von Neumann entropy of
its (total) density operator -- the puzzle then arising because this
quantity, being a unitary invariant, must therefore remain constant
under a unitary time-evolution and, in fact, if the (initial) state is a
vector state, be zero (at all times).

\smallskip

\item [coincidence] 
The problem (raised and solved in \cite{kay1}) of explaining the
coincidence that the thermodynamic entropy
$S_{\hbox{\small{thermodynamic}}}(M)$ of a (assumed for simplicity
spherically symmetric, neutral) black hole of mass $M$  turns out to
have (approximately) the same value -- namely $1/4\pi M^2$, i.e. `one
quarter of the area of the event horizon' --  as the gravitational
entropy
$S^{\hbox{\small{matterless}}}_{\hbox{\small{Gibbons-Hawking}}}(T)$ of a
Gibbs state of matterless gravity at temperature $T$  when $T$ is taken
to be the Hawking temperature $T=1/8\pi M$.

Here, what we mean by the thermodynamic entropy
$S_{\hbox{\small{thermodynamic}}}(M)$ is the value of $S$ implied by
the standard thermodynamic relation 
\[ 
dE=TdS 
\] 
when we identify $E$ with $M$ and $T$ with $1/8\pi M$ and take $S$ to
be $0$ when $M=0$.  This is easily calculated to be $4\pi M^2$.  On the
other hand, what we mean by  the gravitational entropy  
$S^{\hbox{\small{matterless}}}_{\hbox{\small{Gibbons-Hawking}}}(T)$ is
the value of $S$ calculated using the standard 
equilibrium-statistical-mechanics formula
\[
S=\ln Z-\beta\partial \ln Z/\partial\beta 
\]
from the Gibbons-Hawking \cite{Gibbons-Hawking}
Euclidean-quantum-gravity partition function $Z(\beta)$, at temperature
$T=1/\beta$, of a Gibbs state of a {\it matterless}
$<$\ref{Note:tradblack}$>$ quantum gravitational field in a suitable
(spherical) box. In fact, when (cf. \cite{Gibbons-Hawking}) this
partition function is approximated by its zero-loop value
$\exp(-I_{\hbox{\small{classical}}})$  with
$I_{\hbox{\small{classical}}}$ the classical action, $4\pi M^2$, of the
Euclideanized Schwarzschild metric of mass $M=1/8\pi T$ ($=\beta/8\pi$),
the value of $S$ obtained in this way  is again easily seen to be $4\pi
M^2$.  (We note that the matterless gravitational field here is a
mathematical construct and, from a physical point of view of course,
counterfactual.  What we call the Gibbons-Hawking partition function for
this is the quantity one would obtain if one were to remove the terms
representing matter in the equations in \cite{Gibbons-Hawking} -- see
the next Note  $<$\ref{Note:tradblack}$>$)

\end{description}

\smallskip

Other puzzles for which our hypothesis offers natural resolutions
include the {\bf thermal atmosphere puzzle} and a puzzle which we raise
here which we call the {\bf neutron-star entropy puzzle}.  We shall say
what these are and give our proposed resolutions of them at the end of
Note $<$\ref{Note:resolutions}$>$.

\item
\label{Note:low}
In an exact unified theory at the Planck energy and higher, one would ,
of course, not expect there to be a clear demarcation between `matter'
and `gravity'.  But at energies well below the Planck energy, one
expects that there will.  By `matter' here, we of course mean everything
other than gravity.  We remark that our hypothesis entails that entropy
is a quantity which \textit{emerges} at low energies and will cease to have any
meaning at energies around and above the Planck energy.

\item
\label{Note:tradblack}
(This Note is really a footnote to Note $<$\ref{Note:puzzles}$>$.)  To give
more details about the connection with the work of Gibbons-Hawking, and
to explain what the traditional assumptions have to say about our
coincidence puzzle, we emphasize that our statement of our {\bf
coincidence} puzzle refers to the partition function for a Gibbs state
of a matterless gravitational field (i.e. `pure gravity')  whereas in
\cite{Gibbons-Hawking} -- see also the review and further discussion in
\cite{HawkingEuclidean} -- the partition function actually considered is
that for a Gibbs state of gravity together with matter fields.  It turns
out that the zero-loop approximation to this is identical to that for
matterless gravity discussed in the main text.  However the
one-and-higher loop corrections will involve the matter as well as the
gravity.  In  these (and many other, later) references the traditional
assumption is adopted that it is the Gibbs state of gravity together
with appropriate matter which represents the physical total state of a
black hole in equilibrium in a box and therefore they find the total
physical entropy to be the sum of the same (matterless-gravity)
zero-loop contribution (equal to one quarter of the area of the event
horizon) discussed here in the main text with a term arising from the
one-and-higher-loop matter and gravity part of their partition function
which is supposed in those references to represent a correction to the
entropy due to the {\it thermal atmosphere} of the black hole.  It has
become clear since the early work of Gibbons and Hawking that, due to a
quantum-field-theoretic divergence which is, in its turn, due to the
infinite coordinate-distance of the horizon from any point outside the
horizon when the appropriate (i.e. `tortoise') coordinate is used, this
one-loop term is actually  impossible to calculate within the Euclidean
quantum gravity framework without imposing an ad hoc cut-off near the
horizon and its value is therefore in doubt (and, as far as we are
aware, this doubt has still not been convincingly removed by any other
approach to quantum gravity).   For the traditional assumptions to
resolve the coincidence puzzle, one would clearly have to assume, as is
in fact implied in the Gibbons-Hawking references, that the one-loop
`thermal atmosphere' entropy is only a small correction to the
zero-loop term, so that the total entropy of the 
traditionally-described black hole in thermal equilibrium in a box will
be well approximated by what we called in Note  $<$\ref{Note:puzzles}$>$ 
$S^{\hbox{\small{matterless}}}_{\hbox{\small{Gibbons-Hawking}}}(T)$.
This would then formally resolve the {\bf coincidence} puzzle within the
traditional assumptions, since,  as in the resolution on our hypothesis,
both the quantities
$S^{\hbox{\small{matterless}}}_{\hbox{\small{Gibbons-Hawking}}}(T)$ (for
$T=1/8\pi M$) and $S_{\hbox{\small{thermodynamic}}}(M)$ would again be
understood to approximate the physical entropy of a physical black hole
of mass $M$  -- in one case in equilibrium in a box and in the other
case radiating into empty space.   But this is surely a very
unsatisfactory and unconvincing resolution of {\bf coincidence} since on
the one hand, the quantity  $S_{\hbox{\small{thermodynamic}}}(M)$
derives from the thermodynamics of the thermal atmosphere (which, after
all is what would escape as Hawking radiation were the walls of the box
to be removed) while, on the other hand, the entropy of the thermal
atmosphere has to be asserted to be approximately zero in order to make
the resolution work!   This is of course related to the {\bf thermal
atmosphere puzzle} to which, as we explain in Note
$<$\ref{Note:resolutions}$>$, our hypothesis also offers a natural
resolution.  Indeed, on our hypothesis,  the entropy of the matter part
of the thermal atmosphere is predicted to be approximately equal to 
$S^{\hbox{\small{matterless}}}_{\hbox{\small{Gibbons-Hawking}}}(T)$
which, up to one-and-higher loop graviton corrections, is $4\pi M^2$. 
The one-and-higher loop graviton corrections themselves, on the other
hand, are plausibly comparable in magnitude to the entropy of just one
of the matter fields making up the thermal atmosphere, and therefore
might be expected to have a  value of the same order of magnitude as 
$4\pi M^2/N$ -- where $N$ is some appropriate value for the `number of
matter fields in nature' -- and hence plausibly small.

\item 
\label{Note:entangle} In general, given a pure (i.e. vector) state
with density operator $\rho=|\Psi\rangle\langle\Psi|$ on a total
Hilbert space, $\cal H$, which arises as a tensor product, ${\cal
H}_m\otimes {\cal H}_g$ of two other Hilbert spaces (`m' and `g' here
could stand for `matter' and `gravity' or could be given a more general
interpretation) then one can define the two partial density operators
$\rho_m$ on ${\cal H}_m$ (the standard partial trace of $\rho$ over
${\cal H}_g$) and $\rho_g$ on ${\cal H}_g$ (the standard partial trace
of $\rho$ over ${\cal H}_m$) and it is well-known and easy to show that
the two von Neumann entropies $S_m=-\tr(\rho_m\ln\rho_m)$ and
$S_g=-\tr(\rho_g\ln\rho_g)$ are equal.  In modern terminology, an
alternative name for these latter (equal) quantities is the {\it m-g
entanglement entropy} of the state $\rho$.

\item
\label{Note:resolutions}
Our hypothesis may be understood as a specific variant of the
environment-induced decoherence paradigm \cite{Zurek, Joosetal} namely a
variant in which the total state of a closed system is taken to be a
pure state evolving in time according to a unitary time-evolution. But
it has three crucial new features, (1) that  the gravitational field has
a privileged status as a permanent piece (see Note
$<$\ref{Note:open}$>$)  of the total environment, (2) that the entropy
which results by tracing over it is regarded as an attribute of the
total closed matter-gravity system rather than only of the matter, and
(3) that this entropy is identified with the closed system's physical
entropy.  With these three new features, one obtains, as far as we are
aware, for the first time, a definite notion of decoherence even for
closed systems and an objective definition (the matter-gravity
entanglement entropy) for the physical entropy of a general closed
system, thus offering a resolution to the {\bf entropy} puzzle (see Note
$<$\ref{Note:puzzles}$>$).   Note of course that this resolution entails
that one rejects both the option of identifying the physical entropy of
a closed system with the von Neumann entropy of its total density
operator $\rho_{\hbox{\small{total}}}$ (which would of course be zero
if, as our hypothesis assumes, $\rho$ is pure, i.e. a projector onto a
single vector) and that one also rejects the traditional attempt to
define  it in terms of some sort of (subjective)  coarse-graining of
$\rho_{\hbox{\small{total}}}$.)  (For this to be a resolution of the
{\bf entropy} puzzle, we of course need to argue that in the specific
case of a closed system consisting of a black hole, the matter-gravity
entanglement entropy will have the correct value -- i.e. (approximately)
one-quarter of the area of the event horizon.  We postpone our argument
for this to the next paragraph.)  Moreover, by  adopting our new general
definition of entropy, one obtains a plausible  mechanism for an
entropy-increase result for closed systems thus offering resolutions to
the puzzles {\bf second law} and {\bf information loss} as we explained in
the main text. 

\smallskip

Turning to the application of our general hypothesis to the subject of
black hole thermodynamics, our hypothesis goes naturally together with a
radically different description of quantum black holes from the
traditional description.  In this new description,  the total state of a
(say spherically symmetric, neutral) quantum black hole of mass $M$ in
equilibrium with radiation in a suitable box is understood to be a {\sl
pure} state of a quantum-gravitational {\sl closed} system and not, as
would be traditionally assumed, a Gibbs state at the Hawking
temperature.   However, this pure total state is expected to possess, in
the relevant low energy description, just the right sort of
matter-gravity entanglement for its partial trace
$\rho_{\hbox{\small{matter}}}$ over  ${\cal H}_{\hbox{\small{gravity}}}$
to resemble a Gibbs state of the matter fields at the Hawking
temperature $T=1/8\pi M$ in the presence  of a background black hole of
mass $M$, and also for its partial trace,
$\rho_{\hbox{\small{gravity}}}$, over ${\cal
H}_{\hbox{\small{matter}}}$, to  resemble a Gibbs state of (matterless)
gravity at the same Hawking  temperature.  In fact, it is reasonable to
assume, the latter state may be identified with the Gibbs state of a
matterless quantum gravitational field -- whose partition function may
be calculated by the Gibbons-Hawking Euclidean path-integral method
summarized in Note $<$\ref{Note:puzzles}$>$ by removing the matter terms
in \cite{Gibbons-Hawking}. (See  Note $<$\ref{Note:tradblack}$>$.)  We
remark that of course this  extension of our hypothesis entails that the
appropriate `theory of everything' will turn out to have states which,
at suitably low energies admit of such a  description with appropriate
matter multiplets etc.

We are now in a position to complete our resolution of the {\bf entropy}
puzzle, for, by the result mentioned in Note $<$\ref{Note:entangle}$>$,
the matter-gravity entanglement entropy of our black hole state must
equal the von Neumann entropy of  $\rho_{\hbox{\small{gravity}}}$
and therefore, in view of the second  assumed resemblance mentioned
above, must be close to   `one quarter of the area of the event
horizon' ($4\pi M^2$).   Moreover we see that, as a byproduct of this
discussion, our hypothesis has led to a natural resolution to the {\bf
coincidence} puzzle since, having identified the matter-gravity
entanglement entropy with the physical entropy of our black hole, we
would also expect it to be the correct `$S$' in the thermodynamic relation
`$dE=TdS$'.

\smallskip

Our hypothesis also offers a natural resolution to some other puzzling
features of the traditional understanding of black hole entropy.  Thus,
previously, it was unclear to what extent the entropy of a black hole
(say in equilibrium with radiation in a box)  resides in the quantum
geometry of the hole itself, i.e. in the quantum gravitational field, and
to what extent it resides in the matter-fields, i.e. in the thermal
atmosphere.  This is essentially the {\bf thermal atmosphere puzzle}
discussed e.g. in  \cite{Wald}.   It is particularly puzzling because, 
on the traditional understanding, it
was possible to make arguments (cf. Note
$<$\ref{Note:tradblack}$>$) to the effect that the entropy all resides in the
gravitational field but one could also (cf. also e.g. the recent review by
Don Page \cite{Page}) make counter-arguments to the effect that it all
resides in the thermal atmosphere.  As a further byproduct to the
discussion above, we see that our hypothesis also offers a very neat
solution to this puzzle.  For if we identify `the entropy of the
gravitational field' with the von Neumann entropy of 
$\rho_{\hbox{\small{gravity}}}$ and `the entropy of the (matter part of)
the thermal atmosphere' with $\rho_{\hbox{\small{matter}}}$ and the
entropy of the total state with the matter-gravity entanglement entropy,
then, by our resolution to {\bf entropy} (combined with the result
mentioned in $<$\ref{Note:entangle}$>$) we see that our hypothesis
entails that all these three quantities are actually identical.  In
particular it turns out on our hypothesis that both statements `the
entropy resides in the gravitational field' and `the entropy resides in
the thermal atmosphere' (which traditionally were regarded as mutually
exclusive possibilities) become simultaneously correct statements!

\smallskip

Another puzzle in the traditional understanding (which we point out
here) was that while a black hole (formed say from neutron star
collapse) is traditionally understood to have a very  large non-zero
entropy, the entropy of a neighbouring (not necessarily stationary)
state of quantum gravity which consists just of a neutron star (on the
verge of collapsing) would, traditionally, be assigned a zero entropy.  
We call this the {\bf neutron star entropy puzzle}.  With our definition
(matter-gravity entanglement entropy) any state of any closed system
would be expected to have a non-zero entropy and the entropy of such a
neighbouring neutron-star state could, quite plausibly, have a value
which neighbours (but is a bit less than) the entropy of the black hole
it is on the verge of collapsing to.  (For preliminary estimates of the
entropies, on our hypothesis, of certain simple model matter-gravity
systems in the Newtonian limit, see \cite{kay2}.  These have very small
values, but, as discussed in Section \ref{Sect:discussion}, the
Newtonian results are expected only to be ``the  small non-relativistic
tip of a large relativistic iceberg''!)

\smallskip

What our hypothesis tells us about the puzzle {\bf measurement} is of
course the subject matter of the rest of this paper.

\smallskip

Finally, we remark that one might regard the fact that our hypothesis
allows a more satisfactory definition for the entropy of an open system
than traditional ideas do as further evidence for the correctness of our
hypothesis.  See Note $<$\ref{Note:open}$>$.

\item
\label{Note:D}
As explained in \cite{kay2}, $e^{-D({\bf a_1}, {\bf a_2})}$ is the value
of the inner product $\langle g_1| g_2\rangle$, in a suitable Hilbert
space, ${\cal H}_{\hbox{\small{gravity}}}$, for the linearized quantized 
gravitational field, between two (non-radiative) state vectors,  $g_1, 
g_2\in {\cal H}_{\hbox{\small{gravity}}}$ which are the quantum representations
of the static Newtonian gravitational field due to our ball at rest with
its centre of mass located at ${\bf a_1}$, ${\bf a_2}$ respectively. 
(As discussed in \cite{kay2}, $\langle g_1| g_2\rangle$ turns out to be
real.)   To understand how the formula (\ref{1.1}) arises, it is helpful
to first consider a would-be wave function, $\psi$, which consists of two
sharp peaks localized around, say $\bf a_1$ and $\bf a_2$; we shall
write \begin{equation} \label{cat} \psi=c_1\psi_1 + c_2\psi_2
\end{equation} where $c_1, c_2\in {\mathbb C}$, $\psi_1$ is a normalized
wave function  consisting of single sharp peak centred on ${\bf a_1}$
and $\psi_2$ is a normalized wave function consisting of a single sharp
peak centred on ${\bf a_2}$.
 
\smallskip

In this case, it is envisaged that a full quantum gravitational
description is given by a vector in ${\cal H}_{\hbox{\small{matter}}}\otimes
{\cal H}_{\hbox{\small{gravity}}}$ which takes the form 
\[
\Psi=c_1\psi_1\otimes g_1 + c_2\psi_2\otimes g_2. 
\]
$\rho$, obtained by tracing $|\Psi\rangle\langle\Psi|$
over ${\cal H}_{\hbox{\small{gravity}}}$, is then clearly given by  
\begin{equation}
\label{decocat}
\fl\rho=|c_1|^2|\psi_1\rangle\langle\psi_1| + 
c_1c_2^*\langle g_2| g_1\rangle|\psi_1\rangle\langle\psi_2|
+c_1^*c_2\langle g_1| g_2\rangle|\psi_2\rangle\langle\psi_1|
+|c_2|^2|\psi_2\rangle\langle\psi_2|
\end{equation}
and a little thought should suffice to see that, (a) the sharper the two
peaks in the would-be wave function, the closer this will be, in its
position space representation to (\ref{1.1}), (b) for a general would-be
wavefunction, one may argue that the formula (\ref{1.1}) must still hold
by regarding it as a limiting case of a wave function with multiple
peaks.

\item
\label{Note:NGaussLog}
The generalization of the Gaussian asymptotic regime to $N$ balls would
be relevant e.g. to a superposition of two configurations, each
involving $N$ balls, such that, while the balls within each
configuration are possibly well separated, the configurations themselves
differ only slightly so that each ball in the second configuration is
displaced by much less than its radius relative to its position in the
first configuration.  We remark that the formula easily generalizes to
the case of $N$ balls of equal radius $R$ but possibly differing masses
in which case $D$ takes the form
\[
9(M_{\hbox{\small{total}}}^2/R^2)(x_{\hbox{\small{cm}}}-x'_{\hbox{\small{cm}}})^2
\] 
where $M_{\hbox{\small{total}}}$ is sum of the $N$ masses and
$x_{\hbox{\small{cm}}}$ and $x'_{\hbox{\small{cm}}}$ the centres of mass of the
unprimed and primed configurations which constitute the arguments of $D$.  We
also remark that the generalization to the many-ball case of the
logarithmic asymptotic regime has
\begin{equation}
\label{logasympt}
\fl\exp(-D({\bf x}_1, \dots , {\bf x}_N; {\bf x}_1', \dots , {\bf x}_N'))=
\prod_{I=1}^N\prod_{J=1}^N\left ({|{\bf x}_I'-{\bf
x}_J||{\bf x}_I-{\bf x}_J'|\over |{\bf x}_I-{\bf x}_J||{\bf x}_I'-{\bf
x}_J'|}\right )^{-12M_IM_J}
\end{equation}
where it is to be understood that, in the cases $I=J$, the terms in the
denominator 
\break $|{\bf x}_I-{\bf x}_J||{\bf x}_I'-{\bf x}_J'|$ are to be replaced
by $R_I^2$ (This is an easily obtained generalization of the formula
given (`for simplicity') in \cite{kay2} in the case where all the
$M_I$ and  all the $R_I$ are the same.  The robustness of the latter
formula under  changes in shape and graininess of the bodies is
discussed in \cite{Abyaneh-Kay}.  

\item 
\label{Note:timev}
The rule (\ref{1.5}) is an automatic consequence (see Note
$<$\ref{Note:D}$>$) of taking the total state for a would-be wave
function $\psi({\bf x})$, say with a single sharp peak at ${\bf x}={\bf
a}$, to be $\psi\otimes g$ where $g$ is the (non-radiative) quantum
counterpart to the static Newtonian gravitational field of our ball when
its centre of mass is  located at ${\bf a}$ -- irrespective of whether
the would-be wave function itself is static or not.  Obviously, it
amounts to neglecting the radiative part of the gravitational field. 
This approximation and its generalizations to the many-ball case etc.
are however expected to be excellent as long as the classical
gravitational radiation from the relevant systems would be negligible. 
See Note $<$\ref{Note:graviton}$>$ for a quantitative estimate on how
good this approximation is likely to be.

\item 
\label{Note:GKS/L} 
Ignoring some technicalities, the GKS/L form (often known as `Lindblad
form')  for a master equation holds if and only if the master equation
integrates up, for positive times, to a `semi-group of completely
positive maps' (sometimes known as a `quantum dynamical semigroup') on
the relevant space of density operators -- see \cite{GKS}, 
\cite{Lindblad}. (\cite{GKS} treated such semigroups of matrices and 
\cite{Lindblad} of bounded operators.) A prototype (see Section
\ref{Sect:entropy}) is the BLP master equation (\ref{BLP}) \cite{BLP}
$\dot\rho=-i[H,\rho]-c[x, [x, \rho]]$, $c>0$, for a one-particle density
operator.

If one has a total Hilbert space, $\cal H$, which arises as a tensor
product of form  ${\cal H}_m\otimes {\cal H}_g$,  and if the  dynamics
of a  total state vector $\Psi(t)\in {\cal H}$  evolves according to a
(unitary) Schr\"odinger  time-evolution for some Hamiltonian, $H$, then,
if the time-zero state vector $\Psi(0)$ takes the product form 
$\phi_m\otimes\psi_g$, then one knows (\cite{GKS}) that for any
(positive or negative) time $t$, the partial trace, $\rho_m(t)$, of
$\rho(t)=|\Psi(t)\rangle \langle\Psi(t)|$ over ${\cal H}_g$ will arise by the
action on the (pure) time-zero density operator
$\rho_m(0)=|\phi_m\rangle\langle\phi_m|$ of a `completely positive map'
(which will of course depend on $\psi_g$). (See e.g. also
\cite{GKS} for the definition of `completely positive'.)

Moreover,  \textit{for positive times} and if  ${\cal H}_m$ represents
now, say, `charged stuff' and ${\cal H}_g$ now, say, `radiation', and if
$H$ is such that the  `produced radiation' tends to be  emitted from,
and fly away from, the `charged stuff', then one expects, in line with
general expectations (see \cite{GKS} and references therein) that the
time-evolution will be approximable by the action on $\rho_m(0)$  of a
\textit{semi-group} of completely positive maps whose exact solutions 
obey a master equation of GKS/L form. (This semigroup will extend to act
on arbitrary -- i.e. not-necessarily-pure $\rho_m(0)$)  We remark that,
although this is not always mentioned or emphasized, one of course
equally expects the time evolution for negative times to be similarly
governed by a master equation of `anti-GKS/L' form (i.e. a generator of
a completely positive semigroup in the variable $t'=-t$) and this will
obviously be the case if the Hamiltonian $H$ and initial state $\Psi(0)$
mentioned above are both time-symmetric in obvious suitable senses.

Benatti and Narnhofer \cite{Benatti-Narnhofer} have given necessary and
sufficient conditions for the von Neumann entropy of a density operator
evolving in time (for positive times) according to a master equation of
GKS/L form to increase monotonically.  One can easily check that their
conditions hold for BLP and we mention in passing that it is shown in
\cite{Benatti-Narnhofer} that they hold for the time-evolution of GRW
\cite{GRW}. (The same result with $S$ replaced by $S_1$ -- defined in
Section \ref{Sect:entropy} -- is demonstrated in  Section
\ref{Sect:entropy} for BLP and in \cite{GRW} itself for GRW.)  When a
total unitary time evolution as in the previous paragraph leads to an
approximate GKS/L-form master equation governing the behaviour of
$\rho_m$ at positive times which satisfies Benatti and Narnhofer's
conditions, then one clearly expects entropy to increase for positive
times.  Although this is again not always mentioned or emphasized, one
also expects (and again of course this must happen if the total dynamics
and time-zero state has a time-reversal symmetry) the entropy of
$\rho_m$ will increase for increasingly negative times.  We shall call
such an overall time-behaviour for the von Neumann entropy of $\rho_m$
as $t$ ranges over the whole real line a `two-sided entropy increase'
result.  

(The above two paragraphs are particularly relevant to the discussion in
Section \ref{Sect:entropy}.)

\item
\label{Note:MDM} 
To see that the solutions to (\ref{master}) and (\ref{gaussmaster})
specified by the formulae (\ref{1.5}) and (\ref{Gauss}) consist of
density operators at all times, we note that it is clear from the fact
that $D(0)=1$ that $\rho$ will have unit trace.  It is also clear from
the way that (\ref{1.1}) and (\ref{1.5}) etc are derived that $\rho(t)$
will be a positive operator at all $t$, and this can be shown to hold
also when $D$ in (\ref{1.5}) is replaced by either its Gaussian or
logarithmic asymptotic forms.

The failure of the master equations (\ref{master}) and
(\ref{gaussmaster}) to have GKS/L form can be traced back to the fact
that the modes which are traced over in obtaining $\rho_{\hbox{\small
physical}}$ (see Note $<$\ref{Note:D}$>$) are, in the Newtonian case,
non-radiative and, in contrast to what was envisaged for `radiation' in
our discussion of the approximate validity (for positive times) of
master equations of GKS/L form in the situation envisaged in Note
$<$\ref{Note:GKS/L}$>$, instead of `flying away' from our matter, they are
`dragged around with' (or one might say `slaves to') the matter.  (Cf.
the last paragraph of our discussion of the Penrose experiment in
Subsection \ref{Sect:Penrose}  and Note  $<$\ref{Note:furtherinsight}$>$
and also (especially the last paragraph in) Section \ref{Sect:entropy}.

Below, we shall call time-evolutions on spaces of density operators of
form described in the paragraph containing equation (\ref{1.5})
`multiplicative decoherence models' (MDM) and master equations such as
(\ref{master}) and (\ref{gaussmaster}) `MDM master equations'.

It is interesting to compare and contrast MDM master equations with
master equations of GKS/L form (see Note $<$\ref{Note:GKS/L}$>$ and
Section  \ref{Sect:entropy}).  MDM master equations have solutions for
all times, $t$, ranging over the real line while GKS/L master equations
only have solutions for positive $t$.  On the other hand, for positive
$t$, GKS/L master equations have solutions for arbitrary initial density
operators while MDM master equations only have solutions for very
special initial data (but including of course initial data of form
(\ref{1.1}) for equation (\ref{master}) and (\ref{Gauss}) for equation
(\ref{gaussmaster}) etc.)  What goes wrong for GKS/L equations for
negative times, and for MDM master equations for wrong initial data is
that one can find solutions in the class of unit-trace trace-class
operators, but they will fail to stay in the class of density operators
because they will fail to satisfy positivity.  Related to this, one can
prove (BSK unpublished), for our Gaussian model master equation 
(\ref{gaussmaster}), the theorem:

\smallskip

\textit{The time evolute after an arbitrarily small positive time
$\delta t$ of a pure density operator $\rho=|\psi\rangle\langle \psi|$,
where $\psi$ is even,  fails to be positive except when $\psi$ takes the
Gaussian form $\psi=e^{-cx^2}$ with ${\rm Re}c >0$ and ${\rm Im} c < 0$.
 (In the special case with $c=4\kappa$, $\rho=|\psi\rangle\langle\psi|$
evolves into itself for all later times.)}

\smallskip

If, however, we compare and contrast MDM models, not with
time-evolutions generated by master equations of GKS/L form, but rather
with  time-evolutions such as the sort described in
$<$\ref{Note:GKS/L}$>$ which, as we discussed in that note,  are
expected to be approximately of GKS/L form for positive times
\textit{and of anti-GKS/L form for negative times} then, as far e.g. as
the behaviour of entropy is concerned, the two sorts of models can have
qualitatively similar behaviour.  Namely, in both cases (depending on
the details of the dynamics) one can have a two-sided entropy increase
result in the sense explained in  $<$\ref{Note:GKS/L}$>$.  We show such
a two-sided entropy-increase result for our Gaussian (MDM) model with a
free Hamiltonian (and to first order in $\kappa$) in Section 
\ref{Sect:entropy}.

\item
\label{Note:photon} 

As we shall discuss further in Section \ref{Sect:expt}, we expect that our
conclusions will also apply if the probe particles are relativistic 
(e.g. photons).  

\item
\label{Note:open}
The purpose of this Note is to sketch a natural extension of the
hypothesis of \cite{kay1} (see subsection \ref{Sect:background} of the
Introduction and also Notes $<$\ref{Note:puzzles}$>$ and
$<$\ref{Note:resolutions}$>$) to deal with open systems.  In this
extension, the specification of an open  subsystem  of some given closed
system is taken to correspond to a particular way of expressing the
Hilbert space ${\cal H}_{\hbox{\small{matter}}}$ as a tensor product
\[
{\cal H}_{\hbox{\small{matter}}}={\cal H}_{\hbox{\small{matter, system}}}\otimes 
{\cal H}_{\hbox{\small{matter, environment}}}
\]
so that, in view of (\ref{tensorprod}), the Hilbert space,  
${\cal H}_{\hbox{\small{total}}}$ for the total system will arise as a triple
tensor product
\[
{\cal H}_{\hbox{\small{total}}}={\cal H}_{\hbox{\small{matter, system}}}\otimes 
{\cal H}_{\hbox{\small{matter, environment}}} \otimes
{\cal H}_{\hbox{\small{gravity}}}.
\]
For a given $\rho_{\hbox{\small{total}}}$ (which, see (\ref{total}), 
will, according to our hypothesis, take the form
$|\Psi\rangle\langle\Psi|$) of the relevant  total closed system, the
density operator,  $\rho_{\hbox{\small{matter, system}}}$ describing the
partial state of the matter belonging to the open subsystem is then
declared to be the partial trace of  $\rho_{\hbox{\small{total}}}$ over
${\cal H}_{\hbox{\small{matter, environment}}} \otimes
{\cal H}_{\hbox{\small{gravity}}}$ and the physical entropy of the open
subsystem is declared to be the von Neumann entropy of
$\rho_{\hbox{\small{matter, system}}}$.  All this may be represented by
the schematic rectangle-picture in Figure 4a where one may regard the
vertical dividing line as separating the matter
degrees of freedom which belong to the `open subsystem' (marked `matsys' on
the figure) from those which belong to its `environment' (marked `matenv'
on the figure)  while the horizontal line separates all matter from
gravity (indicated by the region marked `grav').  
In the sense that it is always part of what is traced over,
one might say that gravity is regarded as a {\it permanent part of the
environment} so that the horizontal line is fixed, while the
vertical line is slideable.  Sliding it to the right corresponds
to gradual enlargement of our open subsystem of interest until, in the
limit as  it coincides with the right hand boundary of the lower
rectangle, the open system will approach the  total closed system and
its  entropy will approach the entropy of the total closed system which,
of course, will still be non-zero because all of
${\cal H}_{\hbox{\small{gravity}}}$ continues still to be traced over in the 
definition of that.  Indeed, if our total closed system contains a (say
stellar or galactic-centre sized) black hole, then this limiting value of the
entropy will be very large.  As we indicate in the schematic graph Figure
5a, it seems reasonable to expect that the entropy of ever larger (say
nested) subsystems will increase monotonically with the size of the
subsystems.  

\begin{figure}[htbp]
\centering		
\includegraphics[width=1.00\textwidth]{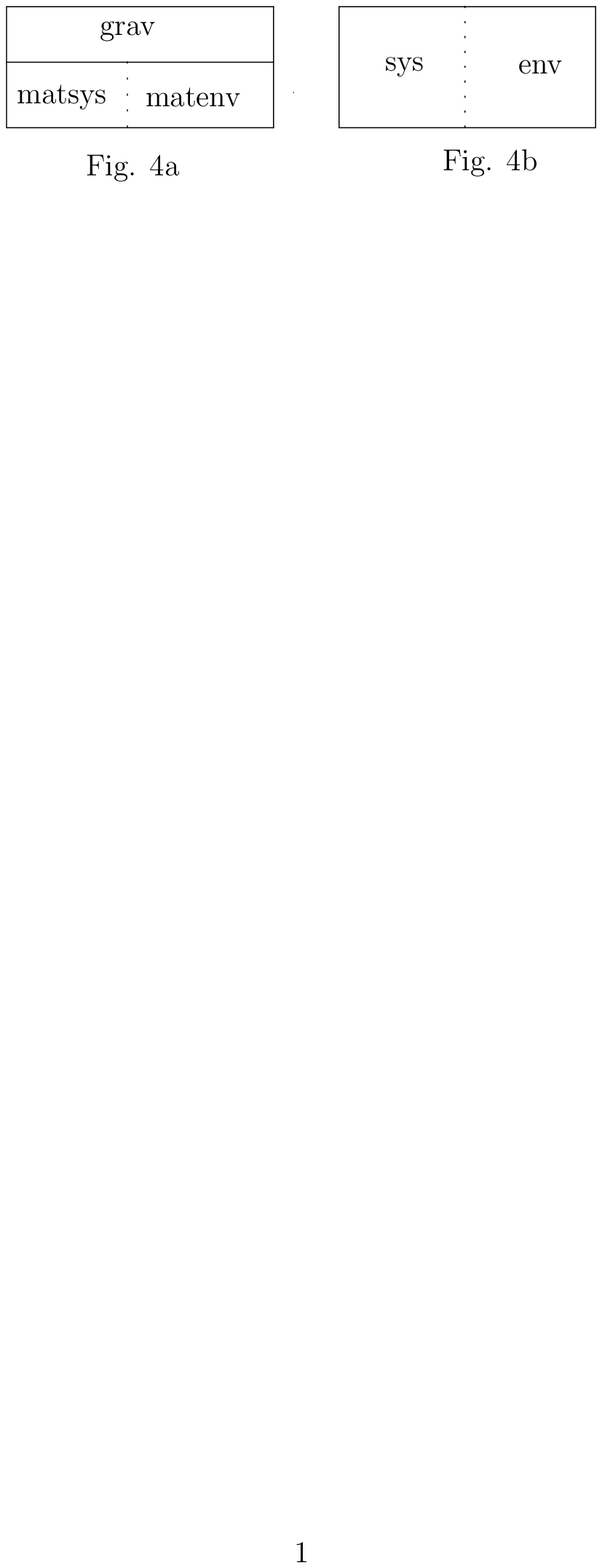}
\caption{Schematic diagrams contrasting our approach to open systems 
(Fig. 4a) with that on the traditional `environment-induced decoherence'
paradigm (Fig. 4b)}
\end{figure}

All this is in contrast to, and we feel, more satisfactory than, the
situation if one were to attempt a definition for the entropy of an open
subsystem of a given closed system on traditional ideas (i.e. on the
assumption that the gravitational field due to the matter of the
subsystem is regarded as belonging to the subsystem) by defining it to
be the von Neumann entropy of the partial trace of the total density
operator over the Hilbert space for the total environment of the
subsystem, (and assuming the total density operator to be pure).  In
other words, by defining it to be the entanglement entropy of the
subsystem with its total environment (and assuming the total density
operator to be pure).  This might be represented by the schematic
rectangle figure 4b.   The entropy, thus defined, for a given open
subsystem would then necessarily (cf. Note $<$\ref{Note:entangle}$>$)
equal the entropy of its environment and therefore, as one considers
ever larger nested subsystems which eventually approach the total closed
system, as  indicated in figure 5b by sliding the vertical dividing line to the
right, the adage  `what goes up must come down' would necessarily
apply and the entropy would approach the value zero! So a typical graph
of entropy against size of region would, in contrast to Figure 5a look
like that sketched in Figure 5b.

\begin{figure}[htbp]
\centering		
\includegraphics[width=0.80\textwidth]{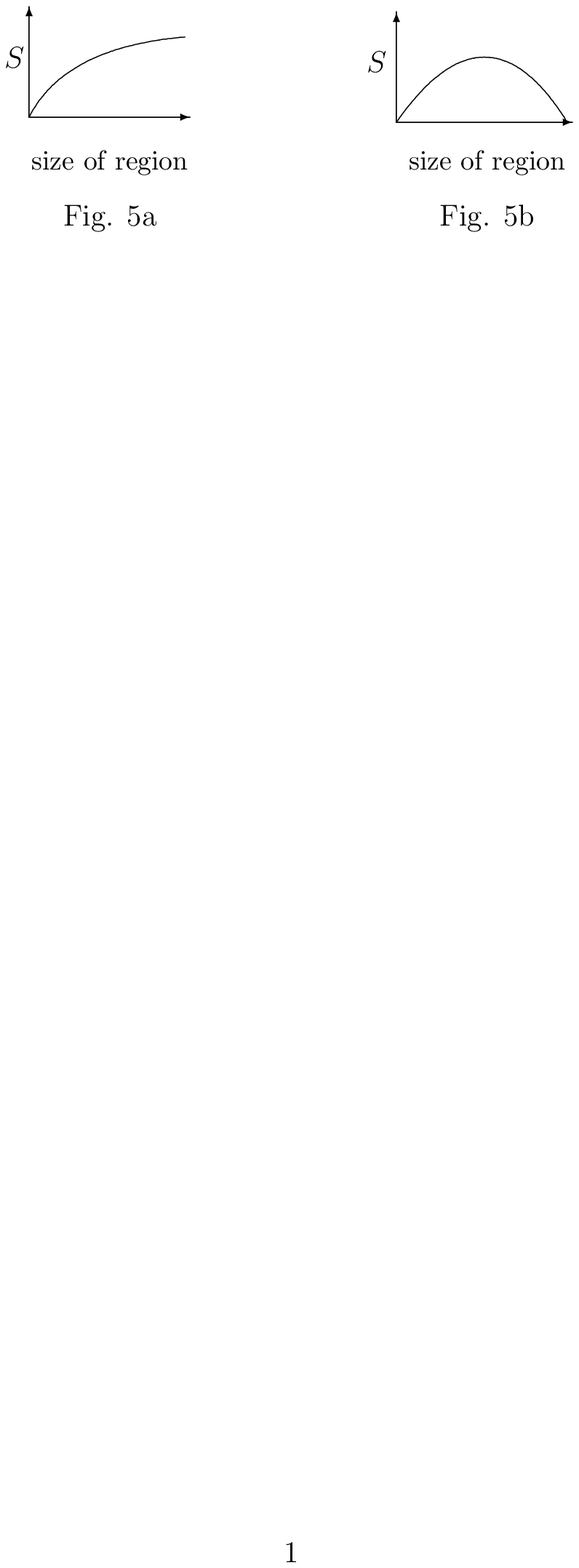}
\caption{Schematic behaviour of entropy against `size of open system' for
our approach to open systems (Fig. 5a) contrasted with the corresponding
behaviour on traditional ideas (Fig. 5b)}
\end{figure}

Finally, we remark that, in defining the density operator of any open 
system, the result of tracing over  ${\cal H}_{\hbox{\small{matter,
environment}}} \otimes {\cal H}_{\hbox{\small{gravity}}}$ is of course
equivalent to first tracing over ${\cal H}_{\hbox{\small{gravity}}}$ and then
tracing over  ${\cal H}_{\hbox{\small{matter, environment}}}$.   So our
proposal is equivalent to a modification of the above-described approach
to defining entropy on traditional lines -- i.e. as pictured in Figure 4b
-- in which one identifies `system' as `matter system' and
identifies `environment' as `matter environment' but, instead of
postulating a total state which is pure, one postulates a total state on
the full matter system which is equal to the (mixed) trace of our 
matter-gravity $\rho_{\hbox{\small{total}}}$ over
${\cal H}_{\hbox{\small{gravity}}}$.

\item
\label{Note:catpuzzle}
In this note, we shall add a number of further remarks about our
resolution to the  Schr\"odinger Cat puzzle discussed in Sections
\ref{Sect:intro} and \ref{Sect:events}:  

\smallskip

We begin by remarking that, in the context of Schr\"odinger Cat-like
situations, something similar to our mechanism whereby a pure state such
as (\ref{catintro}) gets changed to a mixed density operator like
(\ref{decocat}) or (approximately -- see below) like
(\ref{decocatintro}) occurs, of course, in the general context of
`environmental induced decoherence' (see e.g. \cite{Zurek} and
\cite{Joosetal} and references therein). However there are a number
of important differences.  First, the `environmental induced
decoherence' mechanism is only capable of bringing about the replacement
of a pure by a mixed state in the case of an \textit{open} system,
whereas in the theory of \cite{kay1} and \cite{kay2}, by having a
preferred environment (i.e. gravity) and declaring that this must always
be traced over in principle, such a replacement of pure by mixed states
occurs for closed systems.  Secondly, something similar to our proposed
general interpretation (see around the passage labelled (B) in Section
\ref{Sect:intro}  and the beginning of Section \ref{Sect:events}) of our
physical density operators in terms of events which happen is usually
implicity or explicitly assumed when discussing the interpretation of
density operators of form (\ref{decocatintro})  -- but it usually seems
to be assumed that the events which can happen should be identified with
\textit{one-dimensional} diagonalising subspaces of the relevant density
operator (and that the probability with which they happen is the
associated eigenvalue).   In the case of (\ref{decocatintro}) this will
not change the interpretation as long as $|c_1|^2 \ne\frac{1}{2}$.  But
in that special case, on the latter usual interpretation, an ambiguity
occurs since the eigenvalue $\frac{1}{2}$ is degenerate.  The existence
of this ambiguity in this usual interpretation  is sometimes  taken as a
reason to question whether a formalism based on density operators can
really solve the conceptual problems of quantum mechanics. However our
interpretation (B) sidesteps this criticism by entailing that, in this
same special case, there is only \textit{one} possible event which can
happen (with probability 1), namely that corresponding to the subspace
spanned by $\psi_1$ and $\psi_2$.  This is free from ambiguity although
it is admittedly somewhat strange. However, the strangeness is perhaps
mitigated in that, in a typical preparation of a would-be Schr\"odinger
Cat-like state, such as (\ref{catintro}), the quantities $c_1$ and $c_2$
would be expected to vary in time, and the equalities
$|c_1|^2=\frac{1}{2}=|c_2|^2$ to only hold for a single moment of time,
and hence our interpretation would entail that there are two possible
events, each evolving in time, then fusing into one possible event for a
fleeting moment before splitting into two (evolving) possibilities
again.  The ball (or the cat!) is still predicted to be one thing or
another except at that moment and at no time is the statement about the
possible events which can happen (and the probabilities with which they
will happen) ambiguous.

\smallskip
Finally we remark that $\langle g_1|g_2\rangle$ in (\ref{decocat}) can
be small but never actually zero and therefore the density operator
(\ref{decocatintro}) can only be an approximation to (\ref{decocat}) in
Note $<$\ref{Note:D}$>$.   In fact, in general the subspaces belonging
to our two events will clearly be one-dimensional subspaces spanned by
(normalizable) vectors of the form 
\begin{equation}
\label{exactcatevents}
\phi_1=k\psi_1 + s^*\psi_2\quad \hbox{and}\quad \phi_2=-s\psi_1 + k^*\psi_2
\end{equation}
with $|k|^2+|s|^2=1$ and $\rho$ will take the form 
\begin{equation}
\label{catdiag}
A|\phi_1\rangle\langle\phi_1|+B|\phi_2\rangle\langle\phi_2|
\end{equation}
where $A$ and $B$ are positive numbers with sum 1 and the values of $A$,
$B$,  $k$ and $s$ will be determined (by equating (\ref{catdiag}) with
(\ref{decocat})) in terms of $|c_1|^2$, $|c_2|^2$ and $\langle
g_1|g_2\rangle$.    When, the ball is of mass much smaller than the
Planck mass and/or the two ball states are close together, $\langle
g_1|g_2\rangle$ will be close to $1$, $A$ will be close to 1, $B$ close
to zero, $k$ close to $c_1$ and $s$ close to $c_2$ so there will be one
`event' with a probability close to 1 which is close to the subspace
spanned by $\psi$ of (\ref{catintro}) while, at the other extreme,
returning to the case of almost complete decoherence, as our ball gets
larger and/or as the two ball-states get further apart,  $k$ will be
close to 1 and $s$ close to zero, and $A$ will be close to $|c_1|^2$ and
$B$ will be close to $|c_2|^2$ and  $k$ will get closer to 1, and $s$
closer to zero.  We feel that our statement that this ``is just what
one would hope for from a resolution to the Schr\"odinger Cat puzzle''
remains equally arguable with this exact description of the
event-subspaces.

\item
\label{Note:asfarasaware}
As far as we are aware, an interpretation in terms of `events which
happen' along the lines of that proposed here in (B)  has not been
considered previously in the context of models for modified quantum
mechanics which are couched in terms of time-evolving density operators.
However, something similar to (but not quite identical to) our proposal
seems often to be implicitly assumed when considering the  physical
interpretation of density operators which arise in contexts such as 
proposed resolutions to the Schr\"odinger Cat puzzle --  see Note
$<$\ref{Note:catpuzzle}$>$.

\item
\label{Note:unravel}
Here, we reserve the term `collapse models' for theories such as the
`GRW' model \cite{GRW} when these are formulated in terms of
time-evolving density operators.  For a given such collapse model, (and
provided, as is e.g. the case for the GRW model, the time-evolving
density operator arises from a semigroup of completely positive maps --
see Note $<$\ref{Note:GKS/L}$>$ -- acting on some initial density
operator) it is now understood that there are typically several
different ways of representing the dynamics in terms of a time-evolving
stochastic wave function -- each constituting what is (since the term
was coined around 1993 by Carmichael \cite{Carmichael}) now called a
different `unravelling' (see e.g. \cite{GisinBrunRigo}) of the model. 
An early example is the `usual' unravelling of the GRW model which
actually predates the use of the term `unravelling'.  This was
formulated in a simple explicit way by Bell in \cite{Bell:collapse}. 
For a given collapse model, a choice of unravelling often suggests
another way to that proposed here of interpreting the collapse model in
terms of `events' which `happen'. For the `Bell' unravelling of the GRW
model, this is explained in \cite{Bell:collapse}. The proposal made here
for how a notion of `events which happen' may be obtained from a given
time-evolving density operator is offered as an alternative to such
interpretations (see Note $<$\ref{Note:asfarasaware}$>$).   It involves
a more abstract notion of `events' to the `collapse centres' of
Bell/GRW.  It may suffer from a number of problems, but it is free from
the problem of the ambiguity of unravellings.  In any case, the notion
of unravellings is only available for time-evolutions which
\textit{exactly} arise from semigroups of completely positive maps --
i.e., cf. $<$\ref{Note:GKS/L}$>$, which obey master equations of exact GKS/L
form.   Hence, unravellings are anyway unavailable to us since the
master equations, (\ref{Newton}) and (\ref{postNewton}), of relevance to
us are not of this form. (At best, (\ref{postNewton}) is
\textit{approximately} of this form (for positive times).  For similar
reasons, unravellings are unavailable for more traditional situations
where one has a time-evolving density operator arising from some sort of
environment-induced decoherence (\cite{Zurek}, \cite{Joosetal}) since such
time-evolutions also do not constitute semigroups of completely positive maps
(these again only arising, if at all, after some sort of mathematical
limiting procedure).   On the other hand, the events interpretation we
propose here would be available.

\item
\label{Note:objprob} It is of course a tricky philosophical issue just
what one means by `probability' here. (Some common views about what
probabilities are make reference to observers!)  We will not attempt to
say anthing original about this issue here, but we note that more or
less the same issue is present with notions of events such as that
proposed by Bell for the GRW model (see Note $<$\ref{Note:unravel}$>$). 
A recent article which addresses the issue in this latter
context is \cite{FriggHoefer}.  

\item
\label{Note:pencil}
The statements about the way the pencil gets reflected off the mirror
can easily be verified with a simple conservation of momentum argument
making the usual small-angle approximations. In explanation of the
various assumptions: (a) the diameter of the pencil is assumed to be
much bigger than $\delta$ in order to ensure that the angle of widening
of the pencil due to spreading of the wave packet will be much less than
the angles between the various beam components after the reflection. 
(There will also be an initial broadening of all the reflected pencils
by an amount $\sqrt 2 \delta$ because of the uncertainty in the location
of the mirror.); (b) the diameter of the mirror has obviously to be
large enough for the  pencil to be entirely reflected by it, no matter
where along the interval $(0,2\delta)$ of the $x$-axis the centre of
mass of the bead is located; (c) the mass of the probe particles is
taken to be much smaller than $M$ so that reflection by the mirror will
consist simply in reversal of the perpendicular component of momentum in
the centre of mass frame; (d) the assumption $P\gg p$ was made simply so
that we can use the usual small angle approximations to obtain the
uniform spacing in the angles of the reflected beam components.

\item
\label{Note:Bellink}
A similar declaration to that considered in the discussion motivating
our \textit{corrected pragmatic interpretation} in Section
\ref{Sect:expt}  has been made before in the context of other approaches
to the measurement problem in quantum mechanics. Cf. e.g. the discussion
around the following passage in the article \cite{Bellink} by John Bell
concerning what might constitute what Bell calls a `beable': ``Not all
`observables' can be given beable status $\dots$.  What is essential is
to be able to define the positions of things, including the positions of
instrument pointers or (the modern equivalent) of ink on computer
output.'' 

\item
\label{Note:undo}
One could e.g. presumably undo the measurement
of the bead momentum inserting a suitable lens-like device between the
probe particle and the screen so that, however its wave function
reflected off the mirror, only one spot forms on the screen.

\item
\label{Note:felix}
The version described here is similar to the `\small{FELIX}' experiment
described in \cite{PenroseMP2000} and to the `suggested space-based
experiment' described in Appendix 2 of \cite{PenroseLSHM}.  In the
realistic versions of the experiment described in these references,
Penrose envisages the photon to be an X-ray photon, and the movable
mirror to be a M\"ossbauer crystal.  Penrose and collaborators have also
proposed a ground-based experiment (see again Appendix 2 in
\cite{PenroseLSHM} and \cite{Marshalletal}) in which the photon and
movable mirror are optical and the photon is stored in suitable optical
cavities.  The details of the ground-based experiment (see
\cite{Marshalletal}) are slightly different but it is easy to see that
our discussion here and in the following subsection could easily be
adapted to this latter  experiment and the conclusions we draw here
would survive in essentially unaltered form.  We remark that what is
difficult about these experiments (and is discussed in the references
just given) is of course to eliminate all causes of decoherence other
than the possible cause that the experiments are designed to test.

\item
\label{Note:assure}
We have made some extra implicit assumptions and simplifications here. 
In particular, the state of the photon/movable-mirror-cum-oscillator 
system at time $t_m$ might
well be more accurately  modelled by an entangled state, rather than the
single tensor product we have assumed here.   Without extending our
analysis to include such a possibility, we note that we are still
assured by our position measurement theorem that, irrespective of how we
model it, as long as the Penrose experiment ends in a type of position
measurement (and assuming that photons may be treated for this purpose
as non-relativistic particles) the prediction of the theory of \cite{kay2}
will still be identical with the prediction of standard quantum mechanics.

\item
\label{Note:phase}
We should really have allowed for a relative phase in the two equations
(\ref{photonmirror}), (\ref{photon}).  However, for simplicity we assume the
phase happens to be 1 and that, with this phase, one predicts zero
detection rate at the detector.  If either of these were not the case,
one could e.g. imagine modifying the experiment by inserting a suitable
phase-shifting device in one arm of the interferometer in order to
continue to predict zero detection rate at the detector.

\item
\label{Note:Penrose}
In his 1996 article, \cite{PenroseGRG}, Penrose explains that what he
actually envisages is a radical new theory of `quantum state vector
reduction' and he emphasises that the technical form that such a theory
will take is not yet known.
Rather, some qualitative and semi-quantitative arguments are given and
the direction in which such a theory may lie, indicated.  In our
discussion of the Penrose experiment in Subsection \ref{Sect:Penrose} we
perhaps should have written ``Now what is envisaged by Penrose, to the
extent that it may be represented within current quantum mechanical
formalism involving density operators etc. $\dots$''.

\item
\label{Note:allthree}
Interestingly, we easily see that the partial
state of the photon at time $t_m$ is predicted to be the same as that of 
(\ref{rhot1}) by all three theories:  
\[
\rho^{\hbox{\small{Kay}}}_{\hbox{\small{matter}}}(t_m)=
\rho^{\hbox{\small{Penrose}}}_{\hbox{\small{matter}}}(t_m)
=\rho^{\hbox{\small{standard}}}_{\hbox{\small{matter}}}(t_m)
\]
(However, in the Penrose experiment, no
direct measurement is made of this state.)

\item
\label{Note:furtherinsight}
We remark that it is of course the equality of $g(t_f)$ with $g(t_0)$
(or more generally the fact that, at any time, $t$, $g(t)$ depends only
on the matter-configuration at time $t$) in the non-relativistic case
which is responsible (cf. Note $<$\ref{Note:timev}$>$) for the form of
(the relevant generalization to the photon/movable-mirror-cum-oscillator
system of) equation (\ref{1.5}) which (together with the relevant
version of equation (\ref{Ddiag})) is responsible for the validity of
our Position Measurement Theorem (when applied to the Penrose
experiment). 

\smallskip

Further insight into the reasons for the recoherence may be had by
considering the simple model closed system consisting just of a
movable-mirror-cum-oscillator as in the Penrose experiment, but
uncoupled from any photons etc.   The centre-of-mass motion of such an
oscillator is of course modelled in standard quantum mechanics as a
standard quantum-mechanical harmonic oscillator.  Any (Schr\"odinger
picture) wave-function solving this will, of course, be periodic in time
and taking the would-be density operator, $\rho_0(t)$, in 
(a one-dimensional version of) (\ref{1.5}) to be
$|\psi(t)\rangle\langle\psi(t)|$ for such a periodic would-be wave
function, it is clear, from (\ref{1.5}), that $\rho$ will also be
periodic and hence alternately decohere and recohere.  This is in
contrast with the typical prediction of a `collapse model' -- see
Section \ref{Sect:entropy} and Notes $<$\ref{Note:GKS/L}$>$ and
$<$\ref{Note:MDM}$>$ for further discussion.

\item
\label{Note:graviton}
By Einstein's quadrupole formula, we would expect the energy radiated
per unit time to be given by a number of order 1 times
$m^2\ell^4\omega^6$ (if we restore $c$ and $G$, this should be
multiplied by $G/c^5$) where $m$ is a relevant mass (assuming the mass
of the `oscillator' part of the movable-mirror-cum-oscillator [the
`cantilever' in \cite{Marshalletal}] can be neglected, then this would
be the mass of the movable mirror and its mounting) and $\ell$ its
amplitude, and $\omega$ its frequency of oscillation.  Thus the expected
number, $N_{\hbox{\small{graviton}}}$, of gravitons emitted per cycle would be a
number of order 1 times $m^2\ell^4\omega^4$ (times $G/(\hbar c^5)$)
(times $2\pi$).  Putting in the values (similar to or larger than those
in the experiment suggested in \cite{Marshalletal})  $m=10^{-8}$ gram,
$\ell=10^{-11}$ cm, $\omega=3\times 10^3$ sec$^{-1}$, this gives an
$N_{\hbox{\small{graviton}}}$ of around $10^{-80}$ which is utterly negligible!

\smallskip

We remark that there will actually inevitably be another source of
decoherence due to a form of radiation by the oscillating mirror, namely
the radiation by the movable mirror of photons at around the oscillator
frequency due to the effect of the oscillating boundary condition
imposed on the quantized electromagnetic field by the oscillating
mirror.  Ford and Vilenkin \cite{FordVilenkin} have calculated that, for
such an oscillating mirror,  the energy per unit time radiated by this
mechanism for a scalar analogue of the photon is given by (note that a
missing factor of $2\pi$ in \cite{FordVilenkin} has been inserted here)
$E=\frac{1}{720\pi^2}\ell^2\omega^6 A$ (times $\hbar/c^4$)  where $A$ is
the surface area of the mirror, and we shall assume that the same
formula, multiplied by 2 because of the photon's two polarization
states, holds true for true photons.  (It is not actually known whether
or not this assumption is correct, but it will surely give the correct
order of magnitude.)  The expected number,
$N_{\hbox{\small{photon}}}$, of photons emitted per cycle will then be 
$\frac{1}{180\pi}\ell^2\omega^4 A$ (times $1/c^4$).  With the same
values assumed above for $\ell$ and $\omega$ and with a value for $A$
(again taken from the  suggested experiment \cite{Marshalletal} ) of
$10^{-6}$ cm$^2$, this gives an $N_{\hbox{\small{photon}}}$ of around
$10^{-39}$ which is vastly bigger than $N_{\hbox{\small{graviton}}}$ but of
course, for the purposes of the Penrose experiment (in the version
proposed in  \cite{Marshalletal}), reassuringly, still vastly smaller 
than anything which would need to be taken into account in a practical
analysis of the experiment.

\item
\label{Note:beable}
The discussion of the relationship between the notion of parity in the
naive pragmatic interpretation and the notion of parity in our events
interpretation and, in particular, the coincidence explained after
equation (\ref{paritycoincidence}) may clearly be generalized as we now
explain: It seems natural to introduce the following definitions and
terminology: For any $\rho$ to which one intends to apply our events
interpretation, we say that a self adjoint operator, $B$, is a
\textit{beable} for $\rho$ if it commutes with $\rho$ and if all
spectral subspaces of $\rho$ belonging to non-zero eigenvalues of $\rho$
are also spectral subspaces of $B$.  In other words, a beable for $\rho$
is a self-adjoint operator which commutes with $\rho$ and, on the
orthogonal complement of the  kernel of $\rho$, is a function of $\rho$.
We then say that an eigenvalue of $B$ on a given event (i.e. spectral
subspace) of $\rho$ is called the \textit{value} of the beable $B$ when
that event occurs.

\smallskip

With this definition, it is obvious  that, for a given $\rho$, a similar
coincidence to that remarked upon after equation (\ref{paritycoincidence})
will occur quite generally between the expected value of a
beable, $B$, in the sense of \textit{the sum over all possible events of
the probability of each event which can occur multiplied by the value of
the beable $B$ when that event occurs} and the expectation value of $B$
in the conventional sense of $\tr(\rho B)$.

\smallskip

We have taken the word `beable' from the writings of John Bell -- see
e.g. \cite{Bellink} where Bell writes:

\smallskip

``\textit{The beables of the theory are those elements which might
correspond to elements of reality, to things which exist.  Their
existence does not depend on `observation'.  Indeed observation and
observers must be made out of beables.}''

\smallskip

However, we should caution that the candidate for a beable being
proposed here differs from various candidates that Bell himself proposed
(see e.g. Note $<$\ref{Note:Bellink}$>$) and, in particular, we would
remark that, in contrast to other proposals for beables, in our notion,
what is a beable at a given time depends on the state of the system at
that time.

\smallskip

A couple of interesting features of our notion of beable are worth
pointing out.   First if, in the case of an $n$-dimensional Hilbert
space, $\rho$ is $1/n$ times the identity operator then the only beables
will be multiples of the identity operator. This applies, for example to
the Schr\"odinger Cat state (\ref{decocatintro}) when $|c_1|^2={1\over
2}=|c_2|^2$. (See also the discussion of this special situation in
Footnote $<$\ref{Note:catpuzzle}$>$.) At the other extreme, if $\rho$
arises as the projector  $|\psi\rangle\langle\psi|$ onto some vector
$\psi$, then any self-adjoint operator which commutes with $\rho$ is a
beable.

\end{enumerate}

\end{document}